\documentclass[aps,prl,showpacs,preprintnumbers,twocolumn,superscriptaddress,nofootinbib]{revtex4-2}
\usepackage{amsmath,amssymb}
\usepackage{bm}
\usepackage{tipa}
\usepackage{upgreek}
\usepackage{comment}
\usepackage{mathrsfs}
\usepackage{graphicx}
\usepackage{graphics}
\usepackage{braket}
\usepackage{enumitem}
\usepackage{mathbbol}
\usepackage{booktabs}
\usepackage{gensymb}
\usepackage{bbm}
\usepackage[normalem]{ulem}
\usepackage{color}
\usepackage{array}
\usepackage[colorlinks,bookmarks=true,citecolor=blue,linkcolor=red,urlcolor=blue]{hyperref}

\usepackage{pifont}

\usepackage{dsfont}
\usepackage{longtable}
\usepackage{float}

\newcommand{\diagram}[2]{\vcenter{\hbox{\includegraphics[scale=0.45,page=#2]{./#1.pdf}}}}

\def \k{{\mathbf k}}

\def \x{{\mathbf x}}
\def \e{{\mathbf e}}
\def \0{{\mathbf 0}}

\begin{document}

    \title{Finite-Entanglement Scaling of 2D Metals}
    
    \author{Quinten Mortier$^*$}
    \affiliation{Department of Physics, Ghent University, Krijgslaan 281, 9000 Gent, Belgium}
    
    \author{Ming-Hao Li$^*$}
    \affiliation{Rudolf Peierls Centre for Theoretical Physics, University of Oxford, Parks Road, Oxford, OX1 3PU, United Kingdom}

    \author{Jutho Haegeman}
    \affiliation{Department of Physics, Ghent University, Krijgslaan 281, 9000 Gent, Belgium}

    \author{Nick Bultinck}
    \affiliation{Department of Physics, Ghent University, Krijgslaan 281, 9000 Gent, Belgium}
    \affiliation{Rudolf Peierls Centre for Theoretical Physics, University of Oxford, Parks Road, Oxford, OX1 3PU, United Kingdom}
	
    \begin{abstract}
        We extend the study of finite-entanglement scaling from one-dimensional gapless models to two-dimensional systems with a Fermi surface. In particular, we show that the entanglement entropy of a contractible spatial region with linear size $L$ scales as $S\sim L\log[\xi f(L/\xi)]$ in the optimal tensor network, and hence area-law entangled, state approximation to a metallic state, where $f(x)$ is a scaling function which depends on the shape of the Fermi surface and $\xi$ is a finite correlation length induced by the restricted entanglement. Crucially, the scaling regime can be realized with numerically tractable bond dimensions. We also discuss the implications of the Lieb-Schultz-Mattis theorem at fractional filling for tensor network state approximations of metallic states.
    \end{abstract}
    	
    \maketitle	

\def\thefootnote{*}\footnotetext{These authors contributed equally to this work.}
\def\thefootnote{\arabic{footnote}}

    \textit{Introduction}.---In the last two decades it has become increasingly clear that ground states of local lattice Hamiltonians have an interesting and rich entanglement structure. For example, generic ground states of \emph{gapped} local Hamiltonians are observed to satisfy the so-called ``area law'' for the entanglement entropy of a subregion, meaning that the leading term in the entanglement entropy scales with the area of the boundary separating the subregion from the rest of the system (in 1D, the area law has been proven; see Ref.~\cite{Hastings2007}). Furthermore, subleading corrections to the area law contain universal information about the topological phase of the system \cite{Kitaev2006,Levin2006}. Related quantities, such as the entanglement spectrum \cite{Li2008} and multipartite entanglement measures \cite{Zou2021,Siva2022,Liu2022,Liu2023,Kim2022,Tam2022,Sohal2023} have also been shown to reveal topological information.
	
    In the case of gapless local Hamiltonians, the area law is frequently violated. Two notable examples are gapless 1D systems with long-wavelength properties that can be described by conformal field theory (CFT), and systems in higher dimensions with a Fermi surface. In the former case, the entanglement of a subregion with linear size $L$ scales as $S \sim c/3\log L$, with $c$ the central charge of the CFT \cite{Holzhey1994,Vidal2003,Calabrese2004}, whereas $S\sim L^{d-1}\log L$ in a $d$-dimensional system exhibiting a codimension 1 Fermi surface \cite{Wolf2006,Gioev2006,Swingle2010,Ding2012}.
	
    The entanglement structure of ground states has direct practical consequences for classical simulations of quantum systems. The existence of an area law is both a necessary condition as well as a strong motivation to represent the ground state as a tensor network state (TNS) \cite{Fannes1992,VerstraeteCirac2004,Verstraete2004,Cirac2021}. TNSs are compressed representations of quantum states in terms of local tensors which can be stored and manipulated efficiently by classical computers, and therefore present a useful variational space for numerical studies. For systems violating the area law, the theory of ``finite-entanglement scaling'' \cite{Tagliacozzo2008,Pollmann2009,Pirvu2012,Stojevic2015,Sherman2023,Ueda2023} describes how an area-law state (i.e., a TNS) best approximates the non-area-law ground state in the thermodynamic limit. A remarkable result from finite-entanglement scaling is that for 1D critical ground states described by a CFT, the restricted entanglement induced by the finite TNS bond dimension $D$ (i.e., the dimension of the contracted indices of the tensors) results in a finite correlation length $\xi \propto D^\kappa$, where $\kappa$ is a universal number determined by the central charge of the CFT \cite{Tagliacozzo2008,Pollmann2009}. As a result, the entanglement entropy of a region of length $L$ can be expressed in terms of a scaling function which depends only on the ratio $L/D^\kappa$ \cite{Calabrese2009,Tagliacozzo2008,Pollmann2009}.  
	
    In this Letter we extend the finite-entanglement scaling analysis to two dimensions and show that a similar scaling collapse is possible for the entanglement entropy of metals, i.e., states with a Fermi surface, despite the fact that there is no underlying CFT describing the long-wavelength physics. Our results show that by exploiting the scaling collapse, moderate (and numerically tractable) bond dimensions already give rise to a sufficiently high numerical accuracy to reliably access all information about the Fermi surface that is contained in the Widom formula, and hence apply the general approach of ``entanglement spectroscopy'' to metallic systems. Some interesting previous works have studied finite-correlation-length scaling for 2D TNS \cite{Corboz2018,Rader2018,Czarnik2019,Vanhecke2022}, but these works did not consider area-law-violating ground states.	
	
    \begin{figure}
        \includegraphics[scale=0.5]{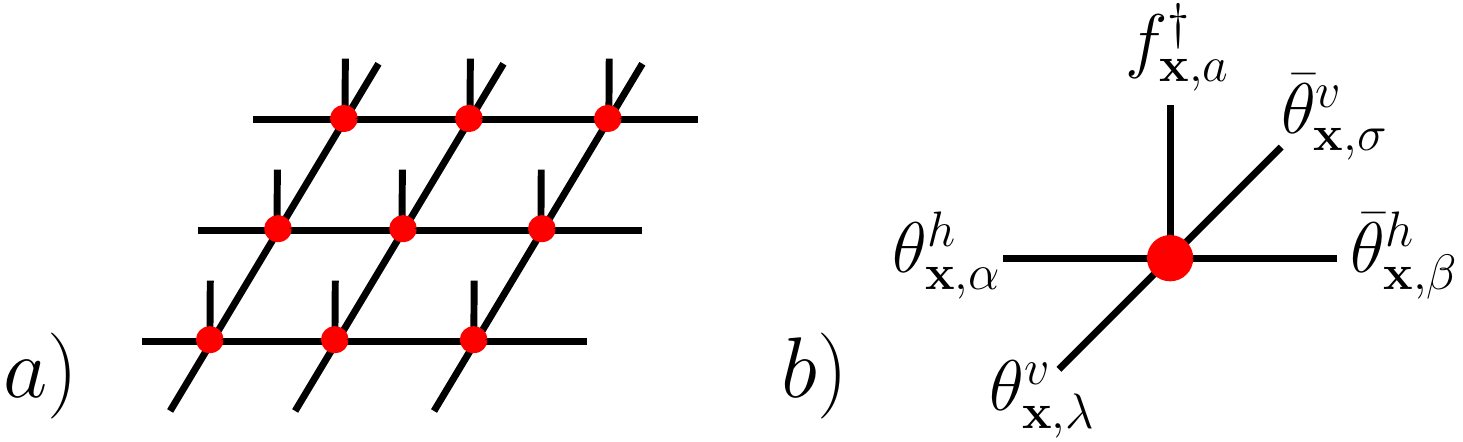}
        \caption{(a) A 2D tensor network on a $3\times 3$ square lattice. (b) The assignment of physical creation operators $f^\dagger_{\x}$ to the physical index, and virtual Grassmann variables $\theta_{\x},\bar{\theta}_\x$ to the virtual indices, of a Gaussian fermionic tensor.}\label{fig:tensors}
    \end{figure}
	
    \textit{Gaussian fermionic TNS}.---For concreteness we will perform our analysis on a square lattice. Since we are interested in 2D states with a Fermi surface we use fermionic projected entangled-pair states (PEPS) \cite{Kraus2010,Barthel2009,Corboz2010,Gu2010,Gu2013,Bultinck2018}. In particular, we will be working with \emph{Gaussian} fermionic tensors, which produce either a Slater determinant or a Bardeen-Cooper-Schrieffer (BCS) pairing (or pfaffian) state after contraction of the virtual indices.  To define the Gaussian tensors, we assign fermion creation operators $f^\dagger_{\x,a}$ ($a \in \{ 1,\dots, N\}$) to the physical index of the tensor at site $\x$, and Grassmann variables $\theta^ h_{\x,\alpha},\bar{\theta}^h_{\x,\beta},\theta^v_{\x,\lambda},\bar{\theta}^v_{\x,\sigma}$ ($\alpha,\beta,\lambda,\sigma \in\{ 1,\dots, M\}$) to the virtual indices, as in Fig.~\ref{fig:tensors}. The Grassmann variables square to zero, are mutually anticommuting, and also anticommute with the creation operators $f^\dagger_{\x,a}$. The Gaussian tensor at site $\x$ is defined as
    \begin{equation}\label{Tx}
	\hat{T}_\x = \exp\left(\frac{1}{2}\chi^T_\x A \chi_\x\right)\,,
    \end{equation}
    where the column vector $\chi_\x \equiv (f^{\dagger}_{\x},\theta^{h}_{\x},\bar{\theta}^{h}_{\x},\theta^{v}_{\x},\bar{\theta}^{v}_{\x})$ collects the $N$ creation operators and $4M$ Grassmann variables assigned to site $\x$ and the antisymmetric matrix $A \in \mathbbm{C}^{(N+4M)\times(N+4M)}$ contains the variational parameters. Note that $A$ is independent of $\x$, which means that we are restricting ourselves to translation-invariant (TI) states. With the definition of the tensors in place, we can now define the (unnormalized) contracted Gaussian fermionic TNS (GfTNS) via the Berezin integral \cite{berezin1966}
    \begin{equation}\label{contraction}
        |\psi\rangle = \int [D\theta] \int [D\bar{\theta}] \prod_{\x} e^{\bar{\theta}_{\x}^{h^T}\theta_{\x+\e_x}^h }e^{\bar{\theta}_{\x}^{v^T}\theta_{\x+\e_y}^v }\hat{T}_{\x}|0\rangle\,,
    \end{equation}
    where $|0\rangle$ is the physical Fock vacuum and $\e_{x/y}$ are unit vectors along the $x/y$-direction. Every Grassmann variable spans a two-dimensional super vector space, so the bond dimension of the GfTNS is $D= 2^M$.	
	
    Because we are considering TI states, the Gaussian Grassmann integral in Eq.\,\eqref{contraction} can be further simplified by going to momentum space. Working with a large but finite system of $N_s = N_xN_y$ sites and (anti-)periodic boundary conditions while defining $\chi_\k = \frac{1}{\sqrt{N_s}}\sum_\x e^{i\k\cdot \x}\chi_\x$, we can write
    \begin{displaymath}
        |\psi\rangle = \int [D\theta] \int [D\bar{\theta}] \exp\left(\frac{1}{2}\sum_\k \chi_{-\k}^T[A+M(\k)]\chi_\k\right)|0\rangle\,.
    \end{displaymath}
    Here, $M(\k)$ is defined as $M(\k) = \0_N\oplus \tilde{M}(\k)$, with $\0_N$ a $N\times N$ zero matrix, and
    \begin{equation}
        \tilde{M}(\k) = \left(\begin{matrix} \0_M & -e^{ik_x}\mathds{1}_M \\ e^{-ik_x}\mathds{1}_M & \0_M \end{matrix}\right)\oplus \left(\begin{matrix} \0_M & -e^{ik_y}\mathds{1}_M \\ e^{-ik_y}\mathds{1}_M & \0_M \end{matrix}\right)\,.
        \label{Mtilde}
    \end{equation}
    Writing $A = \left(\begin{matrix} B & -C^T \\ C & D \end{matrix}\right)$, with $N\times N$ submatrix $B$, $4M\times N$ submatrix $C$, and $4M\times 4M $ submatrix D, we finally obtain
    \begin{equation}\label{BCS}
        |\psi\rangle \propto \mathrm{e}^{\frac{1}{2}\sum_\k f^\dagger_{-\k}(B + C^T[D+\tilde{M}(\k)]^{-1}C)f^\dagger_{\k} }|0\rangle\, .
    \end{equation}
    Here, we have assumed that $D+\tilde{M}(\k)$ is non-degenerate at every $\k$ and refer to Ref.~\cite{Supplement} for the degenerate case.
	
    The construction of GfTNS as presented here was introduced in Ref. \cite{Gu2010}, and has been used in previous studies \cite{Beri2011,Gu2013,Dubail2015,Yin2019}. There also exists an alternative formulation in terms of density matrices \cite{Kraus2010}. For our results presented below we have used both formalisms, each of which has different practical advantages. However, the two formalisms are ultimately equivalent and can be translated into each other \cite{Supplement}.
	
    The state in Eq.\,\eqref{BCS} takes the form of a general BCS pairing state. Given that we set out to study states with a Fermi surface, the reader might worry that we are using TNS which contract to pairing states. The reason for this is simply that a finite-$D$ Gaussian fTNS with explicit charge conservation symmetry always has an integer particle number at every momentum which is constant throughout the Brillouin zone, and therefore cannot represent or even closely approximate a state with a Fermi surface. This is not an embarrassing shortcoming of GfTNS, but a direct consequence of the Lieb-Schultz-Mattis theorem, which states that one cannot have a trivial insulator at noninteger fillings \cite{Lieb1961,Affleck1988,Oshikawa2000,Hastings2004}. In particular, in Ref.~\cite{BultinckCheng2018}, it was shown that if a general, explicitly TI and U(1)-symmetric fTNS is forced to have a filling $\nu = p/q$, with $p$ and $q>1$ coprime integers, then the tensors necessarily have a purely virtual $\mathbbm{Z}_{2q}$ symmetry. The entanglement entropy (EE) in a generic tensor network state with such virtual symmetry scales as $S = \alpha L - \ln 2q + \mathcal{O}(L^{-1})$, which implies that the fTNS has nontrivial topological order \cite{Schuch2010,Bultinck2017,Molnar2022,RuizdeAlarcon2022}. So the incompatibility of Gaussianity and explicit U(1) symmetry at fractional filling $\nu = p/q$ for fTNS is a manifestation of the simple fact that Slater determinants cannot represent states with nontrivial topological order. The only way for a TI GfTNS to introduce finite entanglement in a metallic state is therefore to open a small superconducting gap at the Fermi surface.
	
    \textit{Spinless fermions}.---We first consider the case with $N=1$, i.e., spinless fermions with a single orbital per site. To obtain the GfTNS, we minimize the energy of the following simple hopping Hamiltonian:
    \begin{equation}\label{Hamiltonian}
        H = -t \sum_{\langle ij\rangle} f^\dagger_i f_j -t' \sum_{\langle\langle ij\rangle\rangle} f^\dagger_i f_j + h.c. - \mu\sum_i f^\dagger_i f_i\,,
    \end{equation}
    where the first (second) sum is over nearest (next-nearest) neighbours. We choose the chemical potential $\mu$ such that there is a single electron pocket centered at the $\Gamma$ point. With periodic boundary conditions, the total number of electrons in the spinless Fermi sea is odd for every system size. If a state at momentum $\k$ is occupied, then so is the state at $-\k$. So the electrons appear in pairs, except at the time-reversal invariant momenta (TRIM). Here, the only TRIM which is occupied is the center of Brillouin zone $\k = 0$; hence the overall fermion parity is odd. The tensors defined in Eq.\,\eqref{Tx} have even fermion parity, and therefore the GfTNS also necessarily has even parity (for every system size). This is reflected in the fact that the wave function in Eq.\,\eqref{BCS} with $N=1$ always leaves the states at the TRIM empty. It is possible to fix this discrepancy by inserting an additional Grassmann variable ``on the virtual level'' in Eq.\,\eqref{contraction}, which makes the fTNS have odd parity \cite{Supplement} (this Grassmann variable is identical to the string operators that have appeared in fTNS constructions \cite{Wahl2014} of the $p_x+ip_y$ superconductor \cite{Read2000}). Here, however, we will use a simpler way to sidestep this issue and work with antiperiodic boundary conditions such that the spinless Fermi surface state always has even fermion parity.
	
    \begin{figure}
        \centering
        \includegraphics[scale=0.45]{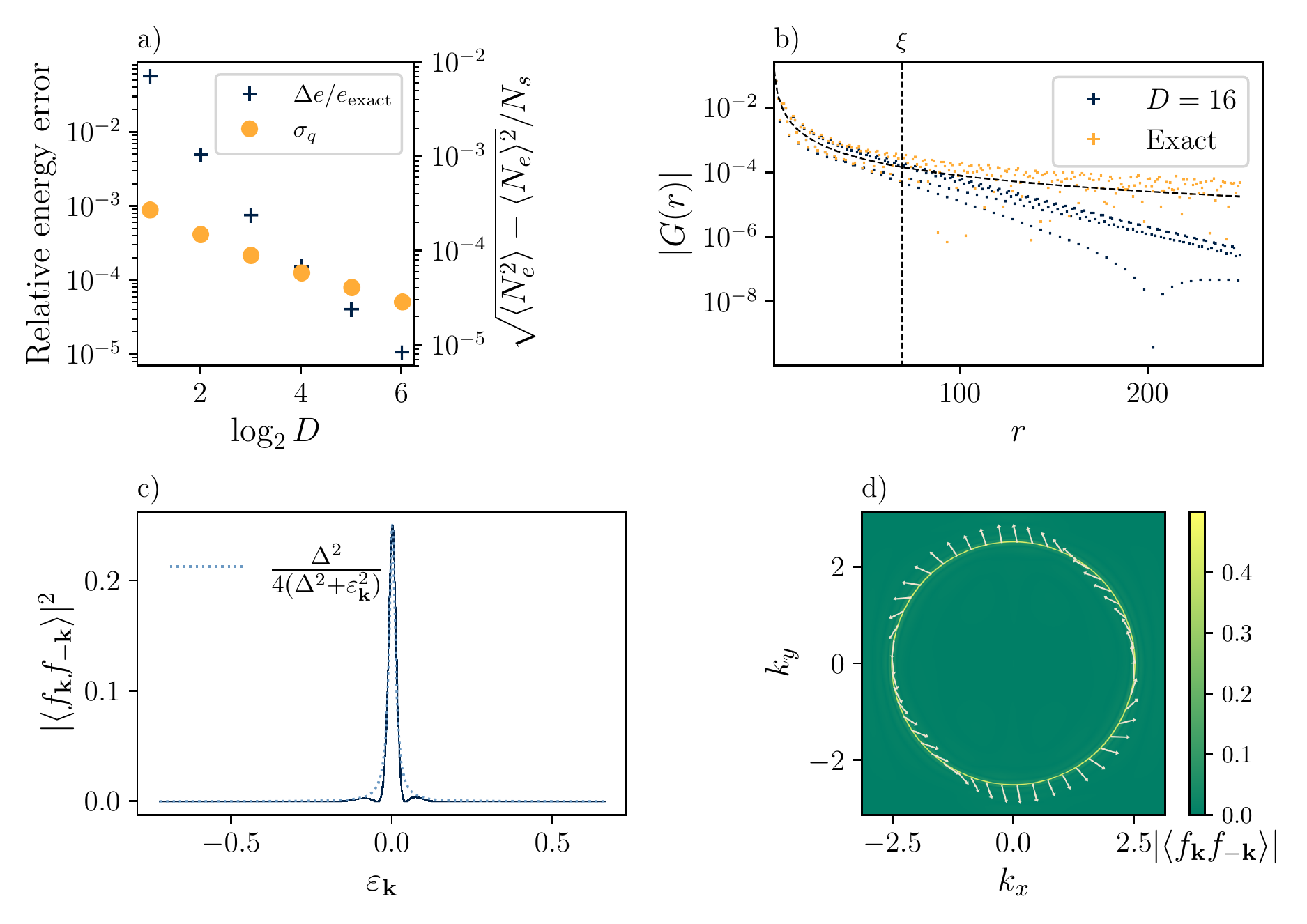}
        \caption{Results for spinless fermions with $t'/t = 0.353$ at half filling of $N_s = 999^2$ sites. (a) Relative difference in energy per site (denoted as $\Delta e/ e_{\mathrm{exact}}$) between the exact ground state and the optimized GfTNS as a function of bond dimension $D$; and the standard deviation $\sigma_{q}$ of the particle number per site of the optimized GfTNS.  (b) $|\langle f_{r} f_{0}\rangle|$ for $D = 16$ GfTNS vs the exact ground state. Near $\xi$, which is extracted from the EE, $|G(r)|$ for GfTNS starts to decay much faster than the power law behaviour for that of the exact ground state. (c) $|\langle f_\k f_{-\k}\rangle|^2$ at $D = 32$ along a radial direction in the Brillouin zone as a function of the single-particle energy $\varepsilon_\k$ of $H$ (Eq. \eqref{Hamiltonian}). (d) $\langle f_\k f_{-\k}\rangle$ at $D=32$ throughout the Brillouin zone. The color map denotes the magnitude, the arrows the complex phase.}
        \label{fig:PEPS_spinless}
    \end{figure}
 
    We choose $\mu$ to fix the total particle number $N_e$ at half filling, i.e., $\nu = N_e/N_s = 1/2$, and use $t'/t = 0.353$ to realise an almost circular Fermi surface. We have optimized the GfTNS to minimize its energy $\langle H\rangle$, at different bond dimensions $D$ (see Ref.~\cite{Supplement} for details of the numerical simulations). Fig.~\ref{fig:PEPS_spinless}(a) depicts the difference in energy of the optimal GfTNS compared the exact result, as well as the standard deviation of the total particle number per site, i.e., $\sigma_{q}\equiv \sqrt{\langle N_e^2\rangle - \langle N_e\rangle^2}/N_s$. The latter is a quantifier for the charge conservation symmetry breaking in the GfTNS. We see that both the energy error and $\sigma_q$ decrease as a function of $D$, indicating---similarly as Ref.~\cite{Mortier2022}---that the optimised GfTNS provides an approximation to the exact metallic ground state which improves systematically with bond dimension. Fig.~\ref{fig:PEPS_spinless}(b) shows the correlation function $G(\mathbf{r})=\langle f^\dagger_{\mathbf{r}+\mathbf{r}'} f_{\mathbf{r}'}\rangle$ of the $D=16$ GfTNS, which agrees with the exact result for $|\mathbf{r}|\lesssim \xi \approx 70$. Note that $\xi \ll N_x=N_y$, such that we can take our results to be representative of the thermodynamic limit. In Fig.~\ref{fig:PEPS_spinless}(c) and \ref{fig:PEPS_spinless}(d) we plot the pairing function $\langle f_\k f_{-\k}\rangle$ for $D = 32$, both along a radial cut, and throughout the entire Brillouin zone. $|\langle f_\k f_{-\k}\rangle|^2$ is peaked at the Fermi surface, and can be approximated by the BCS expression $\Delta^2/4(\Delta^2 + \varepsilon_\k^2)$, where $\varepsilon_\k$ is the single-particle dispersion of $H$. Fig.~\ref{fig:PEPS_spinless}(d) illustrates that the phase of $\langle f_\k f_{-\k}\rangle$ winds by $2\pi$ along the Fermi surface, i.e., it is a $p_x+ip_y$ gap. We explain how the GfTNS deals with the chiral topology of the weak-pairing $p_x+ip_y$ superconductor \cite{Read2000,Wahl2013,Dubail2015} in the Supplemental Material~\cite{Supplement}.
	
    \begin{figure}
        \centering
        \includegraphics[scale=0.45]{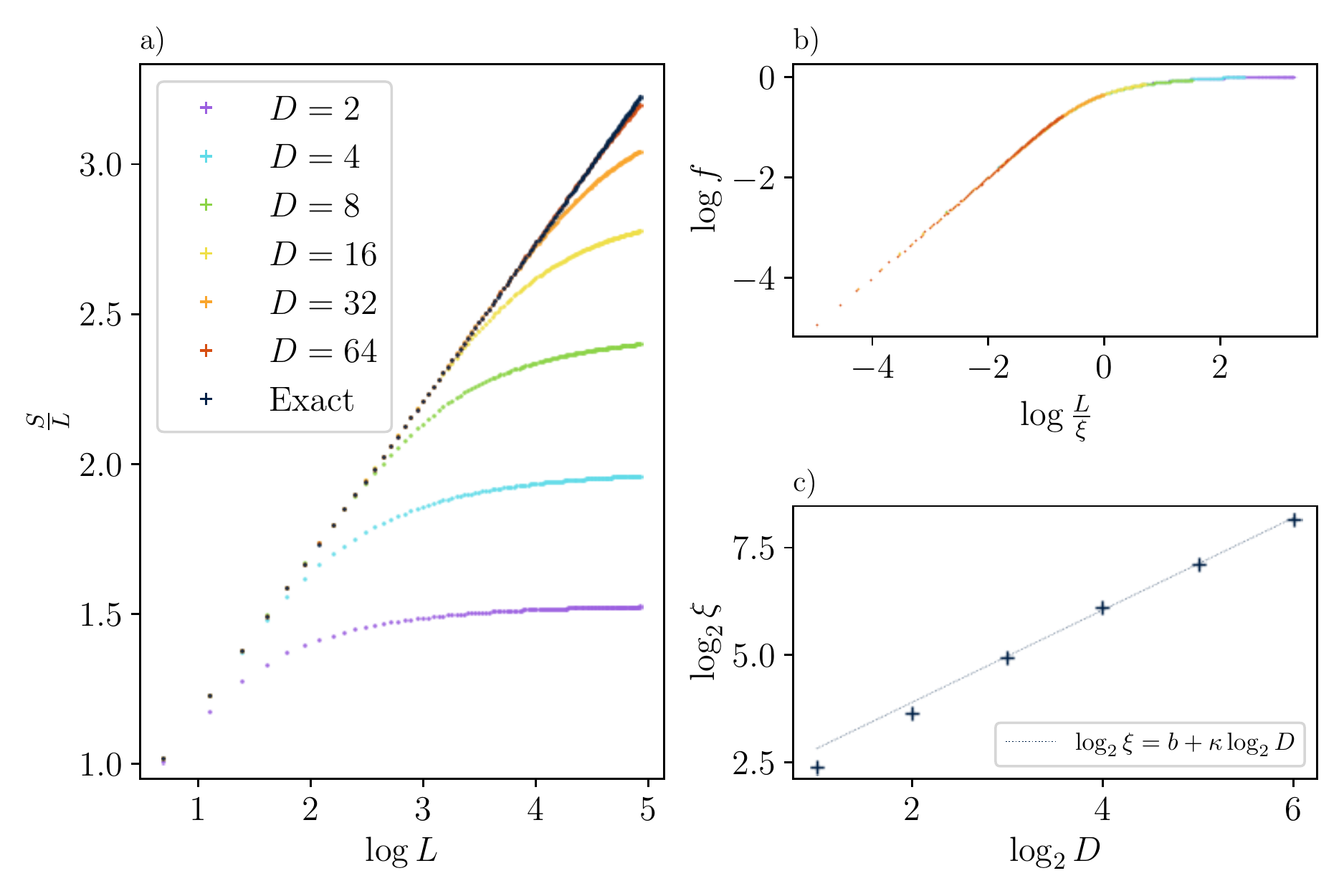}
        \caption{(a) Scaling collapse of the entanglement entropy $S$ of a $L\times L$ square region in the optimized GfTNS for spinless fermions at half filling with $t'/t = 0.353$ and $N_s = 999^2$, obtained at different bond dimensions $D$. (b) Collapse of the GfTNS entanglement entropies using the scaling law of Eq.\,\eqref{scalinglaw}. (c) Linear fit of the correlation length $\xi$ as a function of $D$ obtained from the scaling collapse of $S$, with $\kappa= 1.074$.}
        \label{fig:collapse_spinless}
    \end{figure}
	
    The leading term in the EE of a square $L\times L$ spatial region $R$ in a 2D state with a single spinless Fermi surface is given by	
    \begin{equation}
        \label{Sgen}
        S = \frac{ \log (\Lambda L)}{24\pi} \oint_{\partial R}\oint_{\text{FS}} |\mathrm{d}S_\x\cdot \mathrm{d}S_\k|\, , 
    \end{equation}
    where $\Lambda$ is a nonuniversal inverse length scale, and the integrals are over the boundary of $R$ and over the Fermi surface, and $\mathrm{d}S_\x$ ($\mathrm{d}S_\k$) is a surface element of $\partial R$ (the Fermi surface) \cite{Gioev2006}. For the special case of a circular Fermi surface with radius $k_F$, this general expression evaluates to	$S_{\text{circ}} = \frac{2k_F L}{3\pi}\log \Lambda L$. 
    
    In Fig.~\ref{fig:collapse_spinless}(a), we plot the EE $S$ as a function of $L$, directly calculated from the correlation matrix of the optimized GfTNS at different $D$. This plot shows our main result, which is that the leading contribution to the EE at finite $D$ can be written as	
     \begin{equation}
        \label{scalinglaw}
        S_{\text{fTNS}} = \log\left(\Lambda \xi f(L/\xi) \right) \times \frac{1}{24\pi} \oint_{\partial R}\oint_{FS} |\mathrm{d}S_\x\cdot \mathrm{d}S_\k|\, , 
    \end{equation}
    where $\xi$ is the finite-bond-dimension-induced correlation length, and $f(x)$ is a scaling function which satisfies $f(x\ll 1) \sim x$ and $f(x\gg 1) = $ constant. Fig.~\ref{fig:collapse_spinless}(a) shows how the optimized GfTNS at different $D$ approximate the $L\log L$ scaling of the EE, while Fig.~\ref{fig:collapse_spinless}(b) directly plots the scaling function $f(x)$ onto which the numerical data obtained at different $D$ can be collapsed. Note that to obtain the scaling collapse we have only one tuning parameter $\xi$ if we require that the GfTNS results agree with the exact result at small $L$. The length scale $\xi$ obtained from the EE of the $D=16$ GfTNS ~\cite{Supplement} is indicated as the vertical dashed line in the plot of $G(\mathbf{r})$ in Fig.~\ref{fig:PEPS_spinless}(b). This shows that $\xi$ agrees with the physical correlation length, i.e., the length scale at which the exponential decay of correlations in the GfTNS sets in. Finally, Fig.~\ref{fig:collapse_spinless}(c) confirms that $\xi$ increases monotonically as a function of $D$. For the moderate bond dimensions used in this Letter, $\xi$ seems to follow a power law as a function of $D$. However, based on both analytical \cite{FrancoRubio2022} and numerical \cite{Supplement} results in 1D, which show that Gaussian fermionic matrix product states (GfMPS) cannot reproduce the power-law scaling $\xi \propto D^\kappa$ of generic MPS, we anticipate that for GfTNS, deviations from the power-law relation between $\xi$ and $D$ could occur at higher $D$. For general (i.e., non-Gaussian) fTNS, we nevertheless conjecture that $\xi \propto D^\kappa$.
    
    \textit{Spinful fermions}.---Next, we consider the same Hamiltonian as in Eq.\,\eqref{Hamiltonian}, but now for spinful fermions ($N=2$) created by $f_{\mathbf{x},\sigma}^\dagger$ with $\sigma= \uparrow,\downarrow$. An important difference between the spinless and spinful models lies in the nature of the superconducting gap of the optimal GfTNS approximation. In particular, as we explicitly impose SU$(2)$ spin symmetry on the GfTNS, the superconducting gap will be spin singlet and hence even under inversion. 
    
    We now verify whether the finite-entanglement scaling law (Eq.\,\eqref{scalinglaw}) also holds for the spinful model. In doing so, we have used the density-matrix-based method for GfTNS~\cite{Kraus2010,Supplement}, and a numerical optimization which relies on minimizing the Frobenius norm of the difference between the exact single-particle density matrix and the GfTNS density matrix (see Ref.~\cite{Supplement} for details). Note that the virtual fermion degrees of freedom now carry spin-$1/2$, which means that the bond dimension $D$ is restricted to occur in powers of $4$.	
		
    \begin{figure}
        \centering
        \includegraphics[scale=0.33]{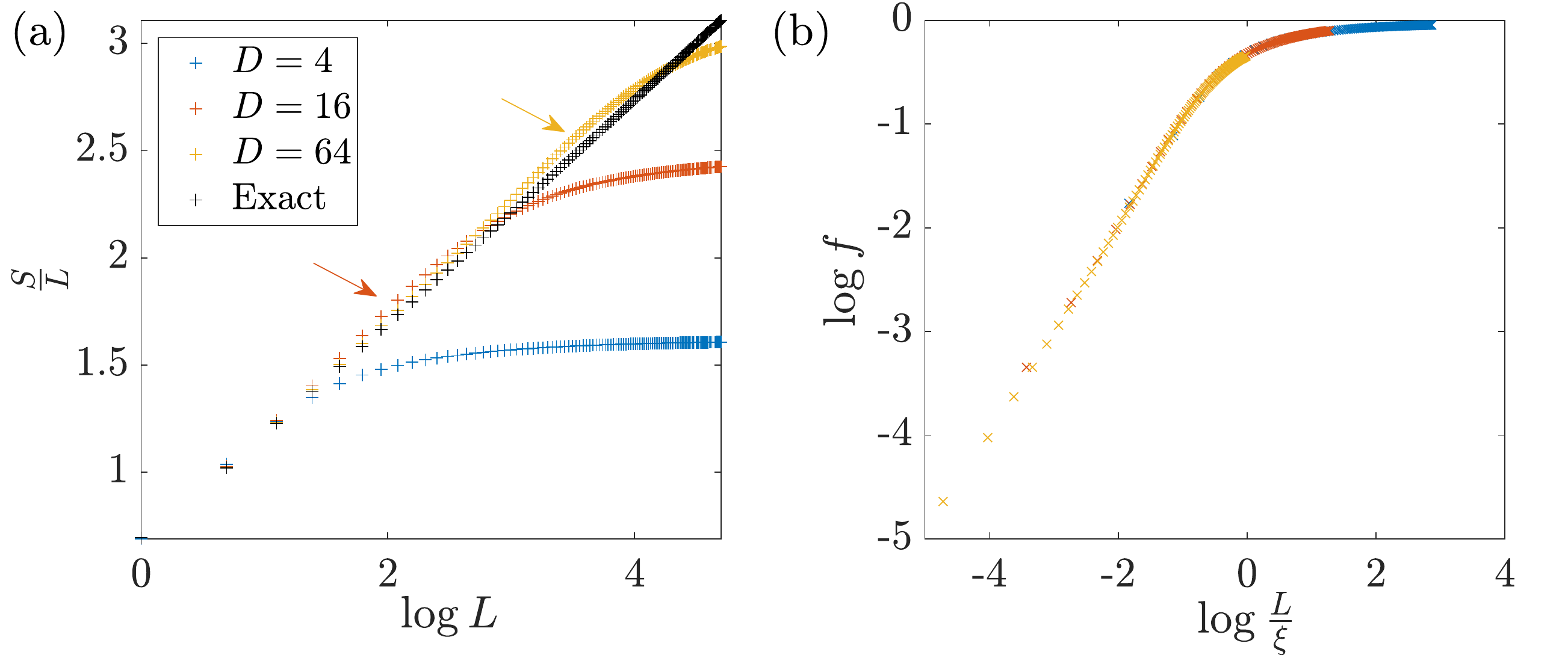}
        \caption{(a) Entanglement entropies for a $L\times L$ square subsystem in the exact ground state and its GfTNS approximations with the indicated bond dimensions. Up to a bulk correlation length $\xi$ the exact profile is reproduced. After the transition region with a characteristic `bump' (marked by the arrows), the GfTNS profiles saturate. (b) Collapse of the GfTNS entanglement entropies using the scaling law of Eq.\,\eqref{scalinglaw}.}
        \label{fig:ent_spinful}
    \end{figure}
    
    As for the spinless model, we have computed the EE of a $L \times L$ region $R$ for the spinful Fermi surface model and its GfTNS approximations. The spinful results displayed in Fig.~\ref{fig:ent_spinful} show a similar behavior as the spinless results in Fig.~\ref{fig:collapse_spinless}, with some minor differences. In particular, right before the EE reaches the area-law regime (signalled by the plateau in Fig.~\ref{fig:ent_spinful}), we discern a small ``bump'' where $S_\text{fTNS}$ rises slightly above the exact value for $S$ (indicated by the arrows in Fig.~\ref{fig:ent_spinful}). We attribute this to the correlations between the different spin flavors induced by the singlet pairing. Another difference are the generally lower $\xi$ values for the same bond dimension. This is a consequence of the increased local Hilbert space dimension. Also, the rate at which $\xi$ increases with $D$ is lower than in the spinless case, which is reminiscent of the 1D case, where an increase in the central charge lowers the exponent $\kappa$ in $\xi \propto D^\kappa$. Besides these minor differences, Fig.~\ref{fig:ent_spinful}(b) confirms our main result, which is that the EE at different $D$ can be collapsed using the scaling law in Eq.\,\eqref{scalinglaw}.
	
    \textit{Properties of the scaling function}.---Similarly to the scaling functions of gapless systems with conformal symmetry in the IR, $f(x)$ is expected to be insensitive to lattice-scale details. We also expect that $f(x)$ will depend on the shape of the Fermi surface, in analogy to the finite-temperature entropy scaling functions for Fermi liquids \cite{swingle2013}. To verify the first expectation we have performed numerical data collapses of the EE obtained at different fillings, while keeping the FS approximately circular. These results~\cite{Supplement}, confirm that the EE at different fillings can indeed be collapsed on the same curve. By tuning away from $t'/t=0.353$ in either direction, which changes the FS to being either more diamondlike or more squarelike, we observe that the results collapse on different scaling functions, thus confirming the dependence of $f(x)$ on the FS geometry \cite{Supplement}.

    \textit{Conclusions}.---We have shown that the theory of finite-entanglement scaling can be generalized from 1D gapless systems to 2D states with a Fermi surface. Our main result is that $S_{\text{fTNS}}(L,D)$, the EE of a $L\times L$ spatial region in the optimal bond-dimension-$D$ fTNS approximation of a metallic state, can be written as $S_{\text{fTNS}}(L,D) \sim L \log(\xi_D f(L/\xi_D)$, where $\xi_D$ is a finite infrared length scale which results from the area-law structure (and thus the finite bond dimension $D$) of the tensor network state, and $f(x)$ is a scaling function which depends on the shape of the Fermi surface, but not on the length scale $k_F^{-1}$, with $k_F$ the Fermi momentum. 

    Fermionic tensor networks are being used in a variety of different ways, e.g., for numerical studies of lattice gauge theories \cite{Zohar2015,Emonts2020,Emonts2023}, as numerically tractable Gutzwiller-projected states \cite{Wu2020,Jin2020,Petrica2021,Jin2021,Li2022,Yang2022,Jin2022}, as a tool for large-scale mean-field calculations \cite{Fishman2015,Schuch2019}, as trial states for topological phases \cite{Bultinck2016,Bultinck2017,Wille2017,Wahl2013,Dubail2015,Hackenbroich2020,Shukla2020}, and as a general class of variational states for numerical simulations of strongly interacting systems \cite{Corboz2010,Wang2014,Zheng2017,Dai2022,ma2023,xu2023}. The results presented in this work show how these applications of fTNS can be extended to metallic states. In particular, being able to perform a scaling collapse for the EE provides solid numerical evidence for the existence of a Fermi surface, and hence can be used to numerically determine whether the ground state of Hubbard-type models (e.g., those obtained from the flat bands of twisted transition-metal dichalcogenides) are metallic or insulating. Similarly, a scaling collapse for the EE can be used to determine whether a frustrated spin model has a spin liquid ground state with a spinon Fermi surface. Furthermore, the scaling collapse significantly enhances the accuracy of the numerically obtained Widom prefactor, and hence provides more reliable information about the fermiology of the metallic ground states.
    
    \textit{Acknowledgments}.~--- 
We acknowledge useful discussions with Hong-Hao Tu, Mike Zaletel, Roger Mong, Ignacio Cirac, Siddharth Parameswaran and Steven R. White. We also thank Siddharth Parameswaran for helpful feedback on an earlier version of the manuscript.
QM and JH have received support from the European Research Council (ERC) under the European Union’s Horizon 2020 program [Grant Agreement No.\ 715861 (ERQUAF)] and from the Research Foundation Flanders. NB was supported by a University Research Fellowship of the Royal Society. ML has received support from the European Research Council under the European Union Horizon 2020 Research and Innovation Programme, through Starting Grant [Agreement No.\ 804213-TMCS], and from a Buckee Scholarship at Merton College.

    \bibliography{bib}
    
    \clearpage
    \newpage
    
    \appendix
    \onecolumngrid
    \begin{center}
		\textbf{\large --- Supplemental Material ---\\ Finite-entanglement scaling of 2D metals}\\
		\medskip
		Quinten Mortier,$^1$ Ming-Hao Li,$^2$ Jutho Haegeman,$^1$ and Nick Bultinck$^{1,2}$\\
    \textit{$^1$Department of Physics, Ghent University, Krijgslaan 281, 9000 Gent, Belgium}\\
    \textit{$^{2}$Rudolf Peierls Centre for Theoretical Physics, Parks Road, Oxford, OX1 3PU, UK}
	\end{center}
    \vspace{10pt}
	
	We provide several appendices to support the main text while maintaining its readability.\\
	
	\begin{tabular}{l l}
		\quad \text{Appendix A} & \quad \text{Degenerate GfTNS and odd parity states}\\	
		\quad \text{Appendix B} & \quad \text{Performing the scaling collapse}\\	
		\quad \text{Appendix C} & \quad \text{Gu-Verstraete-Wen formalism for spinless fermions: optimisation method and additional results}\\
		\quad \text{Appendix D} & \quad \text{Kraus-Schuch formalism for (symmetric) GfTNS}\\
		\quad \text{Appendix E} & \quad \text{Kraus-Schuch formalism for spinful fermions: optimisation method and additional results}\\
		\quad \text{Appendix F} & \quad \text{GfTNS in 1D: power law relation between precision and bond dimension cannot be sustained}\\
		\quad \text{Appendix G} & \quad \text{Chirality of approximate GfTNS}\\
		\quad \text{Appendix H} & \quad \text{Connection between GfTNS and general fTNS defined via super vector spaces}\\
		\quad \text{Appendix I} & \quad \text{Connection between the Gu-Verstraete-Wen and Kraus-Schuch formalisms for GfTNS}\\
	\end{tabular}\\[8pt]
	
	Appendix A discusses how degenerate GfTNS realize odd parity states. In Appendix B we give a detailed explanation of how to perform the scaling collapse both with and without knowing the particular Fermiology. Appendices C-E zoom in on the two Gaussian TNS \textit{Ans\"atze} used in this work and provide additional results. In Appendix F and G we describe some intricacies of these GfTNS. Finally, Appendices H and I relate both Gaussian formalisms to each other, and to generic fermionic tensor network states formulated in the super vector space formalism.\\

	\section{Appendix A -- Degenerate GfTNS and odd parity states}
	
	In the main text we constructed GfTNS via a Berezin integral, and we mentioned two (related) subtleties that can arise in this construction: one when the matrix $D+\tilde{M}(\k)$ as defined in Eq. \eqref{BCS} becomes degenerate, and the other when the state has odd fermion parity for every system size. To see how these two issues are related, we start by discussing the illuminating example of Kitaev's Majorana chain \cite{Kitaev2001}. With periodic boundary conditions, the non-trivial Majorana chain has odd fermion parity, and, as we will shortly see, the matrix $D+\tilde{M}(k)$ which appears in the Gaussian fermionic matrix-product state (GfMPS) representation of the Majorana chain is degenerate at $k = 0$.
	
	For simplicity, we will consider the RG fixed point of the non-trivial Majorana chain (corresponding to decoupled Majorana dimers between neighbouring sites \cite{Kitaev2001}), which is known to have an exact $D=2$ fMPS representation \cite{Bultinck2016}. Here we will write this exact fMPS as a GfMPS using the formalism introduced in the main text. As the fMPS has bond dimension two, we need to assign a single Grassmann number to the virtual indices of the tensors at the different sites labeled by $x$ as follows	
	\begin{equation*}
		\includegraphics[scale=0.7]{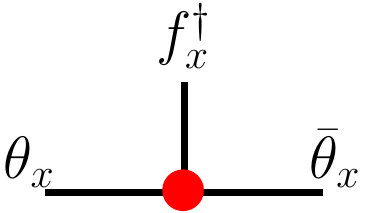} \quad .
	\end{equation*}
	We also assign a single spinless electron creation operator $f^\dagger_x$ to the physical index. Defining $\chi_x = (f^\dagger_x,\theta_x,\bar{\theta}_x)^T$, the Gaussian tensor at site $x$ is given by	
	\begin{equation}
		\hat{T}_x = \exp\left(\frac{1}{2}\chi_x^T A\chi_x\right)\quad \text{with} \quad A = \left(\begin{matrix}
		 0 & 1  & 1 \\
		-1 & 0  & 1 \\
		-1 & -1 & 0\end{matrix}\right) \, .
		\label{degenexp}
	\end{equation}
	Let us now proceed as in the main text and define the unnormalized GfMPS as	
	\begin{equation}
		|\psi\rangle \propto \int [D\theta]\int[D\bar{\theta}] \prod_x e^{\bar{\theta}_{x}\theta_{x+1}} \hat{T}_x|0\rangle \, .
	\end{equation}
	This state has even fermion parity so it cannot be the ground state of the non-trivial Majorana chain with periodic boundary conditions. But let us momentarily proceed with this state, as we will shortly see how to fix this issue.
	
	In momentum space the GfMPS can be written as	
	\begin{equation}
		|\psi\rangle \propto \int [D\theta]\int[D\bar{\theta}] \exp\left[ \sum_k -\left(\theta_{-k} + \bar{\theta}_{-k}\right) f^\dagger_k + \frac{1}{2}\left(\begin{matrix}\theta_{-k} & \bar{\theta}_{-k}\end{matrix} \right) \left(\begin{matrix} 0 & 1 - e^{ik} \\ -1 + e^{-ik} & 0\end{matrix}\right)\left(\begin{matrix} \bar{\theta}_k & \theta_k \end{matrix}\right) \right]|0\rangle \,.
	\end{equation}
	Note that the 2D matrix appearing in the exponent becomes zero at $k=0$. As a result, the momentum-space Grassmann variable $\frac{1}{\sqrt{2}}(\bar{\theta}_0 - \theta_0)$ does not appear in the exponent, and hence the Grassmann integral evaluates to zero. This is a manifestation of the fact that the ground state of the non-trivial Majorana chain needs to have odd fermion parity with periodic boundary conditions -- in Ref.~\cite{Bultinck2016} it was found that when the generic fMPS representation of the Majorana chain is contracted with periodic boundary conditions it evaluates to zero, unless an additional odd matrix (the `$y$ matrix' of Ref.~\cite{Bultinck2016}) is inserted on the virtual level of the fMPS. Here we also have to insert an additional odd object, i.e.~the Grassmann number $\frac{1}{\sqrt{2}}(\bar{\theta}_0 - \theta_0)$, by hand into the Berezin integral in order to obtain a non-vanishing GfTNS. So the correct state is given by	
	\begin{equation}\label{corrected}
		|\psi\rangle \propto \int [D\theta]\int[D\bar{\theta}] (\bar{\theta}_0 - \theta_0) \exp\left[ \sum_k -\left(\theta_{-k} + \bar{\theta}_{-k}\right) f^\dagger_k + \frac{1}{2}\left(\begin{matrix}\theta_{-k} & \bar{\theta}_{-k}\end{matrix} \right) \left(\begin{matrix} 0 & 1 - e^{ik} \\ -1 + e^{-ik} & 0\end{matrix}\right)\left(\begin{matrix} \bar{\theta}_k & \theta_k \end{matrix}\right) \right]|0\rangle \,.
	\end{equation}
	Performing the integral gives	
	\begin{eqnarray}
		|\psi\rangle & \propto & \exp\left[\frac{1}{2}\sum_{k\neq 0} \begin{pmatrix}
			-1 & -1
		\end{pmatrix} \begin{pmatrix}
			0 & 1-e^{ik}\\
			-1+e^{-ik} & 0
		\end{pmatrix}^{-1} \begin{pmatrix}
			-1 \\
			-1
		\end{pmatrix} f^\dagger_{-k}f^\dagger_k  \right] f_0^\dagger |0\rangle \\
		& = & \exp\left[\frac{1}{2}\sum_{k\neq 0} -i \, \text{cotan}\left(\frac{k}{2}\right)f^\dagger_{-k}f^\dagger_{k} \right] f^\dagger_0|0\rangle \, .
	\end{eqnarray}
	After normalizing the state, we finally arrive at
	
	\begin{equation}
		|\psi\rangle = \prod_{k>0}\frac{1}{\sqrt{1+\text{cotan}^2(k/2)}}\left(1 -i \, \text{cotan}\left(\frac{k}{2}\right)f^\dagger_{-k}   f^\dagger_k \right) f^\dagger_0|0\rangle\,,
	\end{equation}
	which is indeed the ground state of the fixed-point Majorana chain Hamiltonian $H = -\sum_x f^\dagger_x f_{x+1} + if^\dagger_x f^\dagger_{x+1} + h.c.$.
	
	Going back to real space, the correct (i.e.~non-zero and parity odd) GfMPS is given by	
	\begin{eqnarray}
		|\psi\rangle & \propto & \sum_{x'}\int [D\theta]\int[D\bar{\theta}] (\bar{\theta}_{x'}-\theta_{x'})\prod_x e^{\bar{\theta}_{x}\theta_{x-1}} \hat{T}_x|0\rangle \\
		& = & \sum_{x'}\int [D\theta]\int[D\bar{\theta}] (\bar{\theta}_{x'}-\theta_{x'+1})\prod_x e^{\bar{\theta}_{x}\theta_{x+1}} \hat{T}_x|0\rangle \, .
	\end{eqnarray}
	This expression is equivalent to	
	\begin{equation}
		|\psi\rangle \propto \int [D\theta]\int[D\bar{\theta}] (\bar{\theta}_{x'}-\theta_{x'+1})\prod_x e^{\bar{\theta}_{x}\theta_{x+1}} \hat{T}_x|0\rangle\, ,
	\end{equation}
	i.e.~we can insert $\bar{\theta}_{x'}-\theta_{x'+1}$ just once on the bond between sites $x'+1$ and $x'$, and the resulting expression is independent of the choice of $x'$. This shows that $\bar{\theta}_{x'}-\theta_{x'+1}$ is exactly the $y$ matrix from Ref.~\cite{Bultinck2016}, which commutes with the (G)fMPS tensors.
	
	Having worked out the Majorana-chain example, we are now naturally lead to the following conclusions for general GfTNS:	
	\begin{itemize}
		\item If the matrix $D+\tilde{M}(\k)$ is not full rank at a particular momentum $\k$, then the state obtained by contracting the Gaussian tensors without additional Grassmann variables is zero, unless the rows of the matrix $C$ (defined in Eq. \eqref{BCS}) span the kernel of $D+\tilde{M}(\k)$. 
		\item Assume that the rows of $C$ do not span the kernel of $D+\tilde{M}(\k)$, and define an arbitrary basis $|v_i\rangle$ for the part of the kernel not contained in the row space of $C$. In order to obtain a non-zero GfTNS, one has to insert the product of Grassmann variables $\prod_i \langle \Theta_\k | v_i\rangle$ ($\langle \Theta_\k | = (\bar{\theta}_{\k,\alpha},\theta_{\k,\beta}))$ in the Berezin integral which defines the GfTNS.
		\item If $D+\tilde{M}(\k)$ at $\k=0$ has a non-trivial kernel (which necessarily has even dimension), and the part of this kernel which is not in the row space of $C$ has an odd dimension, then the corresponding GfTNS with periodic boundary conditions necessarily has odd fermion parity if it is non-vanishing. 
	\end{itemize}
	In the Majorana-chain example, the kernel of $B+\tilde{M}(k)$ at $k=0$ has dimension $2$, and the part which is not spanned by the rows of $C$ has dimension $1$, hence the state necessarily has odd fermion parity. In 2D, the $p_x+ip_y$ superconductor in the weak-pairing phase also has odd parity on the torus with periodic boundary conditions along both cycles \cite{Read2000}. In Refs. \cite{Wahl2013,Dubail2015,Wahl2014}, GfTNS representations of the weak-pairing $p_x+ip_y$ superconductor were studied, and it was found that an additional operator needs to be inserted on the virtual level in order to obtain a non-zero state \cite{Wahl2014}. In our formalism, this operator exactly corresponds to the Grassmann variable $\langle \Theta_\k | v \rangle$ at $\k = 0$, where $|v\rangle$ is the single vector in the kernel of $D+\tilde{M}(0)$ which is not in the row space of $C$. In Ref.~\cite{Bultinck2017}, non-Gaussian fTNS were studied which represent states with non-trivial topological order, and which also have odd parity on the torus with periodic boundary conditions along both cycles. In this case, the proper generalization of the Grassmann numbers $\langle \Theta_\k | v \rangle$ are fermionic matrix-product operators \cite{Bultinck2017}.
	
	\section{Appendix B -- Performing the scaling collapse }
	
\begin{figure}[H]
		\centering
		\includegraphics[scale=0.52]{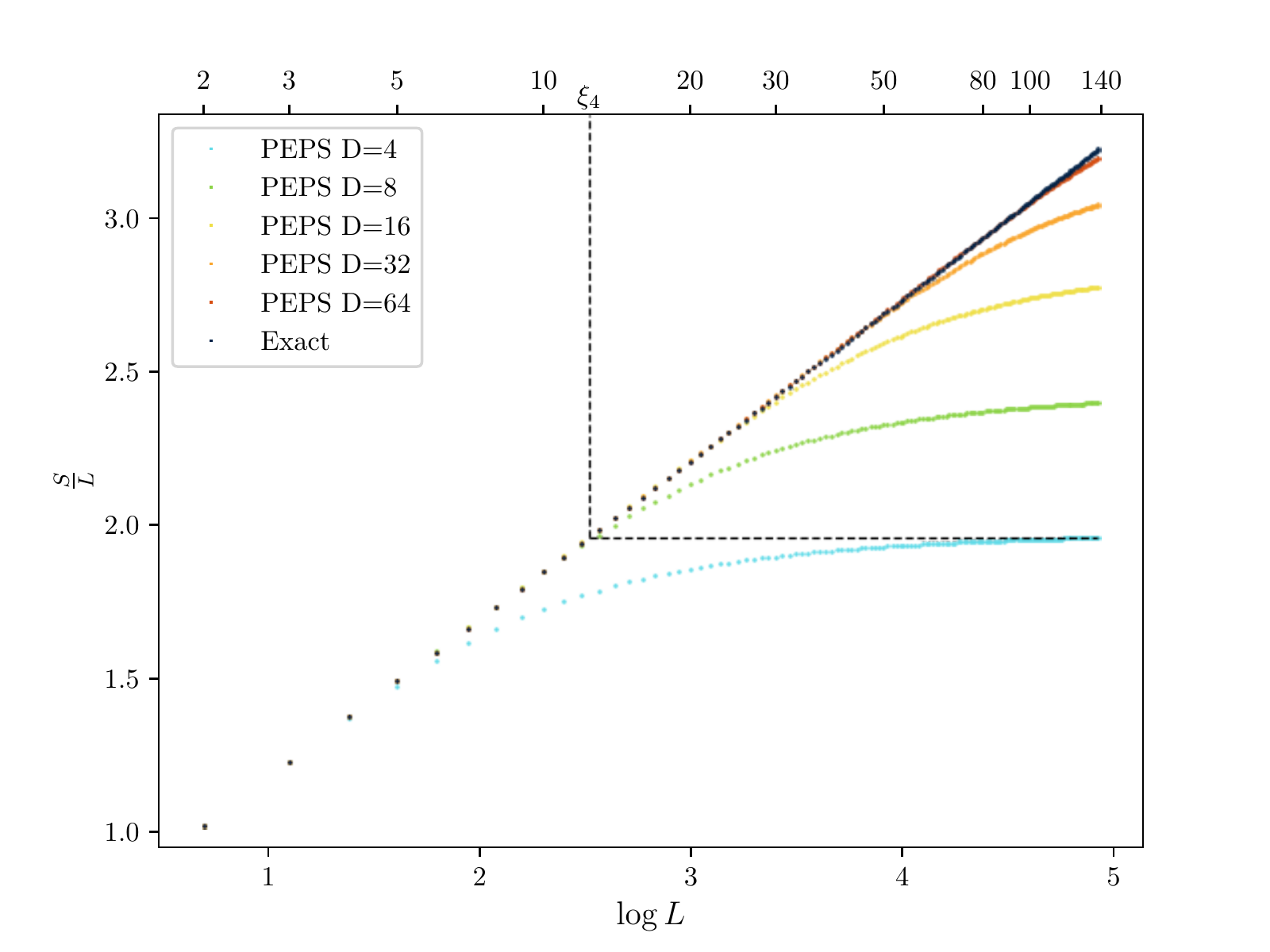}
		\caption{Illustration of the procedure to extract the correlation length from the EE data (here for the same data as in Fig.~\ref{fig:collapse_spinless}). For the GfTNS with bond dimension $D$, the crossover from the $\log L$-Fermi surface regime to the area law regime is long. When the crossover is completed, $S/L$ enters into a plateau with value $y_{\mathrm{plateau}, D}= S_D/L$. We identify the correlation length $\xi_D$ to be the length scale at which the exact Fermi surface $S_{\mathrm{exact}}/L$ reaches the value $y_{\mathrm{plateau}, D}$.}
		\label{fig:extract length}
	\end{figure}

	Once optimised, real-space correlation functions like
	\begin{equation}
		\braket{f^\dag_{\x} f_{\mathbf{y}}} = \frac{1}{N_s} \sum_\k e^{i\k\cdot (\x-\mathbf{y})} \, n(\k) \qquad \text{and} \qquad \braket{f_{\x} f_{\mathbf{y}}} = \frac{1}{N_s} \sum_\k e^{-i\k\cdot (\x-\mathbf{y})} \, x(\k) \, .
	\end{equation}
	(here for $N=1$) can easily be calculated for GfTNS by (inverse) Fourier transforming
        \begin{equation}
		n(\mathbf{k}) = \braket{f^\dag_{\k}f_{\k}} = \frac{|g_{\mathbf{k}}|^2}{1+|g_{\mathbf{k}}|^2} \qquad \text{and} \qquad
		x(\mathbf{k}) = \braket{f_{\k}f_{-\k}} = \frac{g_{\mathbf{k}}}{1+|g_{\mathbf{k}}|^2} \, .
        \label{eq:nx}
	\end{equation}
    One then constructs a real-space correlation matrix $C$, consisting of 
	\begin{equation}
		C_{\x,\mathbf{y}} = C_{\x-\mathbf{y}} = \begin{pmatrix}
			\braket{f^\dag_{\x} f_{\mathbf{y}}} & \braket{f^\dag_{\x} f^\dag_{\mathbf{y}}} \\
			\braket{f_{\x} f_{\mathbf{y}}} & \braket{f_{\x} f^\dag_{\mathbf{y}}}
		\end{pmatrix}
	\end{equation}
	blocks for all combinations of $\x$ and $\mathbf{y}$. Restricting to vectors comprised in an $L \times L$ subregion, the resulting correlation matrix $C_{L \times L}$ can be used to calculate the entanglement entropy. Indeed, for a Gaussian state, $S = -\sum \zeta \log \zeta$ with $\zeta$ the eigenvalues of $C_{L \times L}$. Now recall the scaling hypothesis for the EE from Eq.\,\eqref{scalinglaw}. For a circular Fermi surface it reduces to 
	\begin{equation}
		S = \alpha \, k_F L \log\left( \xi \Lambda f\left( \frac{L}{\xi}\right) \right),
	\end{equation}
	with $L$ a length scale proportional to the circumference of the real space region, $k_F$ the Fermi momentum, an inverse length scale proportional to the circumference of the Fermi surface and $\alpha = \frac{2}{3 \pi}$ a dimensionless prefactor depending on the shape of both. Together with the UV cut-off $\Lambda$, the latter can be determined from the exact ground state so that the only tunable parameters in the scaling law are the correlation lengths. These infrared length scales, depending on the bond dimension $D$ of the GfTNS approximation, should be such that the results are described by a single scaling function $f$. To this end, we rewrite the scaling law as
	\begin{equation}
		\log \left(f\left(\frac{L}{\xi}\right)\right) = \frac{S}{\alpha \, k_F L} - \log (\xi \Lambda) \, .
	\end{equation}
	For the correct $\xi$, plotting the right-hand side of the equation versus (the logarithm of) $\frac{L}{\xi}$ should thus yield a single curve. However, this leaves an overall scale undetermined as setting $\tilde{\xi} = s \xi$ simply leads to a redefined scaling function $\tilde{f}(x) = \frac{1}{s} f(s x)$ with the same shape. This scale can be fixed by requiring that the asymptotic value $\lim_{x\to \infty} f(x) = 1$, so that $\lim_{L \to \infty} S(L) = \alpha \, k_F L \log(\Lambda \xi)$. However, as the cost to compute $S(L)$ grows quite rapidly in $L$, namely as $\mathcal{O}(L^6)$, the calculations are restricted to moderate values of $L$. Hence, in practice, the scale is chosen schematically as demonstrated in Fig.~\ref{fig:extract length}, i.e.~by setting $f(x)=1$ for $x = L/\xi$ for the largest $L$ and for the smallest $D$ in our set, where the asymptotic regime is approximately reached. Once the correlation length for \emph{one} bond dimension is fixed, the correlation lengths for the GfTNS at other values of $D$ can be fixed by the collapse.
	
	The scaling collapse method can also be used to obtain information on an priori unknown Fermi surface. Indeed, if we do posses EE data suggesting a Fermi surface, then we could define 
	\begin{equation}
		\alpha = \frac{1}{|\partial R| |FS|} \oint_{\partial R}\oint_{FS} |\mathrm{d}S_\x\cdot \mathrm{d}S_\k| \label{alphadef}
	\end{equation}    
	where $|\partial R|$ $(|FS|)$ is the circumference of the real-space subregion (Fermi surface) so that $0\leq\alpha\leq1$ again represents a dimensionless form factor depending on the geometry of both. Obtaining a collapse is still possible by not only tuning the $\xi$ values but also the unknown $\alpha \,|FS|$ (always appearing as a product) and $\Lambda$. A simple algorithm could start from a linear regression applied to the linear parts of the $\frac{S}{L}$ vs $\log L$ plots (hinting at the presence of a Fermi surface), yielding approximate values for $\alpha\,|FS|$ and $\Lambda$. The correlation length for the lowest $D$ can be read off via the procedure described in Fig.~\ref{fig:extract length} but now using the intersection with higher $D$ curves. For higher $D$ approximate values for $\xi$ can be obtained in the same way. These initial parameters will result in a sub-optimal collapse after which the following steps can be repeated:    
	\begin{enumerate}
		\item Fit an analytical scaling function (e.g.~a shifted half sigmoid) to the sub-optimal collapse via a least-squares procedure.
		\item Improve the $\alpha\,|FS|$, $\Lambda$ and $\xi$ (except for the lowest $D$) by minimizing the least-squares distance to the analytical scaling function.
	\end{enumerate}
	After a few iterations this alternating least-squares method will yield an improved collapse, not only confirming the presence of a Fermi surface (by the fact that a sharp collapse is possible) but also improving its description in terms of $\alpha\,|FS|$ and $\Lambda$ by making use of all EE data. I.e.~also data for low $D$ are utilized to determine the specific properties of the criticality, thereby increasing the statical precision just like finite-entanglement scaling in 1D. Moreover, one could vary the shape of the real-space region $R$ to get a set of different $\alpha\,|FS|$ values, all describing the same Fermi surface. As such, the double integral in Eq.\,\eqref{alphadef} can be fully exploited to determine the Fermiology of the state. 
 
    We illustrate the alternating least-squares procedure for the scaling collapse of the EE data in Fig.~\ref{fig:collapse_spinless}. For the initial guess we obtain (the already qualitative) $\alpha |FS|=9.617$, $\Lambda=1.272$ and $\xi=5.343,12,31,70,140,280$. The corresponding (rough) collapse is systematically improved, eventually yielding $\alpha |FS|=9.969$, $\Lambda=1.246$ and $\xi=5.392,12.72,30.83,70.26,141.2,282.8$. Note that this indeed approximates the exact $\alpha |FS|=\frac{2}{\pi} 2 \pi k_F= 10.02$ and $\Lambda=1.238$ as well as the correlation lengths obtained with these exact values in Table \ref{tab:correlation list spinless}. The optimal half sigmoid that models the scaling function is given by
    \begin{equation}
        f\left(x=\log \frac{L}{\xi}\right) = \begin{cases}
            x+a-b & x \leq a\\
            \frac{x-a}{(1+|x-a|^k)^{1/k}} + b & x >a
        \end{cases}
    \end{equation}
    where $a = -0.9014$, $b= -0.8963$ and $k=1.367$.
 
	\section{Appendix C -- Gu-Verstraete-Wen formalism for spinless fermions:\\ optimisation method and additional results}
	
	In this section, we describe our numerical method to optimise GfTNS for 2D states with a Fermi surface composed of a single spinless fermion in detail. We apply the Gu-Verstraete-Wen (GVW) formalism for Gaussian tensor networks, as introduced in \cite{Gu2010}. As $N=1$, the $B$ submatrix of $A$ now reduces to zero so that
	\begin{equation}
		\ket{\psi}=\int D[\theta] D[\bar{\theta }]\prod_{\mathbf{x}} \exp \left( \bar{\theta}_{\x}^{h^T} \theta_{\x+\e_x}^h \right) \exp \left( \bar{\theta}_{\x}^{v^T} \theta_{\x+\e_y}^v \right) \exp \left( \frac{1}{2} \theta^T_{\mathbf{x}} D \theta_{\mathbf{x}}\right)  \exp\left( \theta^T_{\mathbf{x}} C  f^\dag_{\mathbf{x}}\right) \ket{0} \, , 
	\end{equation}
	where $\theta_\x \equiv (\theta^{h^T}_{\x},\bar{\theta}^{h^T}_{\x},\theta^{v^T}_{\x},\bar{\theta}^{v^T}_{\x})^T$ collects the virtual Grassmann numbers associated to site $\x$ and $f^\dag_{\x}$ creates the spinless fermion on site $\x$. Furthermore, $D$ is a complex antisymmetric $4M \times 4M$ matrix, whereas $C$ is a complex vector of length $4M$. The number of (complex) variational parameters in this state thus seems to be $8M^2+2M$. However, the state has gauge redundancies which can be eliminated. The origin of the gauge redundancies can be discerned by examining a single term $\exp \left( \bar{\theta}_{\x}^{h^T} \theta_{\x+\e_x}^h \right)$, which is invariant under
	\begin{equation}
		\bar{\theta}_{\x}^{h^T} \mapsto \bar{\theta}_{\x}^{h^T} W^{-1}, \quad \theta_{\x+\e_x}^h \mapsto W \theta_{\x+\e_x}^h \qquad \forall \, W\in GL(M, \mathbbm{C}) \, .
	\end{equation}
	Similarly we can also transform $\theta^{v}$ and $\bar{\theta}^{v}$. To keep the translation invariance explicit, we can apply the same transformation to every bond. Since we are integrating the Grassmann variables out, we can check that the resulting Gaussian state is invariant under these transformations.
	
    To make the discussions more transparent, let us write $D$ and $C$ in terms of blocks,
	\begin{equation}
		D \equiv \begin{bmatrix}
			D_{hh} & D_{h \bar{h}} & D_{hv} & D_{h\bar{v}}\\
			D_{\bar{h}h} & D_{h \bar{h}} & D_{\bar{h}v} & D_{\bar{h}\bar{v}}\\
			D_{vh} & D_{v \bar{h}} & D_{vv} & D_{v\bar{v}}\\
			D_{\bar{v}h} & D_{\bar{v} \bar{h}} & D_{\bar{v}v} & D_{\bar{v}\bar{v}}\\
		\end{bmatrix}, \quad C \equiv \begin{bmatrix}
			C_{h} \\
			C_{\bar{h}} \\
			C_{v} \\
			C_{\bar{v}} 
		\end{bmatrix} \, .
	\end{equation}
	There are two independent gauge transformations, $W_{\mathbf{x}}$ and $W_{\mathbf{y}}$, transforming, for instance, the blocks $D_{hh}$ and $D_{vv}$ in the following way,
	\begin{equation}
		D_{hh} \mapsto W^{T}_{\mathbf{x}} D_{hh} W_{\mathbf{x}}, \quad D_{vv} \mapsto W^{T}_{\mathbf{y}} D_{vv} W_{\mathbf{y}} \, .
	\end{equation}
	Because $D_{hh}$ and $D_{vv}$ are antisymmetric, they admit real normal forms \cite{becker1973}, 
	\begin{equation}
		D_{hh} = U^T_h \left( \bigoplus_{i=1}^{\lfloor\frac{M}{2} \rfloor} \begin{bmatrix}
			0 & \alpha_i\\
			-\alpha_i & 0
		\end{bmatrix} \left( \oplus [0]\right) \right) U_h \, ,\quad D_{vv} = U^T_v \left( \bigoplus_{i=1}^{\lfloor\frac{M}{2} \rfloor} \begin{bmatrix}
			0 & \beta_i\\
			-\beta_i & 0
		\end{bmatrix} \left( \oplus [0]\right) \right) U_v , 
	\end{equation}
	where $U_h$ and $U_v$ are unitary matrices, $\{\alpha_i, \beta_i\}$ are non-negative real numbers and the direct sums $\oplus [0]$ only appear when  $M$ is odd. Therefore, by choosing
	\begin{equation}
		W_{\mathbf{x}} \equiv U^{\dag}_h \left( \bigoplus_{i=1}^{\lfloor\frac{M}{2} \rfloor} \begin{bmatrix}
			 \frac{1}{\sqrt{\alpha_i}} & 0\\
			0 & \frac{1}{\sqrt{\alpha_i}}
		\end{bmatrix} \left( \oplus [1]\right) \right) \, , \quad W_{\mathbf{y}} \equiv U^{\dag}_v \left( \bigoplus_{i=1}^{\lfloor\frac{M}{2} \rfloor} \begin{bmatrix}
		\frac{1}{\sqrt{\beta_i}} & 0\\
		0 & \frac{1}{\sqrt{\beta_i}}
	\end{bmatrix} \left( \oplus [1]\right) \right) \, ,
	\end{equation}
	we bring $D_{hh}$ and $D_{vv}$ into very simple form $J^{\oplus \lfloor\frac{M}{2} \rfloor} \left( \oplus [0]\right)$ with $J=i \sigma_y$. Thus, the gauge degrees of freedom are eliminated by fixing $D_{hh}$ and $D_{vv}$ in our GfTNS, thereby reducing the number of complex variational parameters to $7M^2+3M$. We optimised GfTNS with and without fixed gauge and found matching energies, thus not only confirming the validity of the gauge fixing procedure but also providing confidence that the optimisation procedure (as described in the next paragraph) is not struggling with local minima. However, the optimisation with gauge degrees of freedom removed convergence significantly faster, which was crucial to obtain converged results for the larger bond dimensions $D$ used in this study.
	
	To evaluate the energy of the GfTNS, we go to momentum space,
	\begin{equation}
		|\psi\rangle = \int [D\theta] \int [D\bar{\theta}] \exp\left(\frac{1}{2}\sum_\k \theta_{-\k}^T[D+\tilde{M}(\k)]\theta_\k + \theta_{-\k}^T C f^\dagger_\k \right)|0\rangle = {\prod_{\mathbf{k}}}\frac{1}{\sqrt{1 + |g_{\mathbf{k}}|^2}}\exp \left(\frac{1}{2} g_{\mathbf{k}} \, f^\dag_{-\mathbf{k}}f^\dag_{\mathbf{k}}\right) \ket{0},
	\end{equation}
	where $g_{\mathbf{k}}\equiv C^T [D+\tilde{M}(\k)]^ {-1} C = C^T S(\k)^ {-1} C$ and where $\tilde{M}(\k)$ is defined as in Eq.\,\eqref{Mtilde} in the main text. Consequently, the modal occupation and pairing term expectation values are given by Eq.\,\eqref{eq:nx}. Using this and the fact that we can rewrite the free-fermion Hamiltonian from Eq.\,\eqref{Hamiltonian} as $H = \sum_{\mathbf{k}} H(\k) f^\dag_{\k}f_{\k}$	with $H(\k) = -2 t \left(\cos k_x  + \cos k_y\right) -2t' \left(\cos (k_x+k_y)  + \cos (k_x-k_y) \right) - \mu$, we obtain the energy density,
	\begin{equation}
		e = \frac{1}{N_s} \langle \psi|H\ket{\psi} =\frac{1}{N_s} \sum_{\mathbf{k}} H(\k) \frac{|g_{\mathbf{k}}|^2}{(1+|g_{\mathbf{k}}|^2)}\, .
	\end{equation}
	The above expression serves as our cost function $e\equiv e(z)$ with the variational parameters in $z$ encoding $D$ and $C$. More concretely, the first $7M^2-M$ elements of this complex vector, denoted as $z_D$, encode the gauge independent entries of $D$. The last $4M$ entries, denoted as $z_C$, correspond directly to $C$. The gradient of the cost function w.r.t.~(the complex conjugate of) $z$ can hence be expressed as
	\begin{align}
		2\frac{\partial e}{\partial \bar{z}} & = \frac{2}{N_s}\sum_{\mathbf{k}} H(\k) \frac{g_{\mathbf{k}}(z)}{(1+|g_{\mathbf{k}}(z)|^2)^2} \frac{\partial \bar{g}_{\mathbf{k}}(\bar{z})}{\partial \bar{z}} \, ,
	\end{align}
	 where
	\begin{align}
			\frac{\partial g_{\mathbf{k}}}{\partial z_D} & = - C^T 		S^{-1}(\mathbf{k}) \frac{\partial S(\mathbf{k})}{\partial z_D}  S^{-1}(\mathbf{k}) C\, ,&
			\frac{\partial g_{\mathbf{k}}}{\partial z_C} & = \frac{\partial C^T}{\partial z_C} S^{-1}(\mathbf{k}) C + C^T S^{-1}(\mathbf{k}) \frac{\partial C}{\partial z_C} \, .
	\end{align}
	It should be noted that $\partial S(\mathbf{k})/ \partial z_D$ and $\partial C/ \partial z_C$ are sparse matrices/vectors. Considering $z_D$ as example, each entry in $z_D$ is associated with an ordered pair $(i, j)$, such that $(\partial S(\mathbf{k})/ \partial z_D)_{ij} = - (\partial S(\mathbf{k})/ \partial z_D)_{ji} = 1$, and the rest of the matrix is $0$. Having the analytical form of the cost function and the gradient, we use the quasi-Newton BFGS method, implemented in the \texttt{Optim.jl} package \cite{mogensen2022}, to optimise the cost function. We observed that, for this problem, the BFGS method greatly outperforms alternative methods. We also remark that it is not difficult to work out the analytical expression of the Hessian of our cost function. While this enables the use of the second-order Newton's method, we found that it is too expensive in practice.
	
\begin{figure}[H]
		\centering
		\includegraphics[scale=0.47]{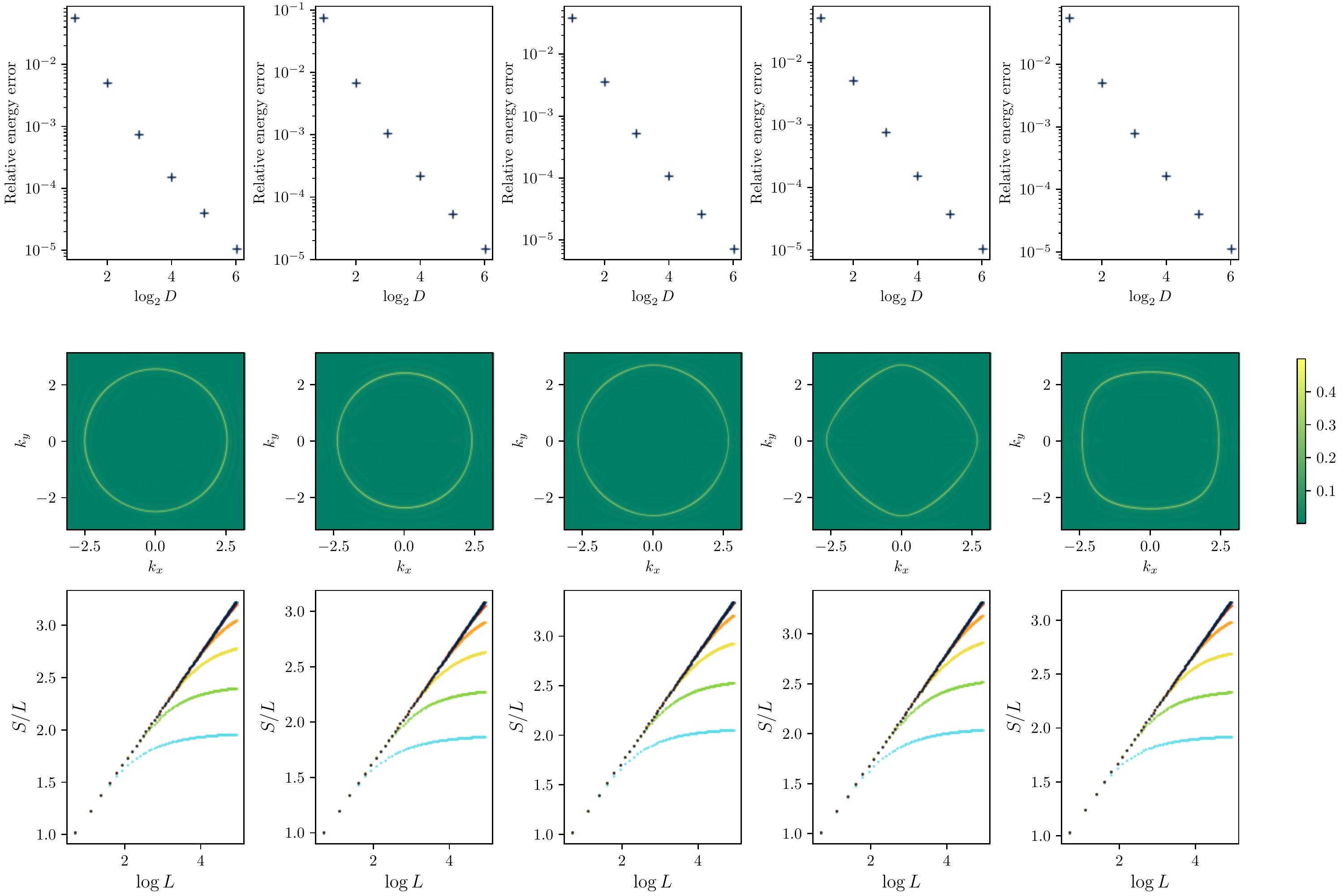}
		\caption{Collection of additional GfTNS results for the spinless model. The columns correspond to different parameter choices for $(t, t', n_{\mathrm{filling}})$, respectively $(1, 0.353, 0.5)$, $(1, 0.353, 0.45)$, $(1, 0.353, 0.55)$, $(1, 0.2, 0.5)$ and $(1, 0.5, 0.5)$. The system size for the optimisation of these GfTNS is $N_s = 1000^2$. The first row collects the relative energy errors for the GfTNS. The second row collects $\langle f_\k f_{-\k}\rangle$ at $D=32$ throughout the Brillouin zone. The color map denotes the magnitude. The third row collects the EE of a $L\times L$ region. The color of the data is chosen to match that of Fig.~\ref{fig:collapse_spinless} for bond dimensions $D=4,8,16,32,64$, while the $D=2$ results have been omitted because they offer a poor description of the Fermi surface.}
		\label{fig:spinless new}
	\end{figure}

\begin{figure}[H]
		\centering
		\includegraphics[scale=0.4]{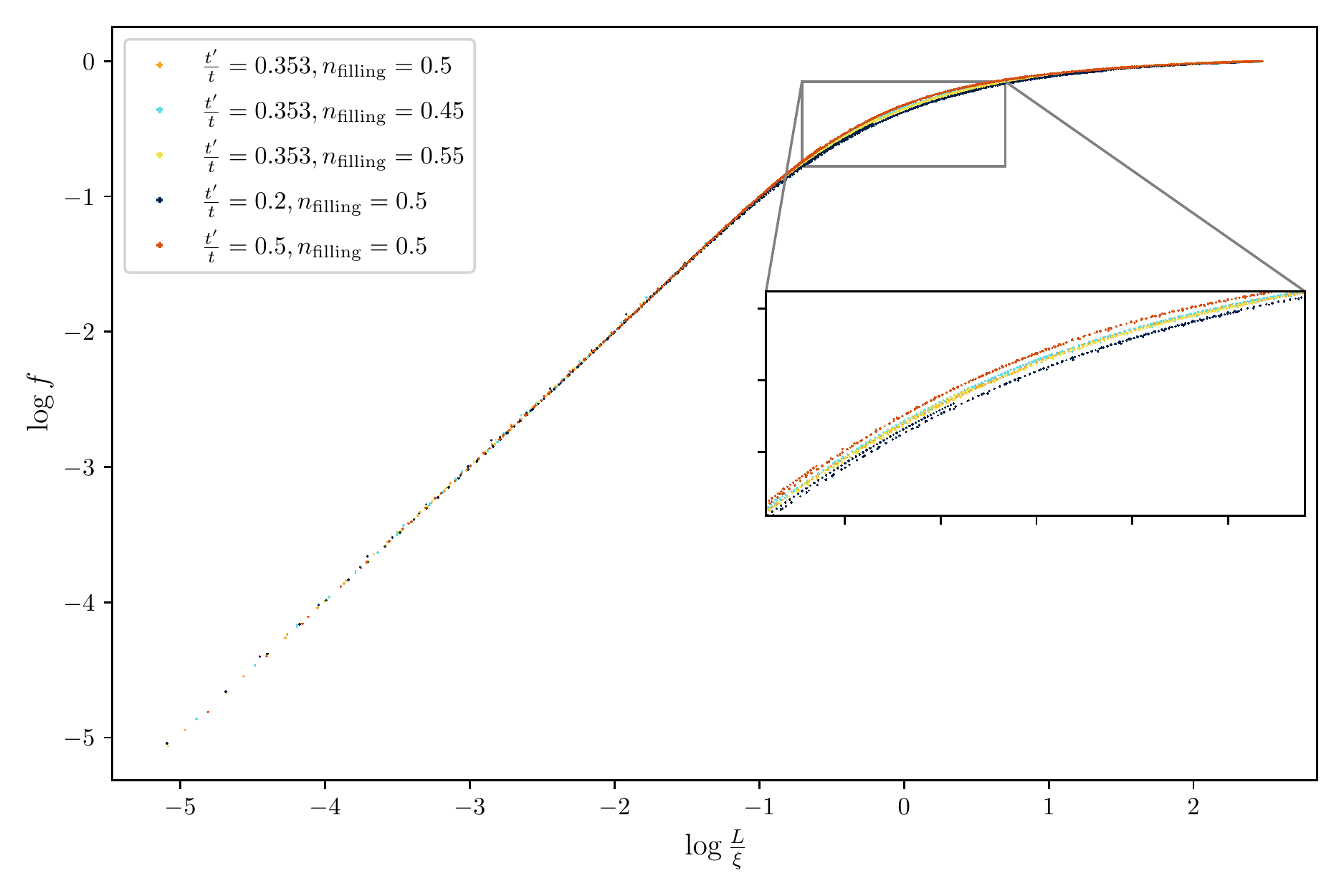}
		\caption{The scaling collapse of EE data. We note that, in contrast with the spinful case, the scaling function for the spinless GfTNS is less sensitive to the change of Fermi surface shape. However, the fuzziness near the transition region indicates that the scaling function for the spinless case could be Fermi surface dependent after all.}
		\label{fig:entropy collapsed spinless}
	\end{figure}

    In addition to the result in the main text, we further optimised GfTNS and performed scaling collapses for Fermi surfaces with different fillings and shapes, the results for which are collected in Fig.~\ref{fig:spinless new} and Table \ref{tab:correlation list spinless}. The four columns in the figure correspond to: 1) $t'/t = 0.353$, with $\mu = 0.754$ that enforces the filling to be 0.5; 2) $t'/t = 0.353$, with $\mu = 0.449$ that enforces the filling to be 0.45; 3) $t'/t = 0.353$, with $\mu = 1.00$ that enforces the filling to be 0.55; 4) $t'/t = 0.2$, with $\mu = 0.476$ that enforces the filling to be 0.5; 5) $t'/t = 0.5$, with $\mu = 0.963$ that enforces the filling to be 0.5. For all of these cases, we observe a similar scaling of energy error/correlation length with the bond dimension. Furthermore, we find that the scaling function is independent of the filling fraction (comparing cases 1, 2 and 3), but does show an (albeit weak) dependence on the shape of the Fermi surface (comparing case 1 with 4 and 5), as demonstrated in Fig.~\ref{fig:entropy collapsed spinless} . The former is quite remarkable. Indeed, deriving a general scaling law for the EE (as in Ref.~\cite{Vanhecke2019}), one could start from the exact ground state EE, given by $S = \alpha \, k_F L \log \Lambda L$ for a circular Fermi surface. Simultaneously scaling $L$ by a factor $\frac{1}{s}$ and $k_F$ by $s$ (thus scaling the periodicity of the Friedel oscillations in real-space by a factor $\frac{1}{s}$) while maintaining the same $\Lambda$ should then decrease $\exp \left(\frac{S}{\alpha \, k_F L}\right)$ by a factor $s$. For an optimal GfTNS, the same reasoning applies but now also the correlation length should be scaled down, yielding the scaling hypothesis,
    \begin{equation}
    	\exp \left(\frac{S(k_F,L,\xi)}{\alpha \, k_F L}\right) = s \exp \left(\frac{S(s k_F,\frac{L}{s},\frac{\xi}{s})}{\alpha \, k_F L}\right) \, .
    \end{equation}
    Setting $s=\xi$, one obtains
    \begin{equation}
    	S(k_F,L,\xi) = \alpha \, k_F L \log \xi + S\left(k_F \, \xi, \frac{L}{\xi},1\right) =\alpha \, k_F L \log\left( \xi \Lambda f\left(k_F \xi, \frac{L}{\xi}\right) \right),
    \end{equation}
    where everything was written down with a single logarithm by defining the scaling function via $S\left(k_F \, \xi, \frac{L}{\xi},1\right) = \alpha \, k_F L \log\left(\Lambda f\left(k_F \xi, \frac{L}{\xi}\right) \right)$. Here the $\Lambda$ prefactor was added to make $f$ dimensionless. We thus retrieve the same scaling law as in Eq.\,\eqref{scalinglaw} but now also depending on $k_F \xi$. Based on Fig.~\ref{fig:entropy collapsed spinless}, this additional dependence drops out. For different Fermi surface shapes on the other hand, the scaling function does change. Drawing intuition from the finite temperature behavior of Fermi surface EE \cite{swingle2013}, this does not come as a surprise. We leave a detailed analysis to a future study.	
    
     \begin{table}
    	\centering
    	\begin{tabular}{c|c|c||c|c|c|c|c}
    		$t'/t$ & $\mu/t$ &$n_{\mathrm{filling}}$ &$D=4$ & $D=8$ & $D=16$ & $D=32$ & $D=64$ \\ 
    		\hline
    		$0.353$ & $0.754$ & $0.50$ & $12.4$ & $30.2$ & $68.9$ & $141.8$ & $287.5$ \\
    		$0.353$ & $0.449$ & $0.45$ & $11.8$ & $28.1$ & $63.4$ & $132.1$ & $265.0$ \\
    		$0.353$ & $1.000$ & $0.55$ & $13.3$ & $33.4$ & $77.6$ & $163.6$ & $321.1$ \\
    		$0.200$ & $0.476$ & $0.50$ & $13.5$ & $34.4$ & $80.9$ & $170.1$ & $322.8$ \\
    		$0.500$ & $0.963$ & $0.50$ & $11.6$ & $27.3$ & $60.1$ & $127.6$ & $244.8$ \\
    	\end{tabular}
    	\caption{The collection of correlation lengths, for the spinless case, obtained via the scaling collapse procedure.}
    	\label{tab:correlation list spinless}
    \end{table}	
			
	\section{Appendix D -- Kraus-Schuch formalism for (symmetric) GfTNS}
	
	In Ref.~\cite{Kraus2010}, Kraus et al.~introduced an alternative to the GVW formulation of fermionic TNS, including a restriction to the Gaussian submanifold based on \cite{Schuch2012}. This formalism was utilized here to optimise GfTNS (with a built-in SU(2) symmetry) for the spinful model. First, we introduce the basic version of this \textit{Ansatz} without the additional symmetry. This is mostly a repetition of \cite{Mortier2022}, which together with its Supplemental Material can be consulted for more details. Next, we discuss how symmetries can be added to the \textit{Ansatz} with a focus on SU(2). Our discussion will be restricted to $d=2$ spatial dimensions for simplicity, but the extension to general $d$ is straightforward.

	Consider again the 2D lattice built up by a periodic repetition of $N_x\times N_y=N_s$ unit cells, spanned by $\mathbf{e}_x$ and $\mathbf{e}_y$. To each vertex we attribute $N$ physical fermionic orbitals with creation (annihilation) operators $f^{\dagger}_{\mathbf{x},j}\,\left(f_{\mathbf{x},j}\right)$ where $j=1,...,N$ is the orbital/flavor index. Corresponding Majorana operators are denoted by $c_{\mathbf{x},2j-1}= f^{\dagger}_{\mathbf{x},j}+f_{\mathbf{x},j}$ and $c_{\mathbf{x},2j}=-i\left(f^{\dagger}_{\mathbf{x},j}-f_{\mathbf{x},j}\right)$. Within this framework, a PEPS \textit{Ansatz} can be obtained by first introducing four sets of virtual Majoranas per site: $\{c^{l}_{\mathbf{x},{i}}\}$, $\{c^{r}_{\mathbf{x},{i}}\}$,  $\{c^{d}_{\mathbf{x},{i}}\}$ and $\{c^{u}_{\mathbf{x},{i}}\}$, respectively in the left, right, down- and upward direction with $i=1,...,\chi$. Next, a maximally correlated pure state $\rho_\text{in}=|\psi_\text{in}\rangle\langle\psi_{\text{in}}|$ is constructed on the virtual level by entangling neighbouring Majoranas in both directions. This is realised by placing the Majoranas in their joint vacuum, essentially creating $\chi$ virtual Majorana chains in each direction. Finally, the maximally correlated state is locally projected onto the physical level by a channel $\mathcal{E}=\bigotimes_\mathbf{x}\mathcal{E}^\text{loc}_\mathbf{x}$ encoding the fermionic PEPS tensor and yielding the (possibly mixed) $\rho_\text{out}=\mathcal{E}\left(\rho_\text{in}\right)$. By increasing the number of virtual Majoranas, the variational set can be enlarged. In the case of a pure state, it can be represented as a fTNS with an effective bond dimension $D=\sqrt{2}^\chi$.
	
	Since $\rho_\text{in}$ is a free-fermion state, Gaussianity of the resulting PEPS can be enforced by restricting the channel $\mathcal{E}$ to be Gaussian as well \cite{Bravyi2005,Schuch2012}. Both the input and the output state can then be fully described in terms of their real and antisymmetric correlation matrices, $\Gamma^{ij}_{\mathbf{x}\mathbf{y}}=\frac{i}{2} \text{Tr}\left(\rho \left[c_{\mathbf{x},i},c_{\mathbf{y},j}\right]\right)$, which are related in the form of a Schur complement, $\Gamma_\text{out}=A+B\left(D+\Gamma_\text{in}^{-1}\right)^{-1} B^T$ where the matrices $A$, $B$ and $D$ parameterize the Gaussian channel $\mathcal{E}$. In fact, because of the local structure of $\mathcal{E}$, these matrices exhibit a decomposition $A=\bigoplus_\mathbf{n}A^\text{loc}_\mathbf{n}$ and similarly for $B$ and $D$ with $A^\text{loc}_\mathbf{x}\in\mathbbm{R}^{2N\times2N}$, $B^\text{loc}_\mathbf{x}\in\mathbbm{R}^{2N\times2d\chi}$ and $D^\text{loc}_\mathbf{x}\in\mathbbm{R}^{2d\chi\times2d\chi}$. Furthermore, $X=\begin{pmatrix}
		A&B\\
		-B^T&D
	\end{pmatrix}$ is antisymmetric and $XX^T\leq\mathbbm{1}$ with the equality holding for a pure state.
	
	For TI Gaussian states, it is more convenient to work in Fourier space, where these states can be described completely in terms of the Fourier transformed correlation matrix,  $G^{ij}_{\mathbf{k}\mathbf{q}}=\frac{i}{2}\text{Tr}\left(\rho\left[d_{\mathbf{k},i},d^{\dagger}_{\mathbf{q},j}\right]\right)$. Herein, $d_{\mathbf{k},i}=\frac{1}{\sqrt{N_s}}\sum_\mathbf{x}e^{-i\mathbf{k}\cdot\mathbf{x}}c_{\mathbf{x},i}$ with momentum modes $\mathbf{k}$. The Fourier transformed correlation matrix is anti-Hermitian, satisfies $GG^\dagger\leq\mathbbm{1}$ (with the equality again holding for a pure state) and, for TI states, decomposes in diagonal blocks $G=\bigoplus_\mathbf{k}G(\mathbf{k})$ with, for instance,
	\begin{equation}
		G_\text{in}(\mathbf{k}) = \begin{pmatrix}
			0 & e^{i k_x}\\
			-e^{-i k_x} & 0
		\end{pmatrix}^{\oplus \chi} \oplus \begin{pmatrix}
			0 & e^{i k_y}\\
			-e^{-i k_y} & 0
		\end{pmatrix}^{\oplus \chi}
		\label{Gin}
	\end{equation}
	for the input state described before. Assuming translation invariance of the PEPS, so that $\mathcal{E}^\text{loc}_\mathbf{x}$ is independent of $\mathbf{x}$, the transition matrix $X$ decomposes into identical blocks and yields $G_\text{out}(\mathbf{k})=A^\text{loc}+B^\text{loc}\left(D^\text{loc}-G_\text{in}(\mathbf{k})\right)^{-1}{B^\text{loc}}^T$ where the purity of the input state was used to replace $G^{-1}_\text{in}(\mathbf{k})$ by $-G_\text{in}(\mathbf{k})$.\\

    To impose symmetries within this framework, we observe that in the current Gaussian context, we have to restrict to symmetries, i.e.\ a set of unitary transformations $\mathcal{U}_g$ (for all $g \in G$ with $G$ the symmetry group) commuting with the Hamiltonian ($\mathcal{U}_g H \mathcal{U}_g^\dagger = H$), that represent a canonical transformation, i.e.\ the operators $\mathcal{U}_g$ map the creation and annihilation operators to a linear combination thereof that preserves the anticommutation relations. Collecting the creation and annihilation operators in the Nambu spinor, $\Upsilon_{\mathbf{x}} = (\{f_{\mathbf{x},i}\} \quad \{f^\dagger_{\mathbf{x},i}\})^T$, we must have
	\begin{equation}
		\mathcal{U} \Upsilon_{\mathbf{x},i} \, \mathcal{U}^\dagger = V^{ji}_{\mathbf{x} \mathbf{y}} \Upsilon_{\mathbf{y},j}\, ,
	\end{equation}
	where the $V^{ji}_{\mathbf{x} \mathbf{y}}$ factors are just numbers. We now focus on on-site symmetries where $\mathcal{U} = \bigotimes_\mathbf{x} \mathcal{U}^\text{loc}_\mathbf{x}$, with $\mathcal{U}^\text{loc}_\mathbf{x}$ independent of $\mathbf{x}$. Equivalently, we have $V^{ji}_{\mathbf{x} \mathbf{y}} = \delta_{\mathbf{x} \mathbf{y}} v^{ji}$ with $v$ a $2N \times 2N$ matrix. In fact, for those symmetries that can be understood at the single particle level, the local matrix $v$ further decomposes as $v = \overline{u} \oplus u$ where $u$ is the local unitary transformation of the modes on a single site, i.e.~the annihilation and creation operators transform as $\mathcal{U} f_{\mathbf{x},i} \mathcal{U}^\dagger = \overline{u}^{ji} f_{\mathbf{x},j}$ and $\mathcal{U} f_{\mathbf{x},i}^\dagger \mathcal{U}^\dagger = u^{ji} f_{\mathbf{x},j}^\dagger$ without mixing. This is true for most symmetries (particle hole transformations excluded) and in particular for the on-site SU(2) that we are targeting here. This requires that the $N$-dimensional mode space associated with each site decompose into a direct sum SU(2) multiplets. For the spinful Hamiltonian that we consider, $N=2$ and the mode space corresponds exactly to a single copy of the spin-$\frac{1}{2}$ irreducible representation. For a spin rotation with angle $\phi$ around axis $\mathbf{n}$, we thus have $u = e^{i \frac{\phi}{2} \hat{\mathbf{n}} \cdot \boldsymbol{\sigma}}$ with $\boldsymbol{\sigma}$ collecting the Pauli matrices. This transformation is obtained by the many-body unitary operator
    \begin{equation}
		\mathcal{U}_{\mathbf{x}}^{\text{loc}} = \exp \left(i \frac{\phi}{2} {f^{ \sigma}_{\mathbf{x}}}^\dagger (\hat{\mathbf{n}} \cdot \boldsymbol{\sigma})_{\sigma \sigma'} f^{ \sigma'}_{\mathbf{x}} \right) = \exp \left(i \phi \,  \hat{\mathbf{n}} \cdot \mathbf{S}_\mathbf{x} \right)
	\end{equation}
	with $\mathbf{S}_\mathbf{x} = {f^{ \sigma}_{\mathbf{x}}}^\dagger (\frac{\boldsymbol{\sigma}}{2})_{\sigma \sigma'} f^{ \sigma'}_{\mathbf{x}}$ the local spin operator.
 
 Now consider a general free-fermion Hamiltonian
	\begin{equation}
		H = \frac{1}{2} \sum_{AB} \Upsilon_A^\dagger H_\text{BdG}^{AB} \Upsilon_B + \tilde{E}\, ,
	\end{equation}  
	where $\Upsilon$ collects the Nambu spinors on all sites with the $A,B$ labels discerning between its upper (annihilator) and lower (creator) half, $\tilde{E}$ is an energy offset and $H_\text{BdG}^{AB}$ is the Bogoliubov-de Gennes single-particle Hamiltonian, which takes the form
 	\begin{equation}
		H_\text{BdG} = \begin{pmatrix}
			\Xi && \Delta\\
			-\overline{\Delta} && -\Xi^T
		\end{pmatrix} \, , \quad \Xi = \Xi^\dagger\, , \quad \Delta = -\Delta^T\, . 
	\end{equation}
and thus satisfies $\left(\sigma_x \otimes \mathbbm{1}_{f}\right) H_\text{BdG}^T \left(\sigma_x \otimes \mathbbm{1}_{f}\right) = -H_\text{BdG}$. Further assuming translation invariance, $\Xi^{ij}_{\mathbf{x}\mathbf{y}}=\Xi^{ij}_{\mathbf{x}-\mathbf{y}}$ and $\Delta^{ij}_{\mathbf{x}\mathbf{y}}=\Delta^{ij}_{\mathbf{x}-\mathbf{y}}$, yields
	\begin{equation}
		H = \frac{1}{2} \sum_{\mathbf{k}} \Upsilon^{\dagger}_\mathbf{k} H_\text{BdG}(\mathbf{k}) \Upsilon_\mathbf{k} = \frac{1}{2} \sum_{\mathbf{k}} \Upsilon^{\dagger}_\mathbf{k} \begin{pmatrix}
			\Xi(\mathbf{k}) && \Delta(\mathbf{k})\\
			-\overline{\Delta(-\mathbf{k})} && -\Xi^T(-\mathbf{k})
		\end{pmatrix} \Upsilon_\mathbf{k}
		\label{hamiltonianbdg}
	\end{equation}
	where $\Upsilon_\mathbf{k} = \begin{pmatrix}
		\left\{f_{\mathbf{k},i}\right\} \,\, , \,\, \left\{f^\dagger_{-\mathbf{k},i}\right\}
	\end{pmatrix}^T$ with $f_{\mathbf{k},i} = \frac{1}{\sqrt{N}}\sum_\mathbf{x} e^{-i\mathbf{k}\cdot\mathbf{x}} f_{\mathbf{x},i}$, $\Xi^{ij}(\mathbf{k}) = \sum_\mathbf{x}  e^{-i\mathbf{k}\cdot \mathbf{x}} \, \Xi^{ij}_\mathbf{x}$ and $\Delta^{ij}(\mathbf{k}) = \sum_\mathbf{x}  e^{-i\mathbf{k}\cdot \mathbf{x}} \, \Delta^{ij}_\mathbf{x}$. As a result, $\Xi^\dagger(\mathbf{k})=\Xi(\mathbf{k})$ and $\Delta(\mathbf{k})=-\Delta^T(-\mathbf{k})$ so that $(\sigma_x\otimes\mathbbm{1}_N) H_\text{BdG}^T(-\mathbf{k}) (\sigma_x\otimes\mathbbm{1}_N) = - H_\text{BdG}(\mathbf{k})$. Invariance of such a Hamiltonian under an on-site single particle transformation $u$ then amounts to
	\begin{equation}
		\begin{pmatrix}
			u & \\
			& \overline{u}
		\end{pmatrix} H_\text{BdG}(\mathbf{k}) \begin{pmatrix}
			u^\dagger & \\
			& \overline{u}^\dagger
		\end{pmatrix} = H_\text{BdG}(\mathbf{k})
	\end{equation}
    and thus requires $u \Xi(\mathbf{k}) u^\dagger = \Xi(\mathbf{k})$ and $u \Delta(\mathbf{k}) \overline{u}^\dagger = u \Delta(\mathbf{k}) u^T = \Delta(\mathbf{k})$. For our case of interest, where $N=2$ and $u$ is an arbitrary SU(2) matrix, this amounts to $\Xi(\mathbf{k}) = \xi(\mathbf{k}) \otimes \mathbbm{1}$ and $\Delta(\mathbf{k}) = \delta(\mathbf{k}) \otimes J$. This shows that pairing terms, opening a superconducting gap in the GfTNS approximation of the ground state of the spinful Hamiltonian, will necessarily entangle both spins as discussed in the main text. 
	
	Having discussed symmetric Hamiltonians, we now have to consider symmetric states. A state $\ket{\psi}$ is symmetric  when $\mathcal{U}_g \ket{\psi} = \lambda_g \ket{\psi}$ for all $g \in G$, with $\lambda_g$ a possible phase factor, which forms a one-dimensional representation of $G$. Expectation values of symmetric operators $O$ ($\mathcal{U} O \mathcal{U}^\dagger = O$) then remain unchanged when performing a symmetry transformation. We can utilize this characterization to work out what the symmetry entails for the correlation matrix of a Gaussian symmetric state. For later purposes, we now generalise to the case where we have $N/2$ spinful fermion modes associated to every site or momentum, the creation operator of which is denoted by $(f^{\sigma}_{\mathbf{k},i})^\dagger$ with $i=1,\dots,N/2$. Imposing SU(2) symmetry, where we can ignore the phase factor as SU(2) does not have non-trivial one-dimensional representations, we then find
	\begin{equation}~
		\begin{split}
			n^{\sigma \sigma'}_{ij}(\mathbf{k}) &= \langle {f^{\sigma}_{\mathbf{k},i}}^\dagger f^{\sigma'}_{\mathbf{k},j} \rangle = \langle \psi | \mathcal{U}^\dagger \mathcal{U} {f^{\sigma}_{\mathbf{k},i}}^\dagger \mathcal{U}^\dagger \mathcal{U}  f^{\sigma'}_{\mathbf{k},j} \mathcal{U}^\dagger \mathcal{U} | \psi \rangle= \bra{\psi}  u^{\tau\sigma} {f^{\tau}_{\mathbf{k},i}}^\dagger \overline{u^{\tau' \sigma'}}  f^{\tau'}_{\mathbf{k},j} \ket{\psi} = (u^T)^{\sigma \tau} n^{\tau \tau'}_{ij}(\mathbf{k})  {\overline{u}}^{\tau' \sigma' }\\
			x^{\sigma \sigma'}_{ij}(\mathbf{k}) &= \langle f^{\sigma}_{\mathbf{k},i} f^{\sigma'}_{-\mathbf{k},j} \rangle = \langle \psi | \mathcal{U}^\dagger \mathcal{U} f^{\sigma}_{\mathbf{k},i} \mathcal{U}^\dagger \mathcal{U} f^{\sigma'}_{-\mathbf{k},j} \mathcal{U}^\dagger \mathcal{U} | \psi \rangle = \bra{\psi}  \overline{u^{\tau \sigma}} f^{\tau}_{\mathbf{k},i} \overline{u^{\tau' \sigma'}} f^{\tau'}_{-\mathbf{k},j} \ket{\psi} = (u^\dagger)^{\sigma \tau} x^{\tau \tau'}_{ij}(\mathbf{k})  {\overline{u}}^{\tau' \sigma' }
		\end{split}
	\end{equation}
	for the Fourier space hopping and pairing term expectation values and this $\forall \, u \in \text{SU}(2)$. As a result, $n(\mathbf{k}) = \overline{u} n(\mathbf{k}) u^T$ and $x(\mathbf{k}) = u x(\mathbf{k}) u^T$ implying that $n(\mathbf{k}) = n_1(\mathbf{k}) \otimes \mathbbm{1}$ and $x(\mathbf{k}) = x_2(\mathbf{k}) \otimes J$, mimicking their corresponding Hamiltonian terms. Noting that $\begin{pmatrix}
		d_{\mathbf{k},2j-1} \\
		d_{\mathbf{k},2j}
	\end{pmatrix} = \begin{pmatrix}
		+1 & +1 \\
		+i & -i
	\end{pmatrix} \begin{pmatrix}
		f_{\mathbf{k},j} \\
		f^\dagger_{-\mathbf{k},j}
	\end{pmatrix}$ the correlation matrix can be expressed as
    \begin{equation}
        G(\mathbf{k}) = J^{\oplus N} +i\left[-W n^T(\mathbf{k}) W^\dagger + \overline{W} n(-\mathbf{k}) W^T + W x(\mathbf{k}) W^T + \overline{W} x^\dagger(\mathbf{k}) W^\dagger \right]\,,
    \end{equation}
    where $W = \mathbbm{1}_N \otimes \begin{pmatrix}
		1 \\
		i
	\end{pmatrix}$. Consequently, the terms containing $n(\mathbf{k})$ will yield contributions of the form $... \otimes \mathbbm{1} \otimes \mathbbm{1}$ and $... \otimes \mathbbm{1} \otimes J$ while those with $x(\mathbf{k})$ give $... \otimes J \otimes \sigma_x$ and $... \otimes J \otimes \sigma_z$. Therefore, a Gaussian TI state is SU(2) symmetric with spin $\frac{1}{2}$ when
	\begin{equation}
		G(\mathbf{k}) = G_0(\mathbf{k}) \otimes \mathbbm{1} \otimes \mathbbm{1} +
		G_1(\mathbf{k}) \otimes \mathbbm{1} \otimes J +
		G_2(\mathbf{k}) \otimes J \otimes \sigma_x +
		G_3(\mathbf{k}) \otimes J \otimes \sigma_z \, .
		\label{quatdecomp}
	\end{equation}
	Together, $\mathbbm{1} \otimes \mathbbm{1}$, $\mathbbm{1} \otimes J$, $J \otimes \sigma_x$  and $J \otimes \sigma_z$ span a real representation of the quaternions so that the correlation matrix of a symmetric state can equivalently be formulated in quaternion form $G^\mathbbm{H}(\mathbf{k}) = G_0(\mathbf{k}) + i G_1(\mathbf{k}) + j G_2(\mathbf{k}) + k G_3(\mathbf{k})$ where ${G^\mathbbm{H}}^\dagger(\mathbf{k}) = -G^\mathbbm{H}(\mathbf{k})$ and ${G^\mathbbm{H}}^\dagger(\mathbf{k}) G^\mathbbm{H}(\mathbf{k}) \leq \mathbbm{1}$.
	
	Knowing how symmetries and in particular SU(2) symmetry are realised in Gaussian models and states, we ask the question of how to build these into the GfTNS \textit{Ansatz}. A natural way to do so is by starting with symmetric input states and projecting these in a symmetric way to the physical level. As we want our fPEPS to be Gaussian, operators on the physical and virtual level are only coupled quadratically, implying that operators transforming according to different irreducible representations cannot couple in a symmetric way. Hence, also the input state should be constructed out of SU(2) multiplets, and its correlation matrix should also decompose in the quaternionic, spin-$\frac{1}{2}$ manner of Eq.\,\eqref{quatdecomp}. As two Majorana operators cannot transform under the spin-1/2 representation of SU(2) (as this representation is not real, whereas a canonical transformation of Majorana operators is real orthogonal), we need at least $\chi=4$ Majorana operators, or thus a full spinful fermion, as elementary building block for the input state. Taking $\chi$ an integer multiple fo four, a symmetric input state can be constructed with
	\begin{equation}
		G_\text{in}(\mathbf{k}) = \left(\begin{pmatrix}
			& e^{- i k_x}\\
			e^{ i k_x} &
		\end{pmatrix} \otimes J \otimes \sigma_x \right)^{\oplus \frac{\chi}{4}} \oplus \left(\begin{pmatrix}
			& e^{- i k_y}\\
			e^{ i k_y} &
		\end{pmatrix} \otimes J \otimes \sigma_x \right)^{\oplus \frac{\chi}{4}} \, .
	\end{equation}
	The GfTNS can be made symmetric by requiring the channel correlation matrix to be quaternionic as well, i.e.~$X \cong X^\mathbbm{H}= X^0+i X^1 +j X^2 +k X^3$  with $X^\mathbbm{H} = -{X^\mathbbm{H}}^\dagger$ and $X^\mathbbm{H} {X^\mathbbm{H}}^\dagger \leq \mathbbm{1}$.
	
	\section{Appendix E -- Kraus-Schuch formalism for spinful fermions: optimisation method and additional results}
	
	In this work, the results for spinful fermions were obtained by approximating the ground state of the spinful model by SU(2) symmetric GfTNS as introduced in Appendix C. In this case, the optimal GfTNS is found by minimising the difference between its correlation matrix $G_\text{out}(\mathbf{k})$ with that of the exact ground state,  $G_\text{ex}(\mathbf{k})$. This procedure has been applied for different bond dimensions ($D=4,16,64$) and for increasing system sizes. To probe the thermodynamic limit, a procedure similar to \cite{Mortier2022} was followed where optimised results for smaller system sizes were used as initial guesses for larger systems reaching up to a $1000 \times 1000$ square lattice. As the model contains both spin up and down fermions, the exact parity is even throughout the full Brillouin zone, allowing to use periodic boundary conditions and even system lengths without parity obstructions. 
	
	For the successive optimisations, a Riemannian limited-memory, quasi-Newton (L-BFGS) procedure, based on \cite{Boumal2014}, was applied to find the minimum of the cost function,
	\begin{equation}
		C_\text{F} = \frac{1}{N_s} \sum_{\mathbf{k}} \left(\frac{1}{\sqrt{2N}} ||G_\text{out}(\mathbf{k})-G_\text{ex}(\mathbf{k})||_\text{F}\right)^2=\frac{1}{2 N N_s} \sum_{\mathbf{k}}||\Delta G(\mathbf{k})||_\text{F}^2 \, ,
	\end{equation}
	where $|| \quad ||_\text{F}$ denotes the Frobenius norm. Its Euclidean gradient w.r.t.\,the variational parameters in $X^\text{loc}$ (we will drop the superscript) is easily computed,
	\begin{equation}
		\begin{split}
			\frac{\partial C_F}{\partial X_{ab}} = \frac{1}{N N_s} \sum_\mathbf{k} \left(\overline{\Delta G}^{ij}(\mathbf{k}) \frac{\partial G_{\text{out}}^{ij}(\mathbf{k})}{\partial X_{ab}} + \Delta G^{ij}(\mathbf{k}) \frac{\partial \overline{G_{\text{out}}}^{ij}(\mathbf{k})}{\partial X_{ab}}\right)
		\end{split} \, ,
	\end{equation}
	where one can use the Schur complement formula to obtain
	\begin{equation}
		\begin{split}
			\frac{\partial G_\text{out}^{ij}}{\partial A_{ab}}(\mathbf{k}) &= \delta_{ia} \delta_{jb} \qquad \qquad \qquad \qquad \frac{\partial G_\text{out}^{ij}}{\partial C_{ab}}(\mathbf{k}) = -\delta_{jb} B_{iu} K_{ua}(\mathbf{k}) \\
			\frac{\partial G_\text{out}^{ij}}{\partial B_{ab}}(\mathbf{k}) &=  -\delta_{ia} K_{bv}(\mathbf{k}) C_{vj} \qquad \qquad
			\frac{\partial G_\text{out}^{ij}}{\partial D_{ab}}(\mathbf{k}) = (BK)_{ia}(\mathbf{k}) (KC)_{bj}(\mathbf{k})
		\end{split}
		\label{dGdX}
	\end{equation}
	with $K(\mathbf{k}) = (D-G_\text{in}(\mathbf{k}))^{-1}$. The parametric $X$ matrix is not fully arbitrary since $X = -X^\dagger$, $X X^\dagger \leq \mathbbm{1}$ in addition to $X \cong X^\mathbbm{H}$ for SU(2) symmetry. Assuming purity and without any additional symmetry, we therefore parametrize $X$ as $X=QEQ^T$ with $Q$ orthogonal and $E$ containing $\pm J$ blocks. All can be set to $+J$ via an orthogonal transformation that can be absorbed in $Q$. As we intend Riemannian optimisation over a manifold of variational parameters, we would like $Q$ to live on a connected Riemannian manifold and thus restrict it to SO$(2(N+2\chi))$. The parity of the number of $-J$ blocks in $E$ does matter now as it cannot be changed by a special orthogonal $Q$. Both parity choices in $E$ should thus be considered. Similarly, SU(2) symmetry can be imposed by parameterizing $X$ as $X=QEQ^T$ where all blocks now have to be (equivalent to) quaternions. For $E$, this implies that only $\pm J$ blocks with pairwise equal signs can appear, corresponding to $\pm \mathbbm{1} \otimes J \cong \pm i$. The sign does not matter as $-i$ can always be transformed to $+i$ (e.g., via $k (-i) \overline{k} = i$). We thus choose $E^\mathbbm{H} = \bigoplus_{n=1}^{\frac{N+2\chi}{2}} (+ i)$. $Q$, on the other hand, has to be orthogonal so that $Q^\mathbbm{H} {Q^\mathbbm{H}}^\dagger = \mathbbm{1}$. Consequently, the parametric manifold is given by U$^\mathbbm{H}(\frac{N+2\chi}{2})$, the unitary quaternion matrices with linear dimension $\frac{N+2\chi}{2}$. For SU(2), there are no different subclasses in the \textit{Ansatz} as the spin labeling the irreducible representations is fully imposed by the physical level and allows no further freedom. We conclude that $X$ is parametrized via an orthogonal $Q$ living on a specific, symmetry-dependent manifold. These manifolds are Riemannian submanifolds of $\mathbbm{R}^{2(N+2\chi) \times 2(N+2\chi)}$ so that the Riemannian gradient of the cost function w.r.t.\,$Q$ can be expressed as $\text{grad } C_F(Q) = \text{Proj}_Q \left(\text{grad } \overline{C_F}(Q)\right)$ where $\overline{C_F}(Q)$ is the smooth extension of $C_F(Q)$ to $\mathbbm{R}^{2(N+2\chi) \times 2(N+2\chi)}$ and $\text{Proj}_Q$ the orthogonal projector of ambient vectors in the Euclidean space to the tangent bundle of the embedded manifold at $Q$ (see \cite{Boumal2014} for more details). Using antisymmetry and the chain rule, the Euclidean gradient can be expressed as $\text{grad } \overline{C_F}(Q) = \frac{\partial \overline{C_F}}{\partial Q} = \frac{\partial C_F}{\partial X} (-2 Q E)$. The orthogonal projector, on the other hand, can always be expressed as $\text{Proj}_Q(V) = V - Q \, \text{sym}(Q^T V)$ where $\text{sym} (X) = \frac{1}{2} (X+X^T)$ so that $\text{grad } C_F(Q) = -2 \frac{\partial C_F}{\partial X} Q E + 2 Q \, \text{sym}(Q^T \frac{\partial C_F}{\partial X} Q E)$. This gradient and projector were combined with a suitable (here, $QR$-based) retraction to build the L-BFGS solver.
	
\begin{figure}[H]
		\includegraphics[scale=0.2]{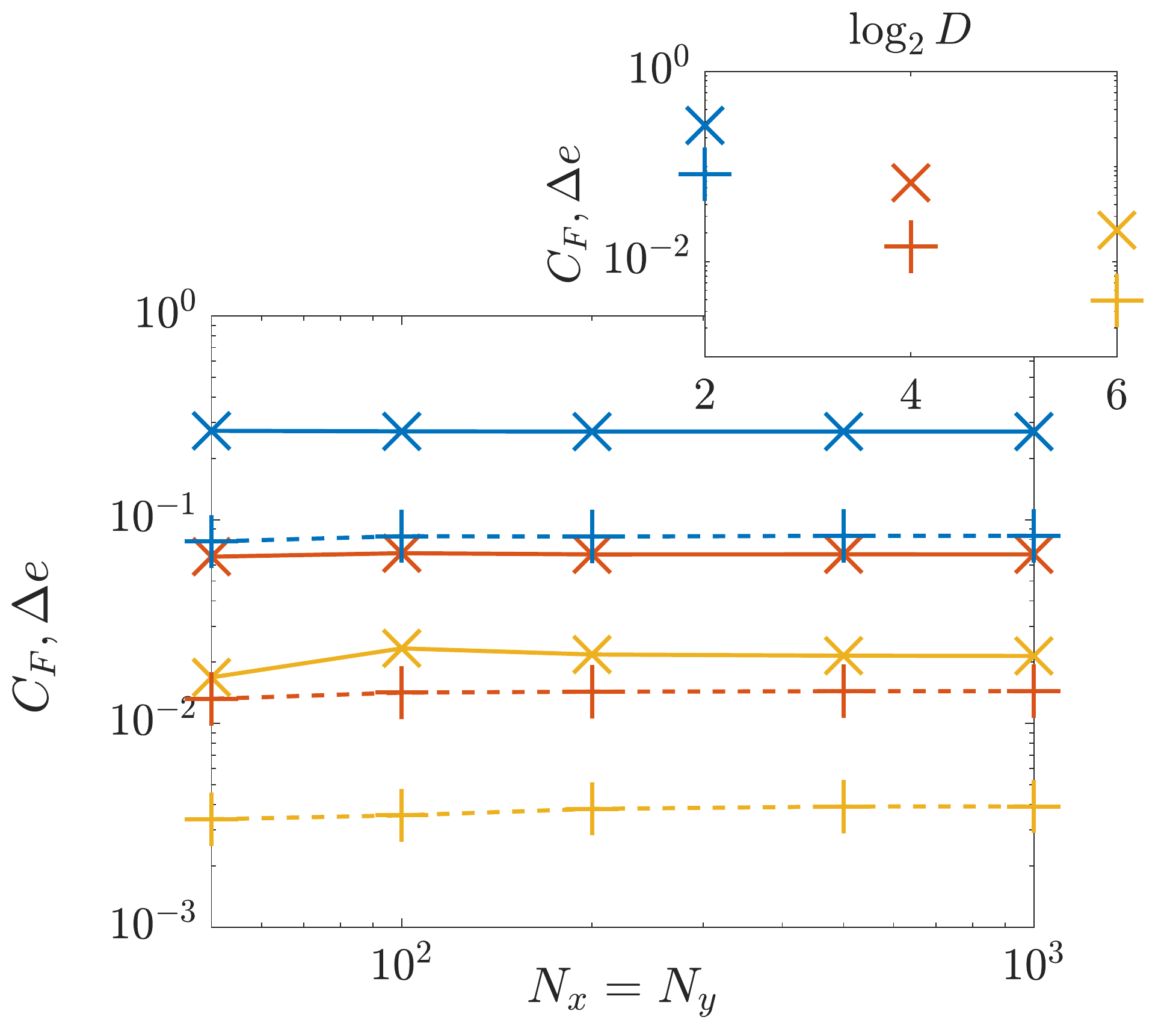}
		\includegraphics[scale=0.2]{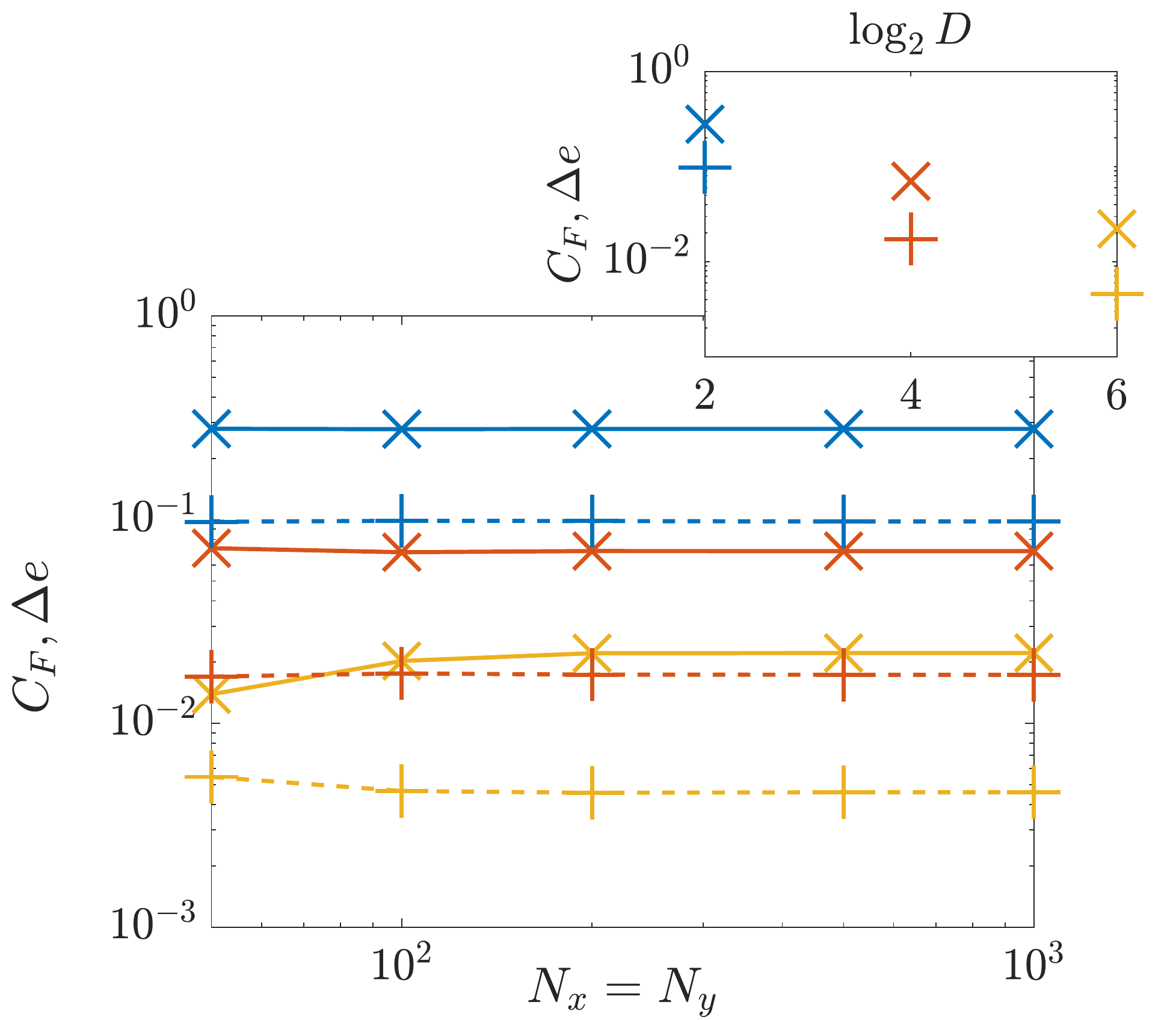}
		\includegraphics[scale=0.2]{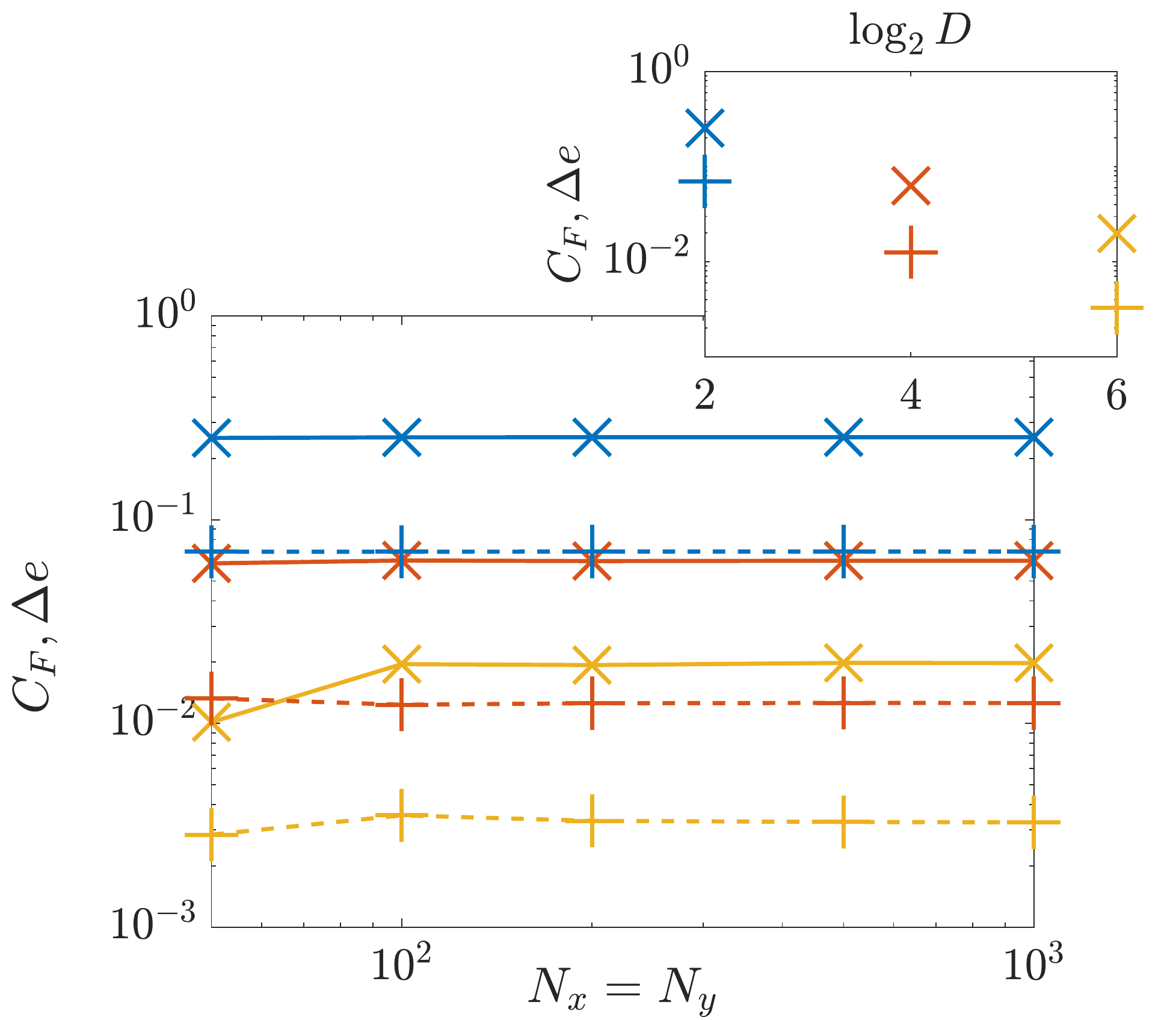}
		\includegraphics[scale=0.2]{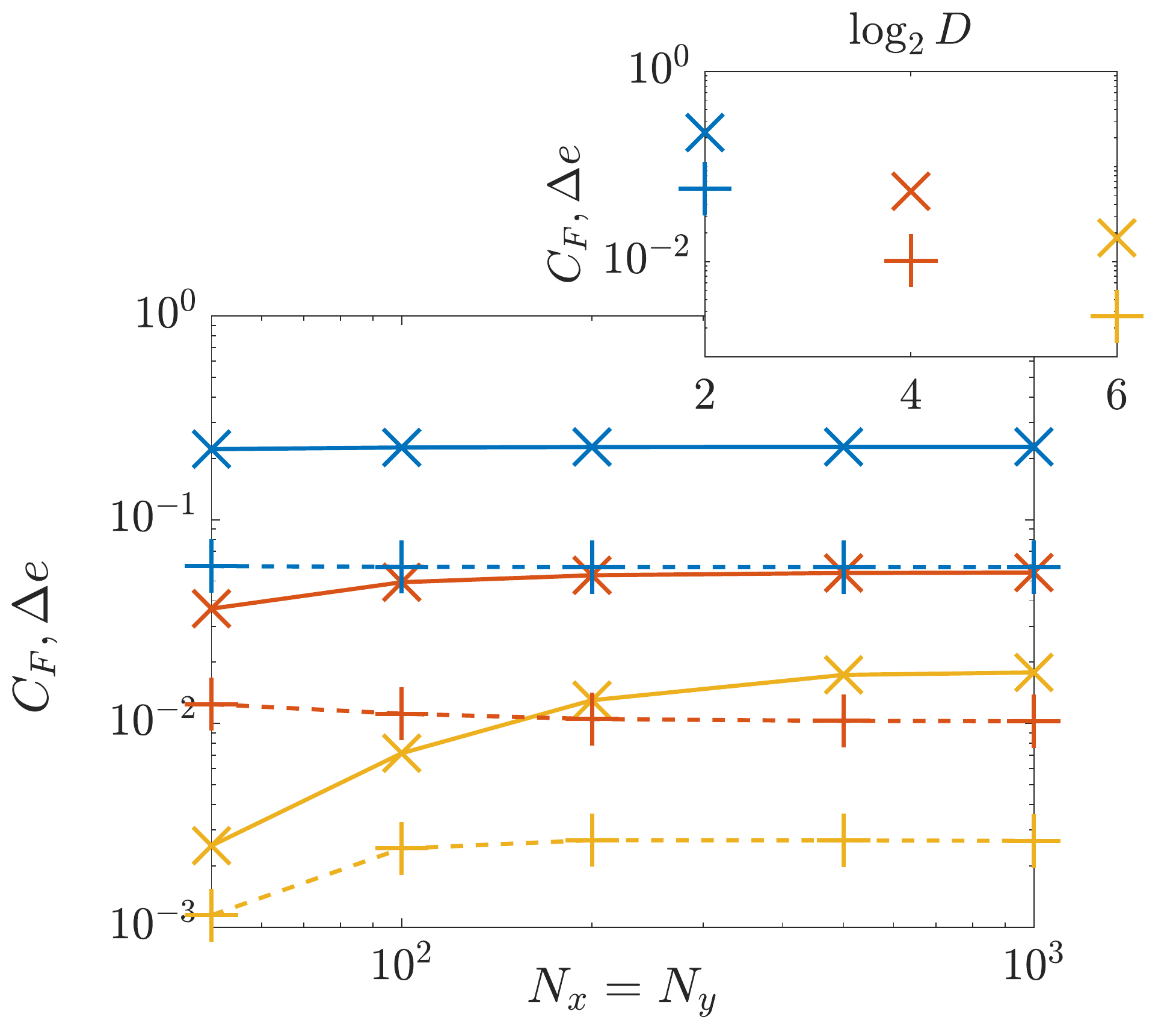}
		\includegraphics[scale=0.2]{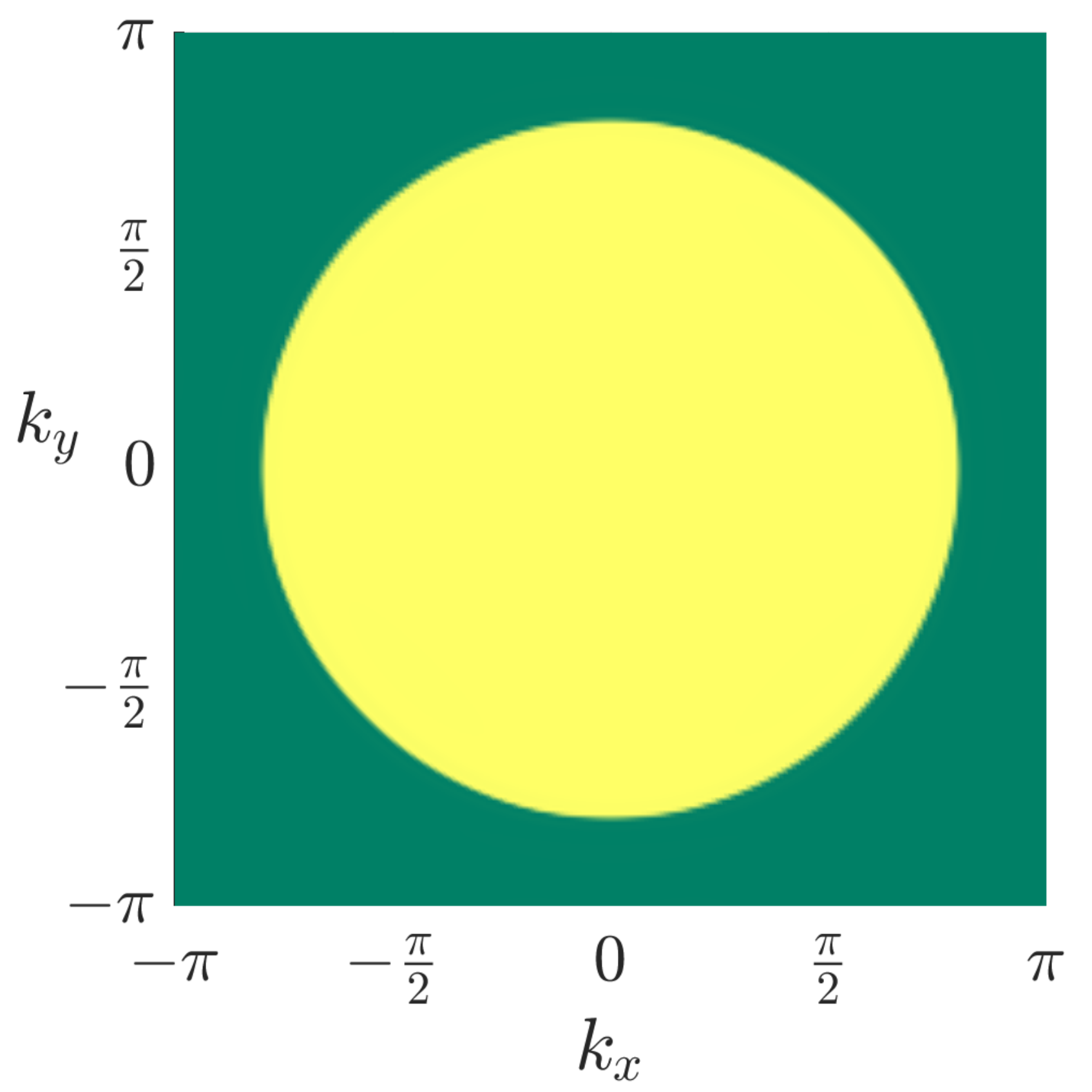}
		\includegraphics[scale=0.2]{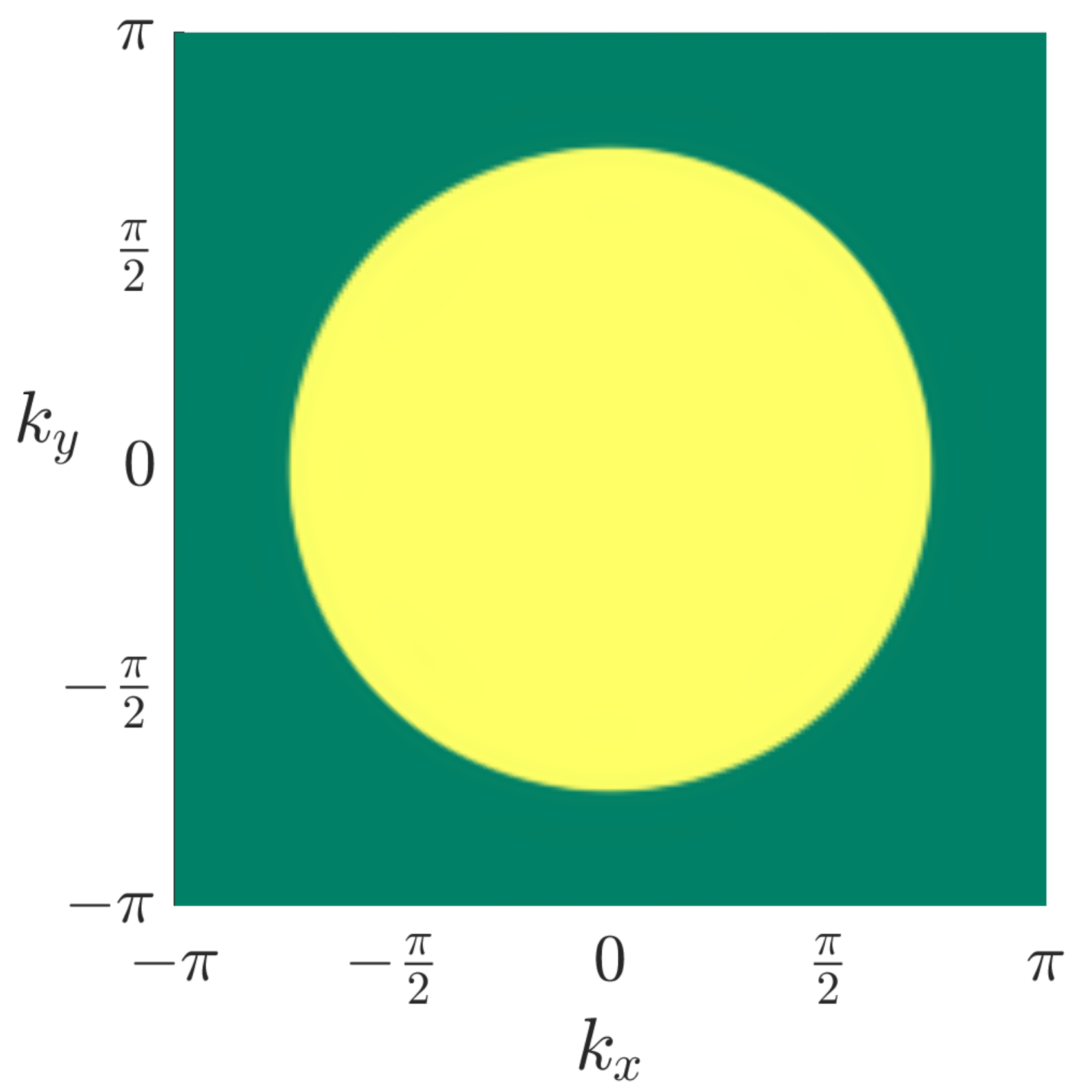}
		\includegraphics[scale=0.2]{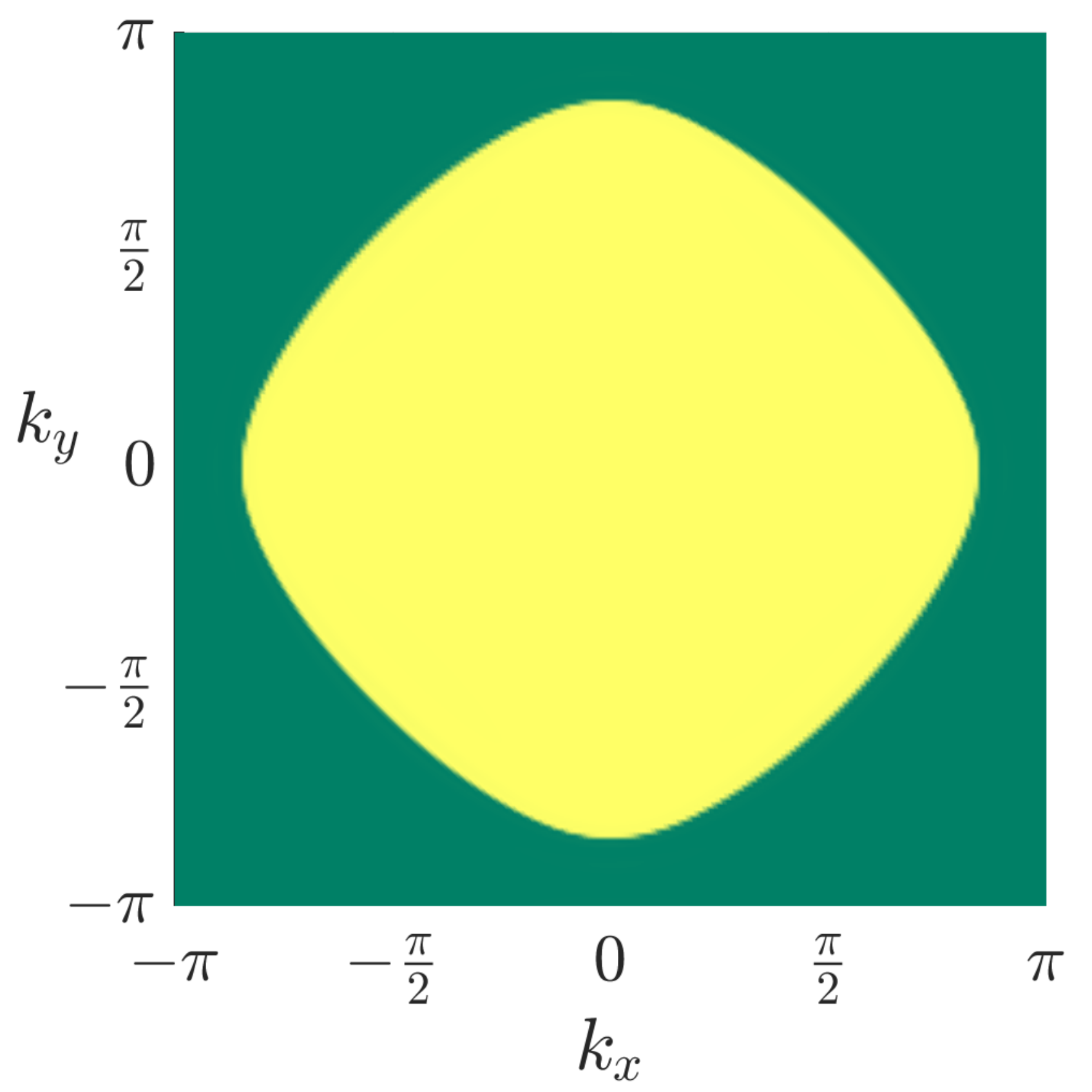}
		\includegraphics[scale=0.2]{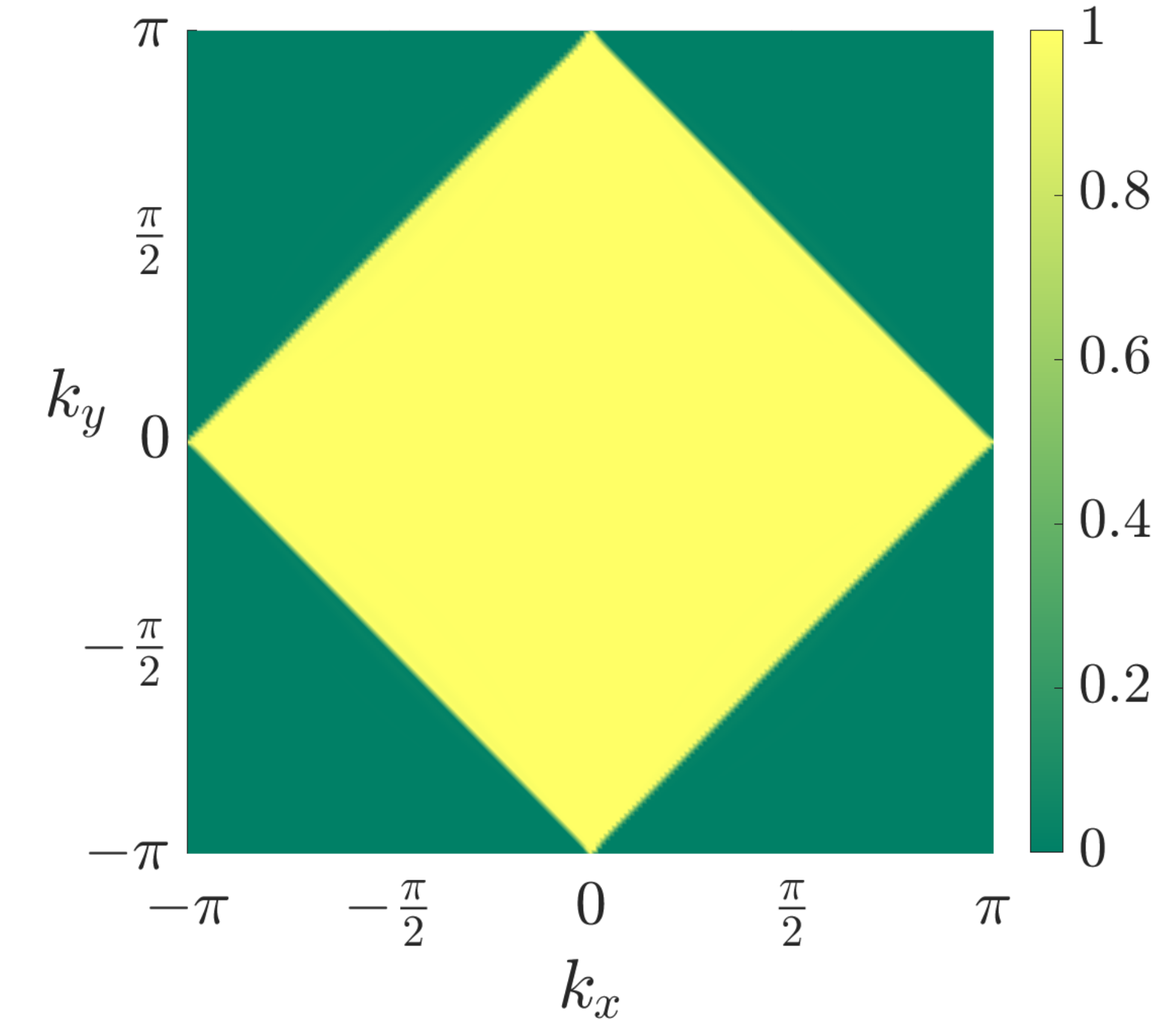}
		\includegraphics[scale=0.2]{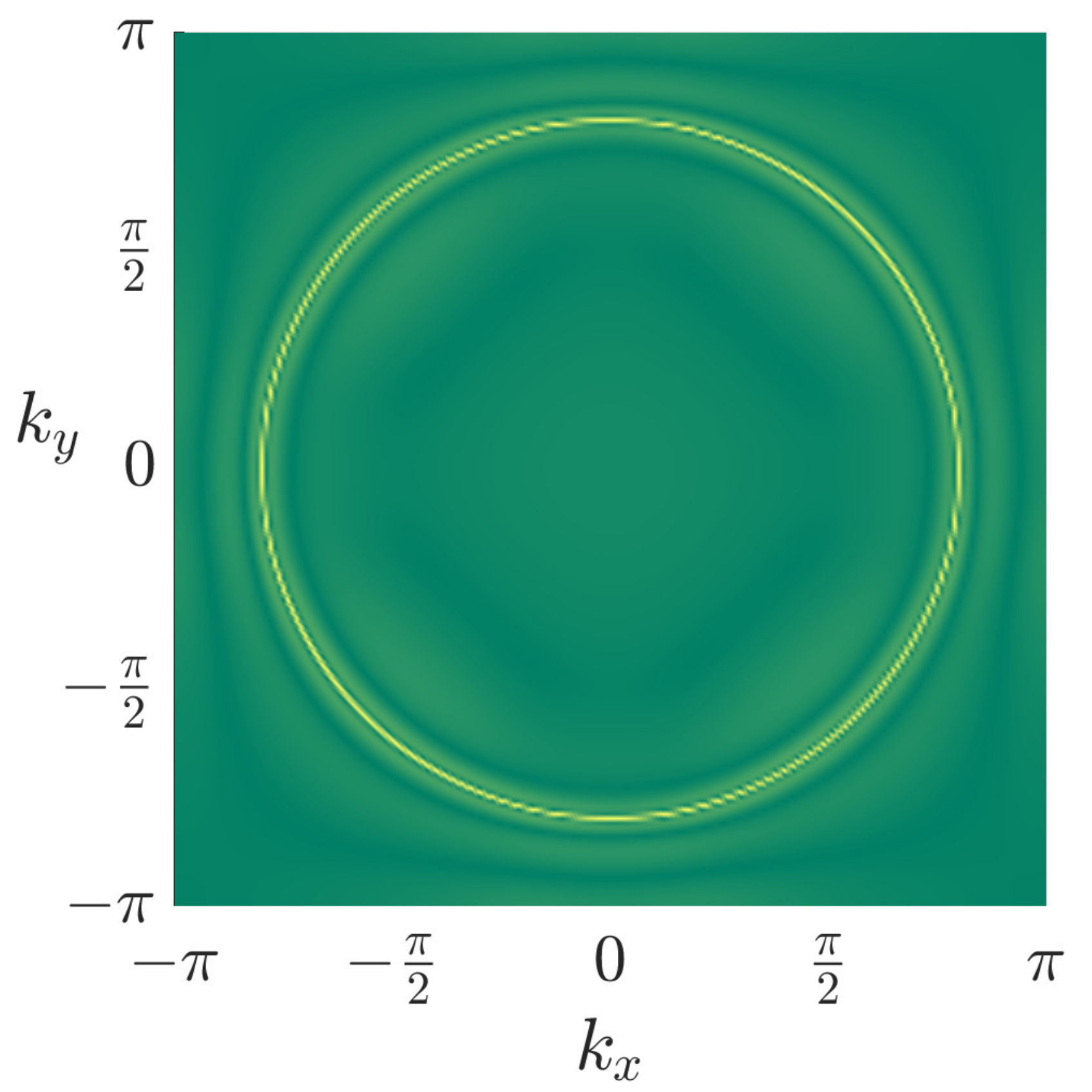}
		\includegraphics[scale=0.2]{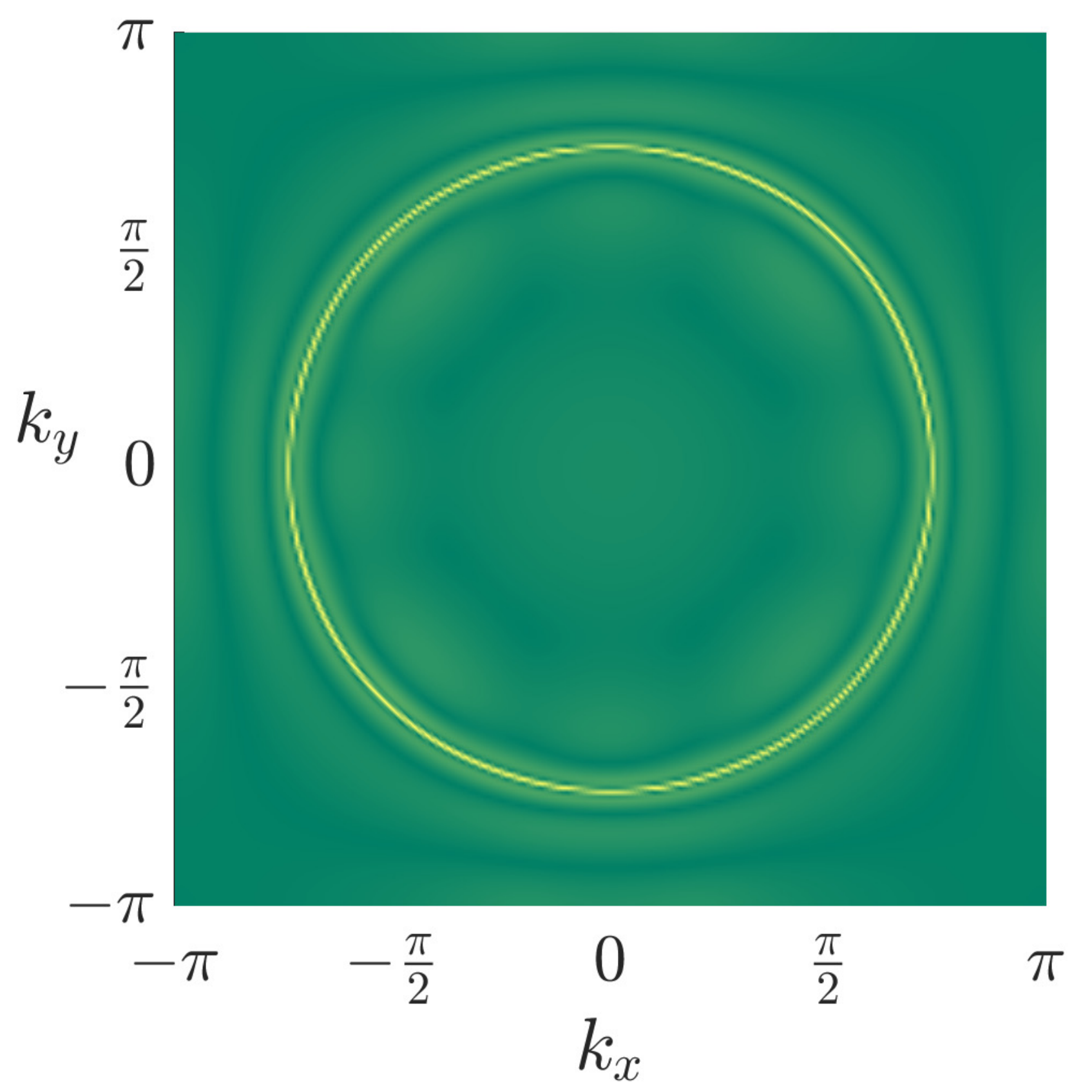}
		\includegraphics[scale=0.2]{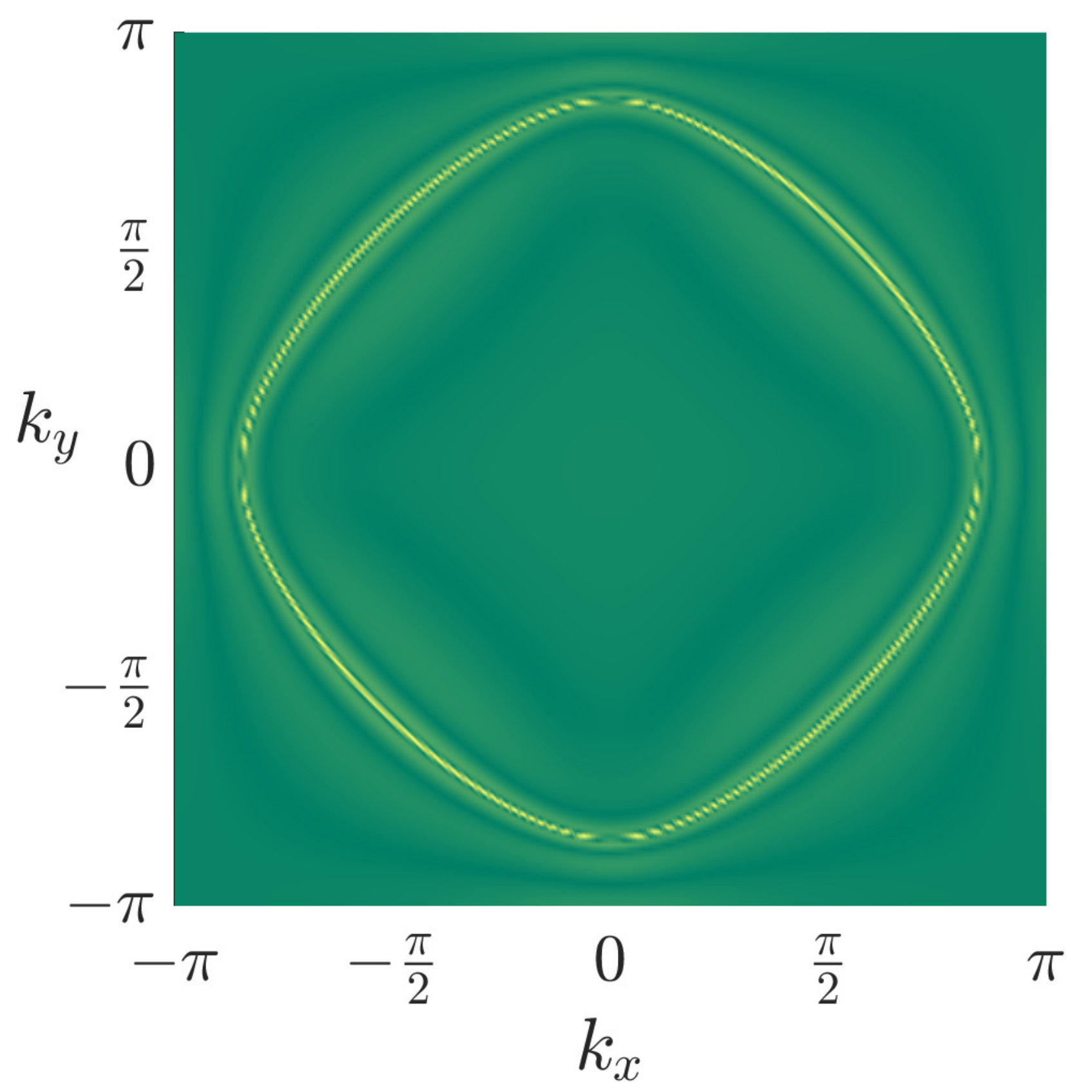}
		\includegraphics[scale=0.2]{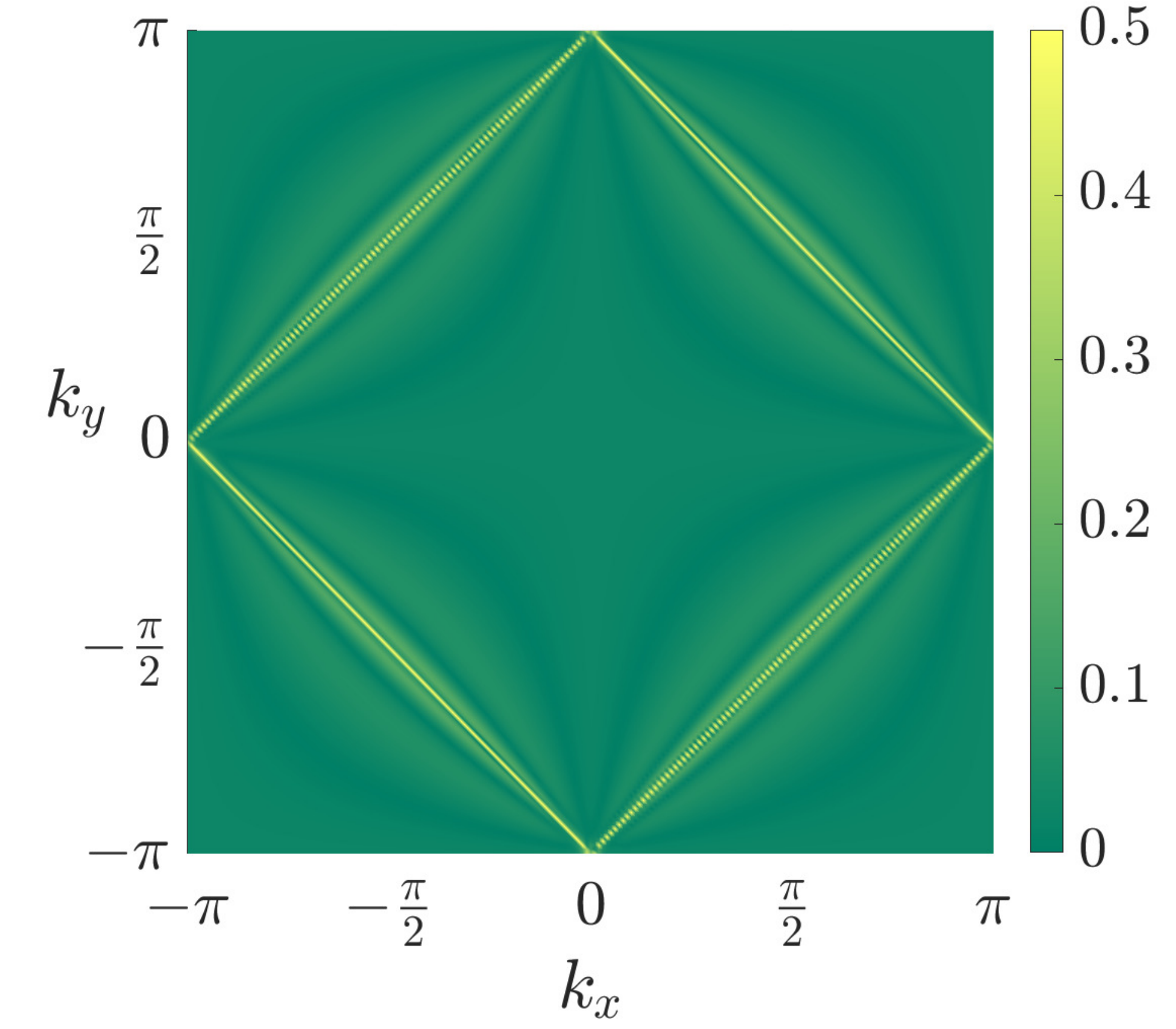}
		\includegraphics[scale=0.21]{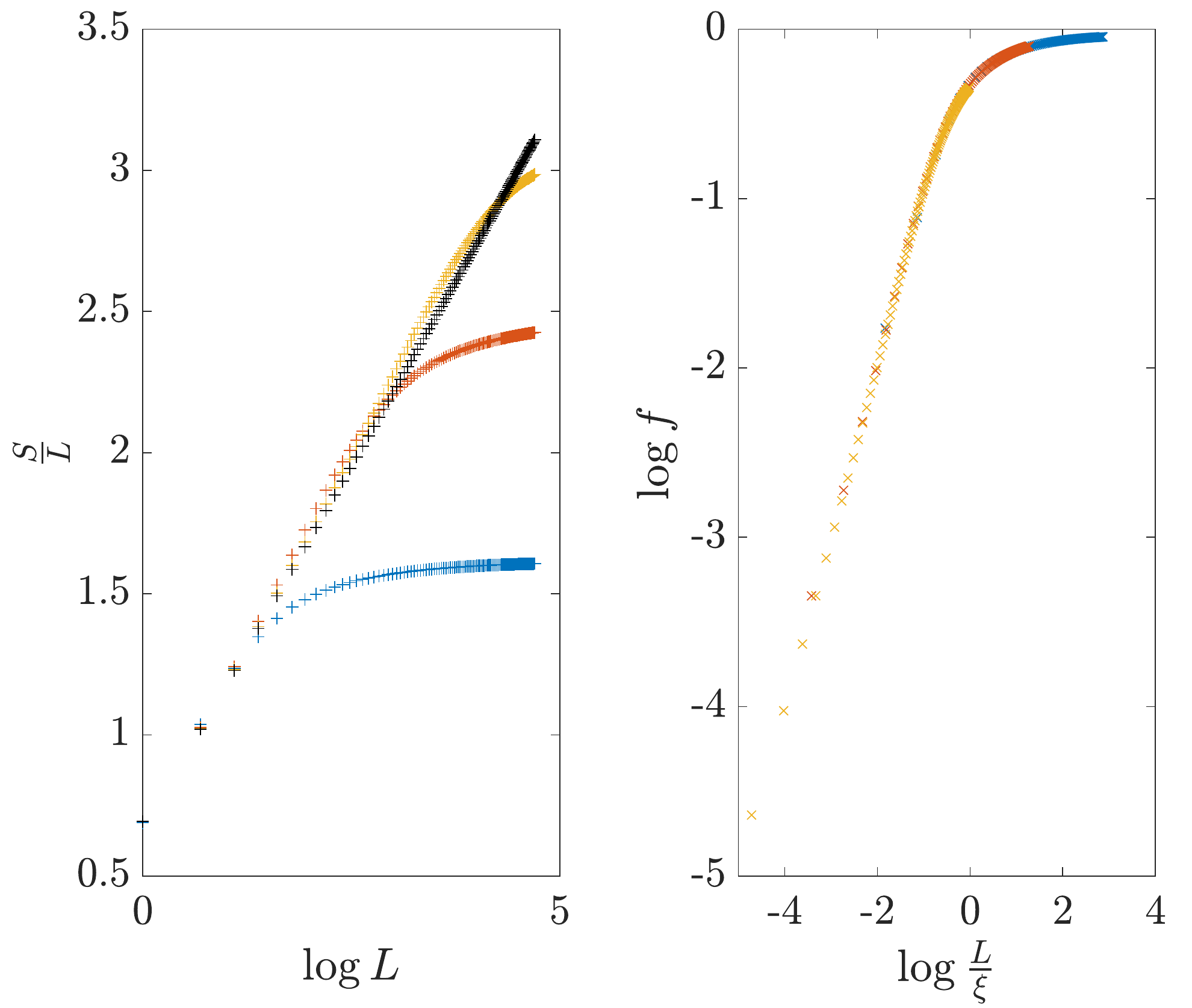}
		\includegraphics[scale=0.21]{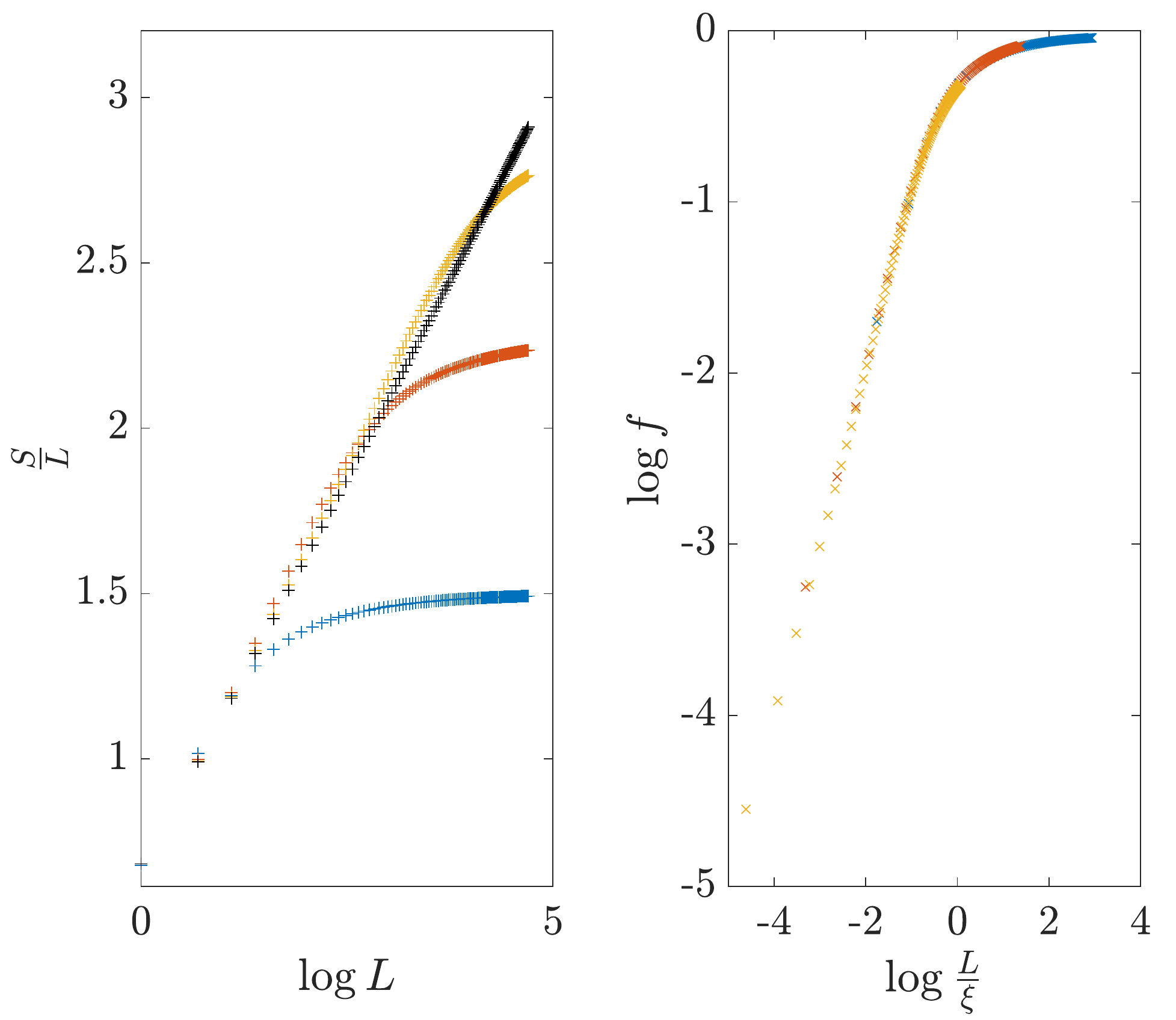}
		\quad
		\includegraphics[scale=0.21]{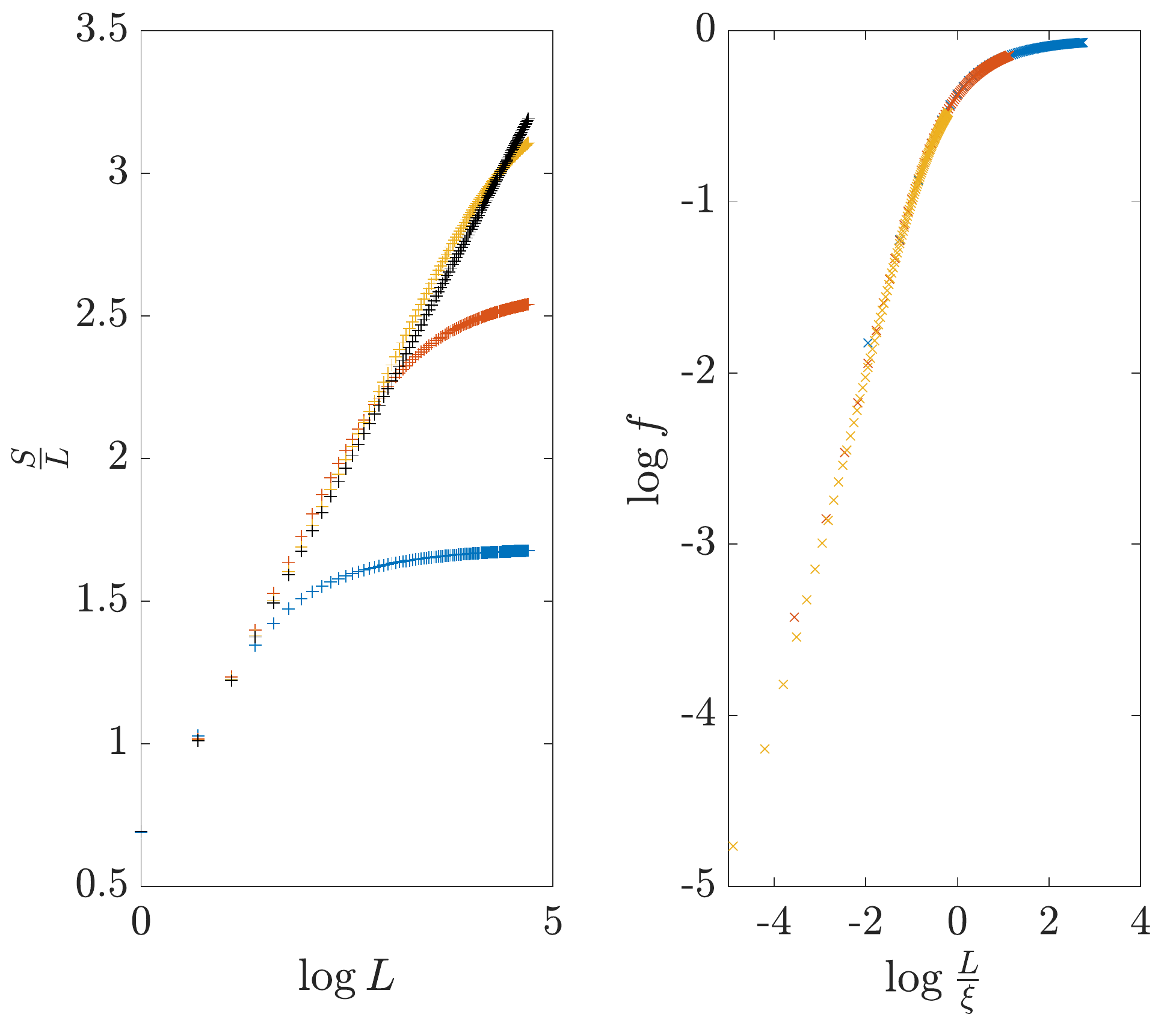}
		\includegraphics[scale=0.21]{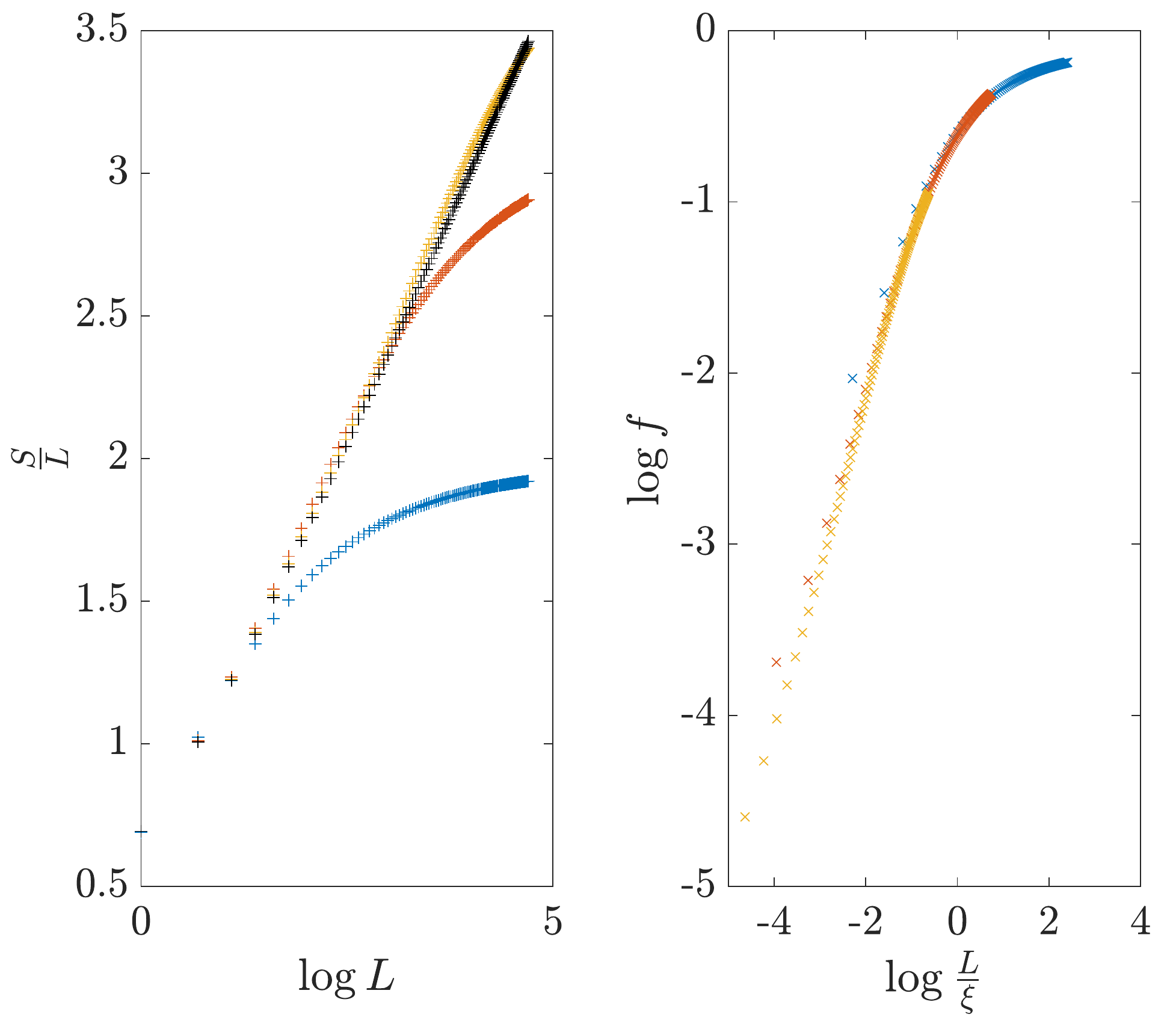}
		
		\caption{Collection of GfTNS results for the spinful model. The columns correspond to different parameter choices for $(t,t',\mu)$, respectively $(1,0.353,0.754)$, $(1,0.353,0.3)$, $(1,0.2,0.476)$ and $(1,0,0)$. The first row collects convergence results of the GfTNS optimisation for different linear system lengths $N_x=N_y$ with blue, red and yellow markers for $D=4,16,64$. Cross (plus sign) markers correspond to the optimised values of the cost function $C_F$ (energy density error $\Delta e$). The full, respectively dashed, lines were added as guide for the the eye and indicate that the thermodynamic limit was probed for the largest system sizes. These results were also repeated in the inset and show a steady decrease as a function of the bond dimension. The second and third row display the modal occupation $n_1(\mathbf{k})$ and spin-coupled pairing $x_2(\mathbf{k})$ throughout the Brillouin zone for the $D=64$ GfTNS at $N_x=200$. While the former quantity shows that the TNS can reproduce a sharp transition from fully occupied to vacated at the Fermi surface, the latter quantity maximises it magnitude there as in Fig.~\ref{fig:PEPS_spinless}(d). In the bottom row, the left panels display the exact EE of a $L\times L$ subregion in black and compare it to analogous GfTNS results for different bond dimensions using the color scheme of the first row. GfTNS results were collapsed according to Eq.\,\eqref{scalinglaw} on the right panels. }\label{fig:extra_spinful}
	\end{figure}

\begin{figure}[h]
		\centering
		\includegraphics[scale=0.25]{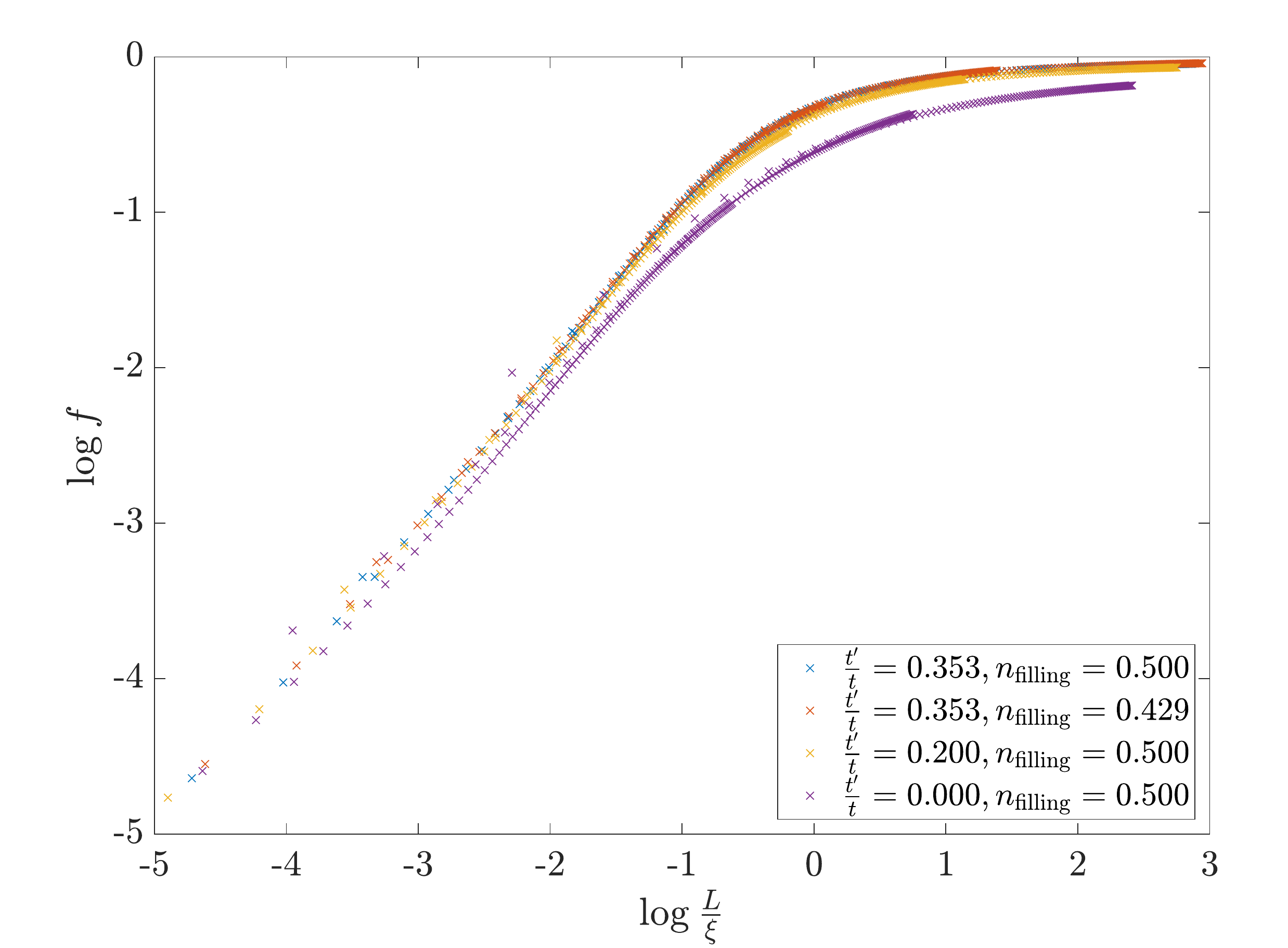}
		\caption{Scaling function for the different parameter choices. While the scaling function depends on the Fermi surface geometry with the $f(x)$ shifting to lower values for the nested case, it is independent of the filling fraction.}
		\label{fig:collapse_spinful}
	\end{figure}

	In the main text, we introduced the finite-entanglement scaling law (Eq.\,\eqref{scalinglaw}). Although the scaling function $f(x)$ is not identical for every model, some degree of universality is present. As in the spinless case, the scaling function does not seem to depend on the filling fraction. The shape of the Fermi surface on the other hand does have an influence. A more nested, square-like shape seems to widen the transition region between the asymptotic behaviors $f(x)\approx x$ for small values of $x$ (subject to subleading corrections to the entanglement entropy for small $L$ and thus small $x$) and $f(x) \approx 1$ for large values of $x$. To verify this, we considered several cases, including the spinful equivalent of Eq.\,\eqref{Hamiltonian} with $(t,t',\mu) = (1,0.353,0.754)$, a non-half-filled version with filling fraction $n_\text{filling}=0.429$ where $(t,t',\mu) = (1,0.353,0.3)$, a non-circular Fermi surface variant with $(t,t',\mu) = (1,0.2,0.476)$, and a model without next-nearest-neighbour hopping terms with $(t,t',\mu) = (1,0,0)$, leading to a highly nested, square Fermi surface. In all four cases, we confirmed the scaling law in Fig.~\ref{fig:extra_spinful}. The entanglement entropies of the circular models with different filling fractions can be collapsed onto a single, \textit{universal} scaling function, while this is not the case for the Fermi surfaces with different shapes, as shown in Fig.~\ref{fig:collapse_spinful}.
	
	We used the SU(2) symmetric optimisation method for $D=4,16,64$, gradually increasing the system size to obtain the results. The first row of Fig.~\ref{fig:extra_spinful} shows the cost function $C_F$, based on the correlation matrix, and the energy density error $\Delta e$ for the optimised states at linear system length $N_x=N_y$. These results indicate that for the largest system sizes, $N_x$ no longer affects the outcome, suggesting that the thermodynamic limit is being properly probed. In the insets, we compare the limit values for the different bond dimensions and find a consistent decrease. Comparing these plots with those in Fig.~\ref{fig:PEPS_spinless} (a), we observe that the larger magnitude in the former can be attributed to the model having twice as many physical degrees of freedom. Additionally, optimising $C_F$ instead of $e$ allows $\Delta e$ to be slightly higher than its minimum minimorum. Finally, the enforced symmetry results in a significantly smaller number of parameters at the same bond dimension. In the second and third row of Fig.~\ref{fig:extra_spinful}, the Fourier space occupation $n_1(\mathbf{k})$, respectively, pairing $x_2(\mathbf{k})$ (with $n(\mathbf{k}) = n_1(\mathbf{k}) \otimes \mathbbm{1}$ and $x(\mathbf{k}) = x_2(\mathbf{k}) \otimes J$)  are displayed for the largest bond dimension and for $N_x=200$. The sharp transition in the modal occupation at the Fermi surface is clearly reproduced by the GfTNS with the same circle at $k_F=\sqrt{2\pi}$ as in Fig.~\ref{fig:PEPS_spinless}(d) appearing for the half-filled model. For $\mu=0.3$, on the other hand, one obtains a smaller circle at $k_F=2.314$ while the third column with $t'=0.2$ transitions to the square Fermi surface in the fourth column. Similar to Fig.~\ref{fig:PEPS_spinless}, the magnitude of the pairing terms maximises at the Fermi surface to facilitate the transition from filled the empty in line with the Lieb-Schultz-Mattis theorem. However, it is spread out more than in the spinless case which can again be attributed to the double amount of physical flavors. 		
	
	The left panels in the final row of Fig.~\ref{fig:extra_spinful} display a comparison between the exact and the GfTNS EE for linear subsystem lengths, $L$, varying from 1 to 110 and at different $D$. The exact $\sim L \log L$ profile is followed up to a certain correlation length $\xi$, after which the finite entanglement results in a saturation of the curves. In the right panels these data are collapsed using the scaling law of Eq.\,\eqref{scalinglaw}. As was mentioned in the main text, the Widom factor for the circular Fermi surface is $\frac{2}{3 \pi} k_F L$ while for the square Fermi surface  Eq.\eqref{Sgen} yields $\frac{2}{3} L$. When $t'/t=0.2$, the double integral has to be calculated numerically, yielding $0.563 L$. The utilized correlation lengths were determined in the same way as in Appendix B and all collapsed results were combined in Fig.~\ref{fig:collapse_spinful}, confirming the scaling function depends on the Fermi surface morphology but not on the filling fraction.
	
	\begin{table}[h]
		\centering
		\begin{tabular}{c|c|c||c|c|c}
			$t'/t$ & $\mu / t$ & $n_{\mathrm{filling}}$ &$D=4$ & $D=16$ & $D=64$ \\ 
			\hline
			$0.353$ & $0.754$ & $0.500$ & $6.255$ & $30.63$ & $111.8$ \\
			$0.353$ & $0.300$ & $0.429$ & $5.797$ & $27.58$ & $101.1$ \\
			$0.200$ & $0.476$ & $0.500$ & $7.032$ & $35.16$ & $134.0$ \\
			$0.000$ & $0.000$ & $0.500$ & $9.868$ & $52.06$ & $206.2$
		\end{tabular}
		\caption{Bulk correlation lengths $\xi$ determined from the bottom row of Fig.~\ref{fig:extra_spinful} as described in Appendix B.}
		\label{tab:corr_lengths}
	\end{table}
	
	\section{Appendix F -- GfTNS in 1D cannot reproduce the power law relation between correlation length and bond dimension}
	
\begin{figure}[H]
		\centering
		\includegraphics[scale=0.4]{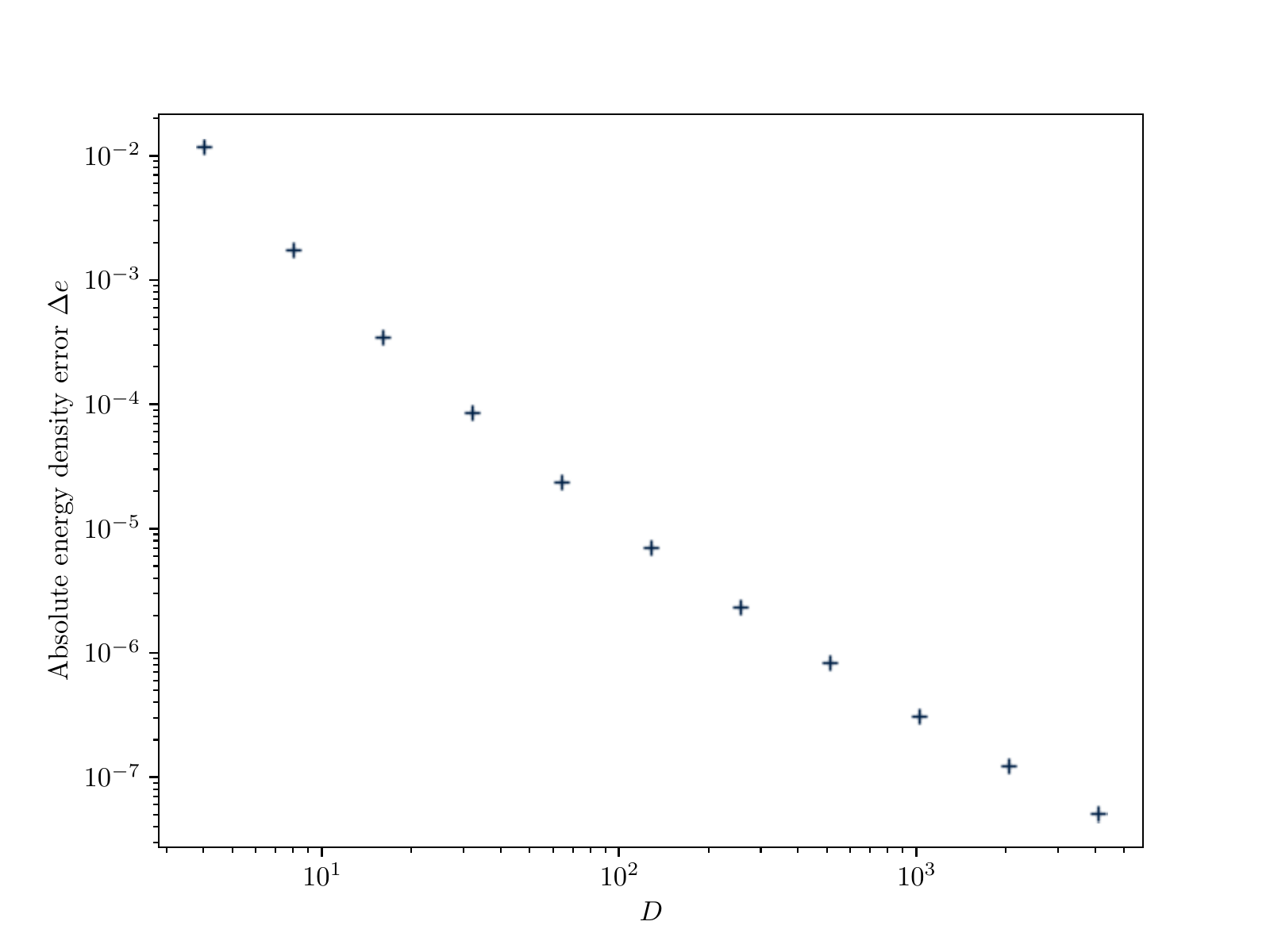}
		\caption{The energy density error of optimised GfMPS.}
		\label{fig:gfmps energy error}
	\end{figure}

	In Ref.~\cite{FrancoRubio2022}, the authors showed that GfMPS approximate critical states significantly less efficiently than generic MPS. Here we back this claim from another perspective, namely by showing that GfMPS cannot reproduce the power-law relation $\Delta e \sim D^{-\omega}$ (recall that $\Delta e$ is the error in the energy density), which is known to hold for generic MPS approximations to one-dimensional critical ground states \cite{Tagliacozzo2008, Pollmann2009, Pirvu2012}. Here we study a one-dimensional fermion system, with antiperiodic boundary condition, described by the following hopping Hamiltonian, 
	\begin{equation}
		H = -\sum_{\langle ij \rangle} \left(f^\dag_i f_j + f_j^\dagger f_i\right)\,.\label{eq:onedimensionalfreefermionshopping}
	\end{equation}
	We work with system size $N_s = 100\,000$, and $\mu = 0$ to ensure half-filling. Using the GVW formalism, we optimised GfMPS of $D = 4,\dots, 4096$. For the optimised GfMPS, the energy density errors are plotted in Fig.~\ref{fig:gfmps energy error}.	A closer examination of the data shows a clear deviation from the power-law behavior. Since our bond dimensions satisfy $D\equiv 2^M$, $\frac{\Delta e (M+1)}{\Delta e (M)}$ would be a constant if the power law relation is satisfied. However, as we can see in Table.~\ref{tab:gfmps energy error}, this is clearly not the case.
	Moreover, in the generic MPS case, $\frac{\Delta e (M+1)}{\Delta e (M)} = 5.28$ for the model we wrote down. Thus the data presented in Table.~\ref{tab:gfmps energy error} shows that GfMPS are not able to approximate critical states as nearly as well as generic MPS with comparable bond dimension.
	
        \begin{table}[h]
		\centering
		\begin{tabular}{lllllllllll}
			\hline
			$M$ & 2 & 3 & 4 & 5 & 6 & 7 & 8 & 9 & 10 & 11  \\
			\hline
			$\frac{\Delta e (M+1)}{\Delta e (M)}$ & 6.6962 & 4.9788 &  4.1240 & 3.6087 & 3.2619 & 3.0111 & 2.8206 & 2.6707 & 2.5500 & 2.4519 \\
			\hline
		\end{tabular}
		\caption{The energy density error ratios between adjacent GfMPS.  \label{tab:gfmps energy error}}
	\end{table}
 
	This behaviour is quite well understood for GfMPS. For a bipartition of the ground state (with open boundary conditions) into a left and right half, the reduced density matrix is also Gaussian, and can be diagonalised to take the form $\exp\left(- \sum_{m} \nu_m b_m^\dagger b_m\right)$. For the Hamiltonian in Eq.\,\eqref{eq:onedimensionalfreefermionshopping} in particular, the values $\nu_m$ are proportional to the nonnegative integers in the thermodynamic limit, i.e.~$\nu_m \sim m$. The entanglement spectrum (logarithm of the spectrum of Schmidt coefficients) can then be scaled to also correspond to the integers (now including zero), where each value $n$ has a degeneracy given by the number of strict partitions $q(n)$ of that integer. A GfMPS approximation of the true ground state with $M$ virtual fermions (or Grassmann numbers) corresponds to an MPS with bond dimension $2^M$, but can only capture the values $\nu_m \sim m$ for $m=1,\dots,M$. As a consequence, the distribution of the entanglement spectrum of the GfMPS captures the right degeneracy only up to the value $n=M$, even though it contains values all the way up to $n=M (M+1)/2$. Hence, only the first $\sum_{n=1}^{M} q(n) \ll 2^M$ Schmidt coefficients are correct, with in particular $\lim_{M\to 0} 2^{-M} \sum_{n=1}^{M} q(n) = 0$, i.e.\ the relative fraction of correct Schmidt coefficients decreases for increasing $M$.  A generic MPS of bond dimension $2^M$ is not bound by this free-fermion structure of the entanglement spectrum and can capture all Schmidt coefficients up to its full bond dimension correctly.
 
    Whether this behaviour extends to the GfPEPS case is unclear. As there is no clear entanglement interpretation to a single PEPS bond, it is impossible to make an informed prediction about the role of the free-fermion structure in the PEPS truncation.
	
	\section{Appendix G -- Chirality of approximate GfTNS}
	
	In the main text and Appendix A, we discussed that the even parity of the local tensors in a GfTNS induces an equal filling of all TRIMs. As a result, any (non-degenerate) GfTNS will leave the $\k=0$ mode empty, thus yielding a fixed energy error. These errors tend to render the numerical optimisation less stable. Therefore, we worked with antiperiodic boundary conditions, avoiding the problematic TRIMs and the corresponding stability issues. While this easily solves the numerical problem, the conceptual one remains. Indeed, directly using a parameter matrix $A$, optimised for a large but finite system size $N_s$, in the thermodynamic limit, the $\mathbf{k}=0$ mode will be sampled again and yield an erroneous occupation. While this will have a vanishing influence on the energy density as the corresponding energy error will be suppressed by dividing it by the diverging $N_s$ factor (therefore not contradicting the claims made in Appendix B and D about energy densities `probing the thermodynamic limit'), conceptual issues also arise in the assessment of the chiral features of GfTNS. Indeed, these are characterized by the Chern number, $C$, a topological invariant that can be computed for all 2D TI BdG Hamiltonians (Eq.\,\eqref{hamiltonianbdg}) as
	\begin{equation}
		C = \frac{i}{2 \pi} \int_\text{BZ} \text{tr} \, \mathcal{F}
		\label{cherndef}\, .
	\end{equation}
	One thus has to integrate over the Brillouin zone and is therefore obliged to work with a $\k$ continuum, i.e.~to work in the thermodynamic limit. We first expand on the integrand where $\mathcal{F}$ is the Berry curvature, a differential form defined as $\mathcal{F} = \text{d}\mathcal{A} + \mathcal{A}\wedge\mathcal{A}$ \cite{chiu2016}. Herein is $\mathcal{A}^{\alpha \beta} = A^{\alpha \beta}_i(\mathbf{k}) \text{d}k_i$ the non-Abelian Berry connection with
	\begin{equation}
		A^{\alpha \beta}_i(\mathbf{k}) = \langle u_-^\alpha(\mathbf{k})|\frac{\partial}{\partial k_i}|u_-^\beta(\mathbf{k})\rangle \, .
	\end{equation}
	where $\{|u_-^\alpha(\mathbf{k})\rangle\}$ are the $N$ eigenvectors of $H_\text{BdG}(\mathbf{k})$ with a negative energy. The trace of the Berry curvature can be expressed as
	\begin{equation}
		\begin{split}
			\text{tr} \, \mathcal{F} &=  \sum_{\alpha, \beta = 1}^{N} \text{d} \mathcal{A}^{\alpha \alpha} + \mathcal{A}^{\alpha \beta} \wedge \mathcal{A}^{\beta \alpha} \\
			&= \sum_{\alpha, \beta = 1}^{N} \left(\frac{\partial}{\partial k_i} A^{\alpha \alpha}_j(\mathbf{k}) + A^{\alpha \beta}_i(\mathbf{k}) A^{\beta \alpha}_j(\mathbf{k})\right)\text{d} k_i \wedge \text{d} k_j
			= \left(\frac{\partial}{\partial k_x} \text{tr} \left(A_y(\mathbf{k})\right) - \frac{\partial}{\partial k_y} \text{tr} \left(A_x(\mathbf{k})\right) \right) \text{d} k_x \text{d} k_y \, .
		\end{split}
	\end{equation}
	Note that there is some freedom in the eigenvectors as they can always be altered by a phase, $|v^\alpha(\mathbf{k})\rangle = e^{i\theta^\alpha(\mathbf{k})} |u^\alpha(\mathbf{k})\rangle$, resulting in gauge dependence for the Berry connection
	\begin{equation}
		{A'}^{\alpha \alpha}_i(\mathbf{k}) = \langle v^\alpha(\mathbf{k})|\frac{\partial}{\partial k_i}|v^\alpha(\mathbf{k})\rangle = {A}^{\alpha \alpha}_i(\mathbf{k}) + i \frac{\partial \theta^\alpha}{\partial k_i}(\mathbf{k})
	\end{equation}
	and thus also for its trace, derivatives \textit{etc}. One could also collect the occupied eigenvectors in a $2N \times N$ matrix, $ U_-(\mathbf{k}) = \left(
	\begin{array}{c | c | c}
		|u_-^1(\mathbf{k})\rangle & \hdots &|u_-^{N}(\mathbf{k})\rangle
	\end{array} \right)$
	so that $\text{tr}(A_i) = \text{tr} \left(U^\dagger_-(\mathbf{k}) \frac{\partial}{\partial k_i} U_-(\mathbf{k}) \right)$, showing that gauge independent properties like the Chern number will not depend on the exact form of the $\{|u_-^\alpha(\mathbf{k})\rangle\}$ but rather on the occupied space spanned by these vectors and how this space behaves throughout the Brillouin zone. Therefore, an even more general gauge transformation is allowed, $V_-(\mathbf{k}) = U_-(\mathbf{k}) g(\mathbf{k})$ where $g(\mathbf{k})$ is an $N \times N$ unitary matrix. For the corresponding change in the Berry connection and curvature, one obtains $\mathcal{A}' = g^{-1}\mathcal{A}g+g^{-1}\text{d}g$ and $\mathcal{F}'=g^{-1}\mathcal{F}g$, manifestly showing that the Chern number is indeed gauge independent. Another way to capture the occupied space is via the spectral projector
	\begin{equation}
		P(\mathbf{k}) = \sum_{\alpha \text{ occ.}} |u^\alpha(\mathbf{k})\rangle \langle u^\alpha(\mathbf{k})| = U_-(\mathbf{k}) U^\dagger_-(\mathbf{k})
	\end{equation}
	or equivalently $Q(\k) = 1-2P(\k)$. Due to gauge independence, both belong to the Grassmannian manifold $\text{Gr}(N,2N) = \text{U}(2N)/\left(\text{U}(N) \times \text{U}(N)\right)$. Topologically distinct maps on this manifold are classified via their second homotopy group, $\pi_2\left(\text{Gr}(N,2N)\right)$, and are hence characterized by an integer topological invariant. This is again the Chern number which can be expressed as
	\begin{equation}
		C = -\frac{i}{16 \pi} \int_\text{BZ} \text{tr} \left(Q \, \text{d}Q \wedge \text{d}Q\right) \, .
	\end{equation}	
	Now consider a pure, Gaussian and TI state, fully characterized by its Fourier transformed correlation matrix $G(\k)$ as discussed in Appendix C. Purity implies that $G^2(\k)=-\mathbbm{1}$ so that the eigenvalues of $G(\k)$ are $\pm i$. Consequently, a flat-band TI BdG parent Hamiltonian can always be constructed with $H_\text{BdG}(\mathbf{k}) = -i V^\dagger G(\k) V$ and $V=\mathbbm{1}_N \otimes \begin{pmatrix} +1 & + 1\\ +i & -i\end{pmatrix}$ the constant matrix transforming the Fourier space Nambu spinor in the $\{d_{\k,i}\}$ operators (with $V^{-1} = \frac{1}{2} V^\dagger$). Indeed, when 
	\begin{equation}
		\begin{split}
			H = \frac{1}{2} \sum_{\mathbf{k}} \Upsilon^{\dagger}_\mathbf{k} H_\text{BdG}(\mathbf{k}) \Upsilon_\mathbf{k} = \frac{1}{8} \sum_{\mathbf{k}} d^{\dagger}_\mathbf{k} V H_\text{BdG}(\mathbf{k}) V^\dagger d_\mathbf{k} \, ,
		\end{split}
	\end{equation}
	the energy of a Gaussian state characterized by $\tilde{G}(\k)$ can be evaluated as 
	\begin{equation}
		E\left[\tilde{G}(\k)\right] = \frac{i}{8} \sum_\k \text{Tr} \left(V H_\text{BdG}(\mathbf{k}) V^\dagger \tilde{G}(\k)\right) = \frac{1}{2} \sum_\k \text{Tr} \left(G(\k) \tilde{G}(\k)\right)
	\end{equation}
	which is minimal for $\tilde{G}(\k) = G(\k)$ with $E = -N N_s$. Denoting the orthonormal eigenvectors of $G(\k)$ with eigenvalue $\pm i$ as $\ket{w_\pm^\alpha(\k)}$ we have $H_\text{BdG}(\k) \frac{1}{\sqrt{2}} V^\dagger \ket{w_\pm^\alpha(\k)} = (\pm 2) \frac{1}{\sqrt{2}} V^\dagger \ket{w_\pm^\alpha(\k)}$, showing that the single particle energy bands are indeed flat and that $\ket{u_\pm^\alpha(\k)} = \frac{1}{\sqrt{2}} V^\dagger \ket{w_\pm^\alpha(\k)}$. We thus get access to the occupied space and find $G(\k) = i \left(W_+(\k) W^\dagger_+(\k) -W_-(\k) W^\dagger_-(\k) \right) = \frac{i}{2} V Q(\k) V^\dagger$. The Chern number can hence be evaluated directly from the Fourier transformed correlation matrix as 
	\begin{equation}
		C = \frac{1}{16 \pi} \int_\text{BZ} \text{Tr} \left( G(\k) \left[\frac{\partial G}{\partial k_x}(\k), \frac{\partial G}{\partial k_y}(\k) \right]\right) \text{d} k_x \text{d} k_y \, .
		\label{chernG}
	\end{equation}

\begin{figure}[h]
		\centering
		\includegraphics[scale=0.28]{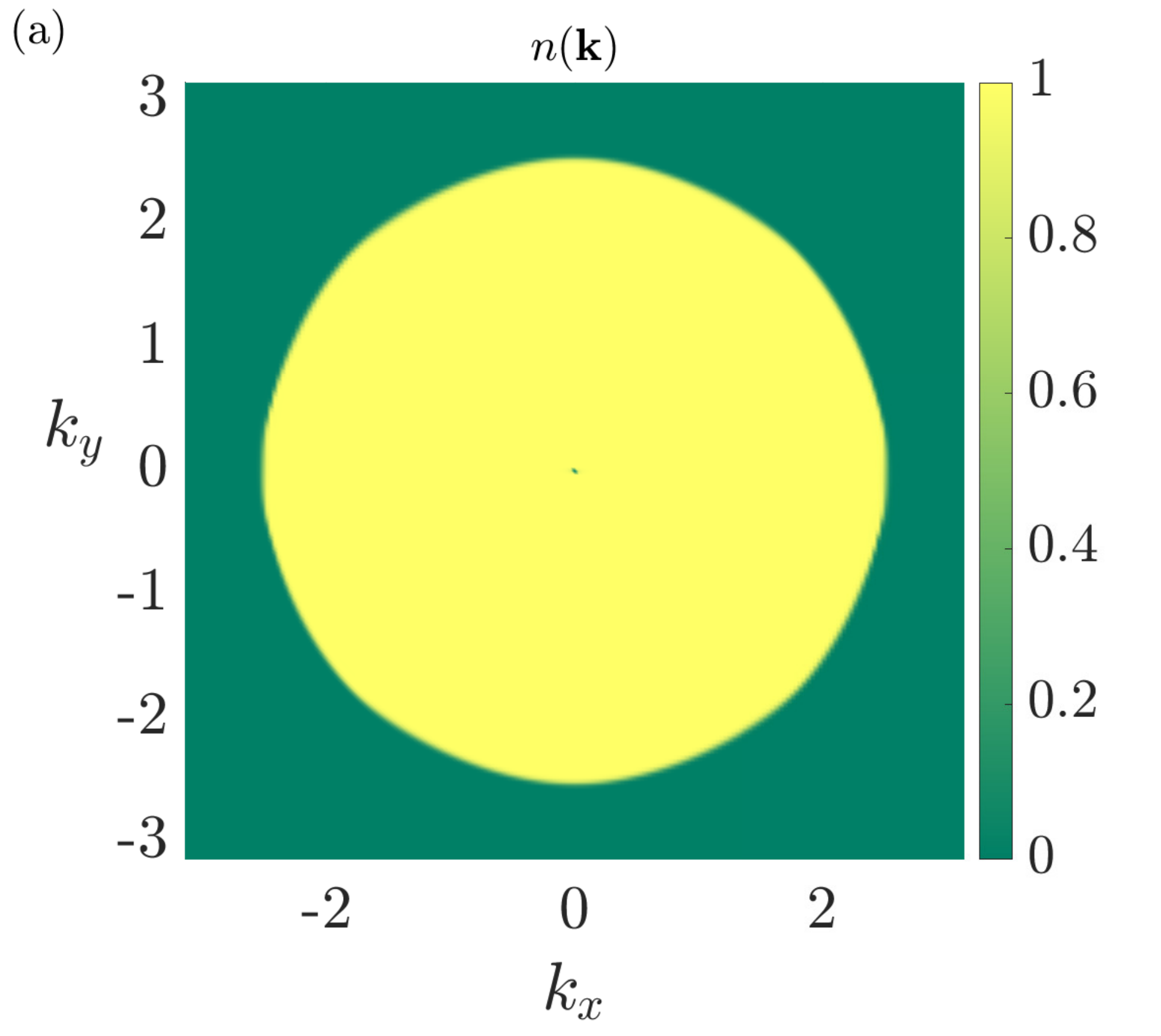} \quad
		\includegraphics[scale=0.28]{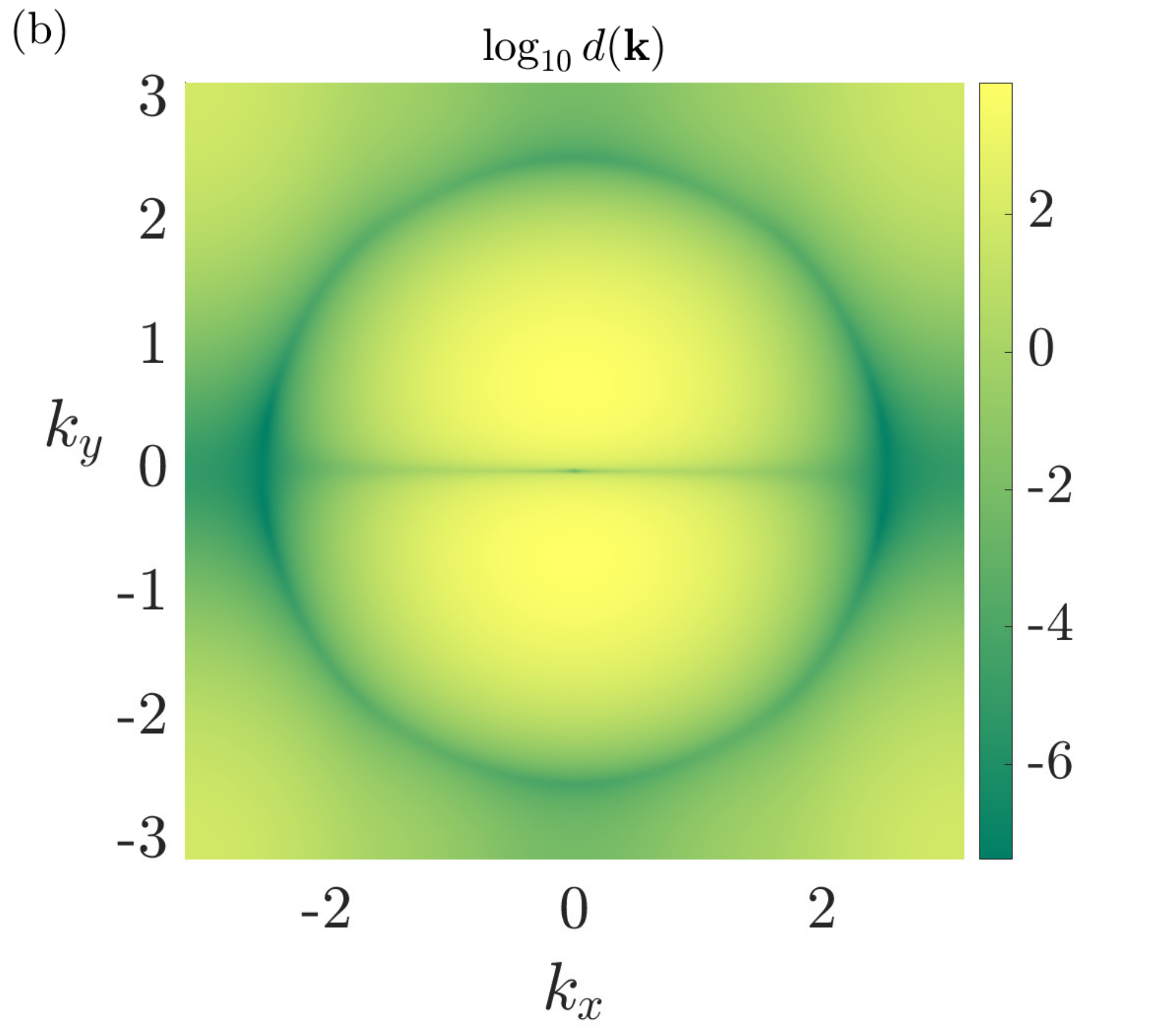} \quad
		\includegraphics[scale=0.28]{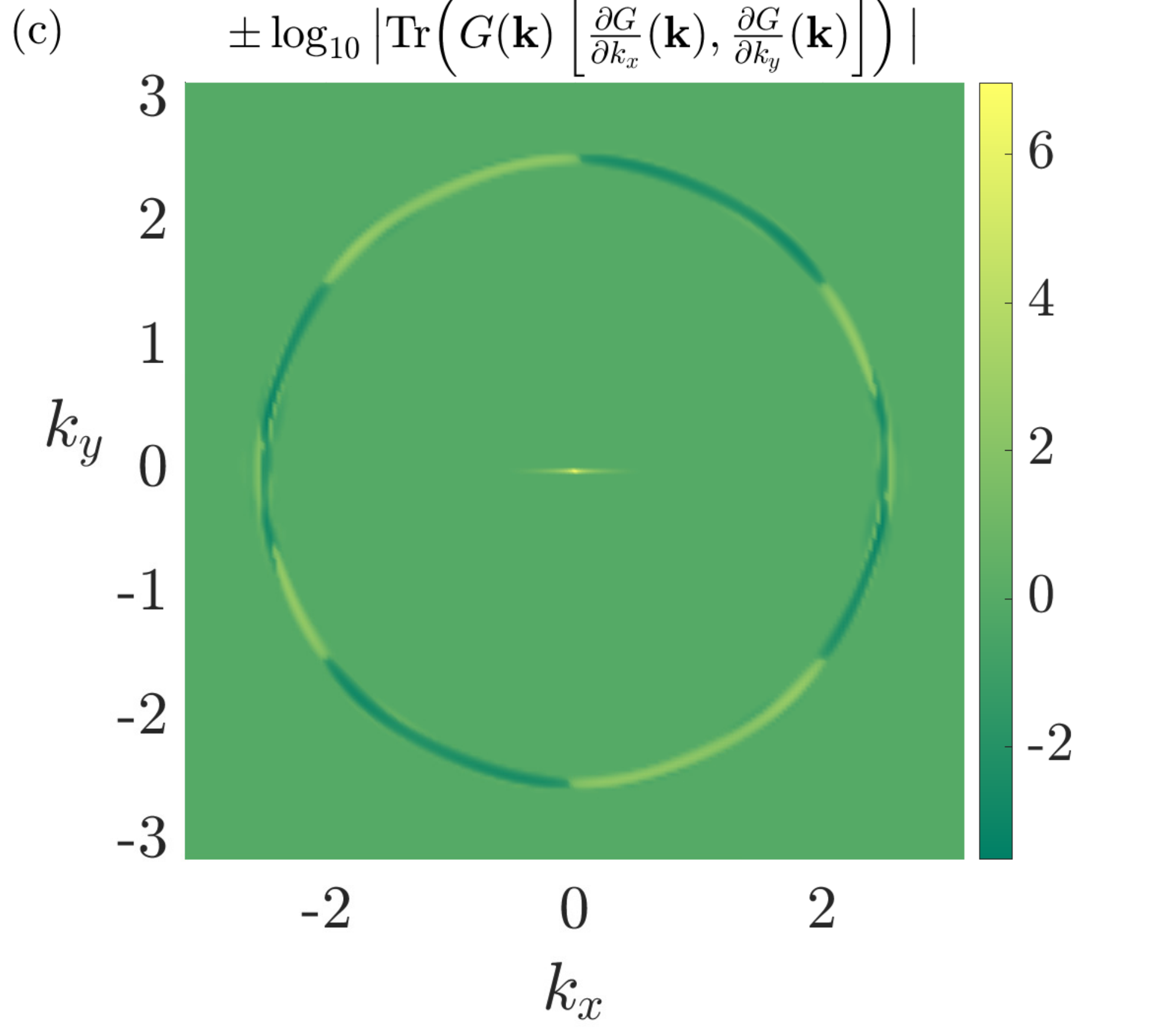}
		\caption{A GfTNS optimised for the spinless model (Eq.\,\eqref{Hamiltonian}) on a finite lattice ($N_s=50^2$) with antiperiodic boundary conditions is examined on a finer mesh in $\k$ space ($N_s=100^2$) with periodic boundary conditions. Panel (a) shows that the average occupation in the zone center is not reproduced correctly due to the well-known parity obstructions for non-degenerate GfTNS. Panel (b) displaying $\log_{10} d(\k)$ confirms that this state is indeed non-degenerate and thus non-chiral. In (c) the latter is made visible by displaying the integrand in Eq.\,\eqref{chernG} throughout the Brillouin zone  (i.e.~the trace of the Berry curvature up to a prefactor). The (mostly negative) Berry curvature generated near the Fermi surface is compensated near the zone center by a positive peak. Note that we took the logarithm of the integrand due to the large absolute values. Signs were added again afterwards and the color axis should be read accordingly.}
		\label{fig:chiral}
	\end{figure}

	We conclude that to mathematically assess the topological features of optimised GfTNS, we have to go to the thermodynamic limit in order to compute this integral. Here, we will be confronted again with the conceptual issues near the zone center. Indeed, take a GfTNS with $D=16$, optimised for the spinless model (Eq.\,\eqref{Hamiltonian}) on a finite lattice with antiperiodic boundary conditions and $N_s=50^2$. Fig.~\ref{fig:chiral} (c) displays the integrand in Eq.\,\eqref{chernG} for this state when evaluated on a finer mesh ($N_s=100^2$) with periodic boundary conditions (for instance in an attempt to numerically approximate the integral in Eq.\,\eqref{chernG}). We observe that the Berry curvature generated at the Fermi surface is compensated near the zone center to yield a topologically trivial state. Indeed, the no-go theorem of Read and Dubail \cite{Dubail2015} requires a degenerate state to realise chirality. We observe that the numerically optimised GfTNS never becomes truly degenerate, i.e.\ the quantity $d(\k)=\det\left(D-G_\text{in}(\k)\right)$ depicted in Fig.~\ref{fig:chiral}(b) becomes small ($\sim 10 ^{-7}$) but never reaches zero. Hence, our numerically optimised GfTNS realises $C=0$ and an equal occupation on all TRIMS (see Fig.~\ref{fig:chiral}(a)). In the strict mathematical sense, GfTNS optimised on a finite lattice are therefore not chiral.\\
	
\begin{figure}[h]
		\centering
		\includegraphics[scale=0.36]{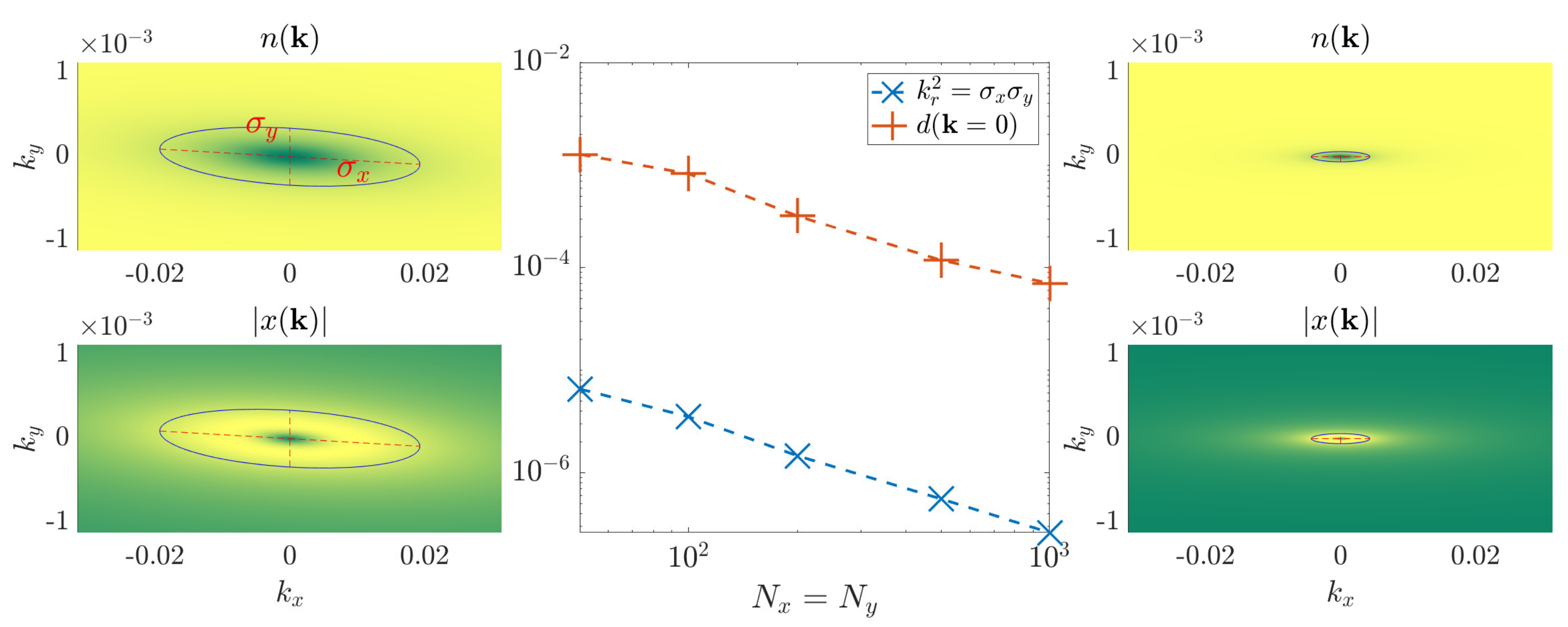}
		\caption{Comparison of the near-zone-center behavior between GfTNS optimised for the spinless model (Eq.\,\eqref{Hamiltonian}) on finite lattices with antiperiodic boundary conditions and increasing system sizes. The left panels display the average occupation $n(\k)$ and pairing term expectation values $|x(\k)|$ for optimisation on the smallest lattice ($N_s=50^2$). The right panels show the same but for $N_s=1000^2$. The underoccupation regions are strongly anisotropic and spread out in one spatial direction (notice the different scales on both axes). Furthermore, they clearly become smaller when the optimisation lattice is increased in size. To quantify this, a Gaussian profile with standard deviations $\sigma_x, \sigma_y$ was fitted to the underoccupation peaks with $k_r = \sqrt{\sigma_x \sigma_y}$ as a relevant momentum scale. As $N_x=N_y$ increases, $k_r^2$ (proportional to the area of the ellipses spanned by $\sigma_x$ and $\sigma_y$) decreases $\sim N_x^{-1}$. A similar power law can be observed in the determinant $d(\k)$ in the zone center. We conclude that by increasing $N_s$ the GfTNS decreases the size of the erroneous region, simultaneously making the state more degenerate.}
		\label{fig:chiralscaling}
	\end{figure}

	In Fig.~\ref{fig:chiral} we show that the region at the zone center, compensating the Berry curvature generated near the Fermi surface, is very small in $\k$ space. Optimising the GfTNS at increasingly larger system sizes shows that this region steadily shrinks. Indeed, from a numerical perspective, we understand that when more small $\k$ modes are sampled and optimised towards complete occupation, the size of the region around $\k=0$ where $n(\k)\neq1$ decreases. This is confirmed in Fig.~\ref{fig:chiralscaling}, where we compare the average occupation for a zoomed-in region around the zone center for GfTNS optimised at different $N_s$. Not only is the dip in the occupation smeared out in one spatial direction (i.e.\ the $k_x$ direction here, as could already be observed in Fig.~\ref{fig:chiral}) to avoid as many sampled $\k$ points as possible, it also becomes smaller when the GfTNS optimisation is performed at higher $N_s$. Repeating this process for increasing $N_s$ values, the central panel of Fig.~\ref{fig:chiralscaling} shows that the characteristic radius $k_r=\sqrt{\sigma_x \sigma_y}$ of the underoccupation region (obtained by a Gaussian fit to the peak with standard deviations $\sigma_x$ and $\sigma_y$) steadily decreases. If this process were continued for even higher $N_s$, one would eventually end up with a state where $k_r$ decreases to zero in the thermodynamic limit, or thus, the underoccupation region reduces to a single point. Of course, to accommodate these smaller underoccupation regions, the determinant $d(\k)$ will become smaller as well (see again the central panel of Fig.~\ref{fig:chiralscaling}) to eventually yield a degenerate state with a nonzero Chern number $C\neq0$. Finally, note that the pairing terms in the neighbourhood of the zone center are again maximised where the occupation changes.
	
	\begin{figure}[h]
		\raggedright
		(a) \\
		\centering
		\quad $D=16$ \qquad $N_s = 100^2$\\
		\includegraphics[scale=0.36]{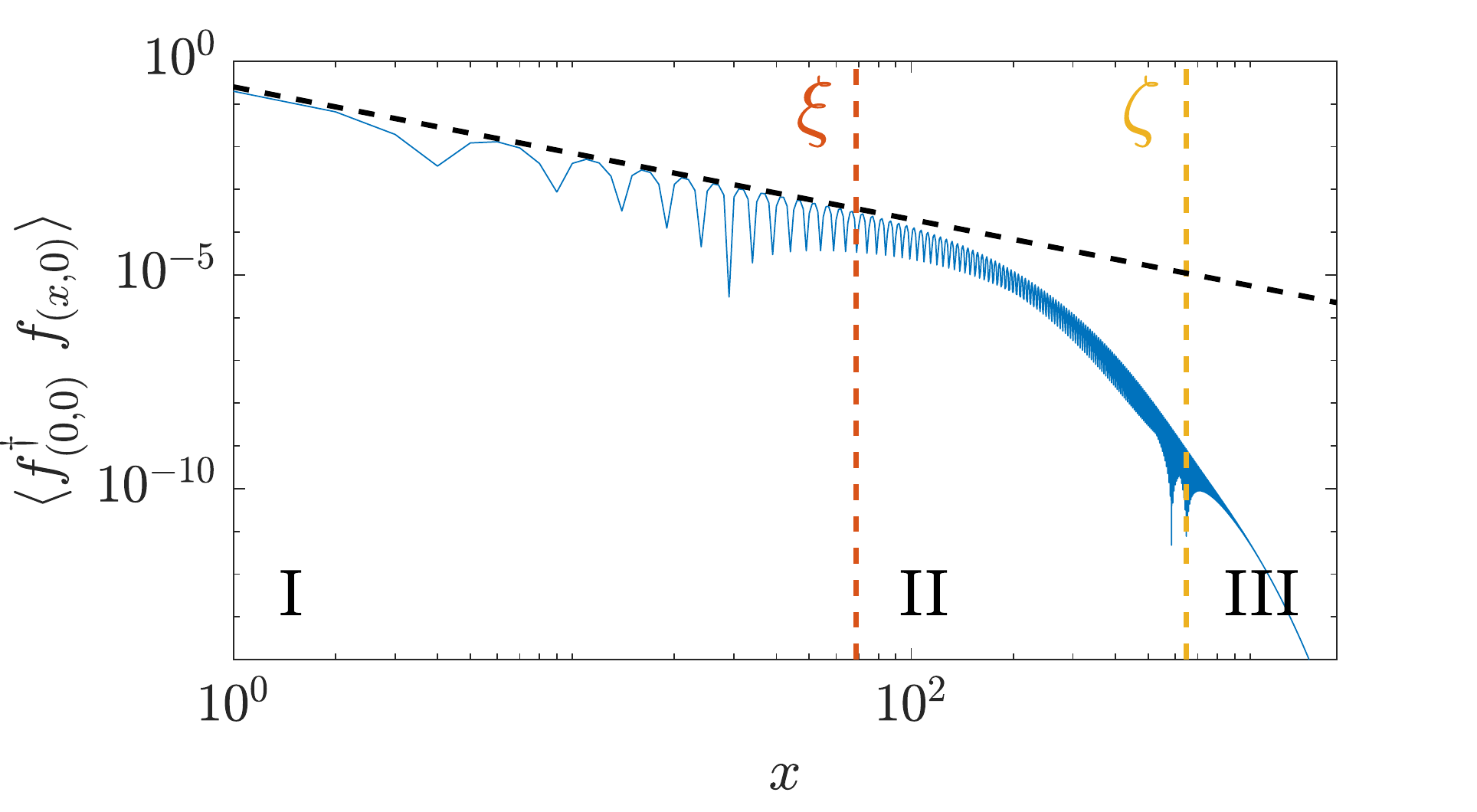}  \includegraphics[scale=0.36]{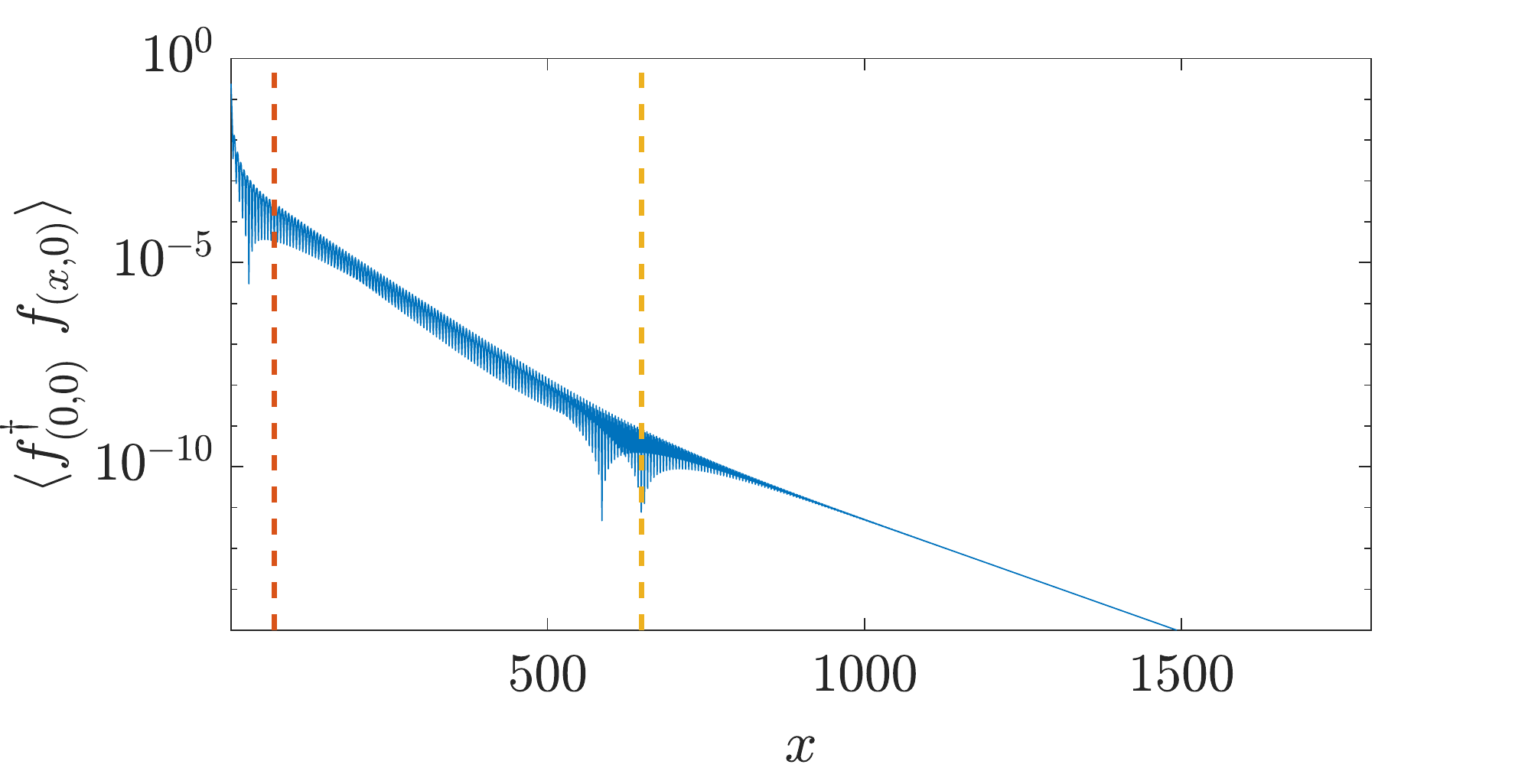}\\
		\includegraphics[scale=0.36]{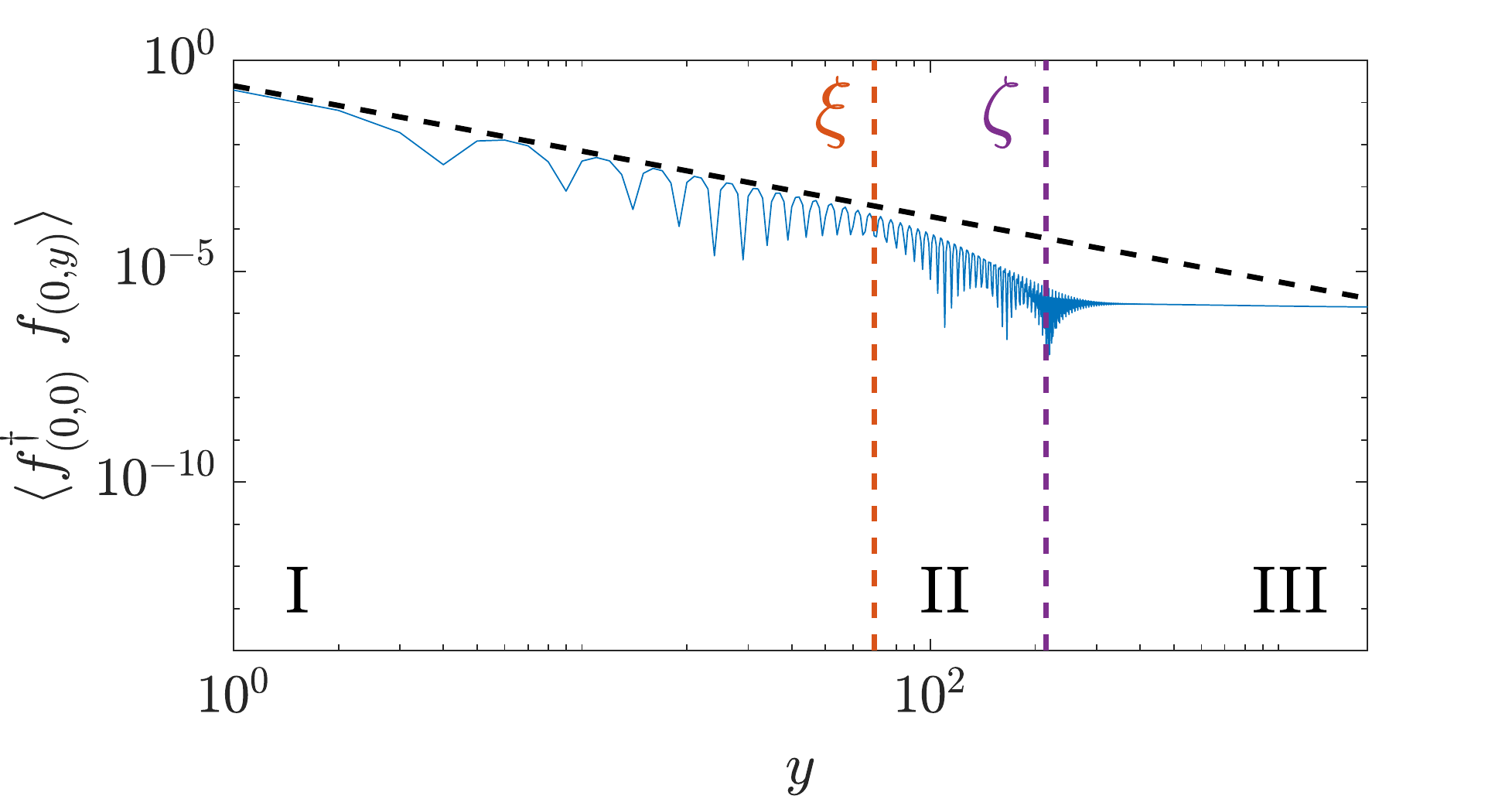}  \includegraphics[scale=0.36]{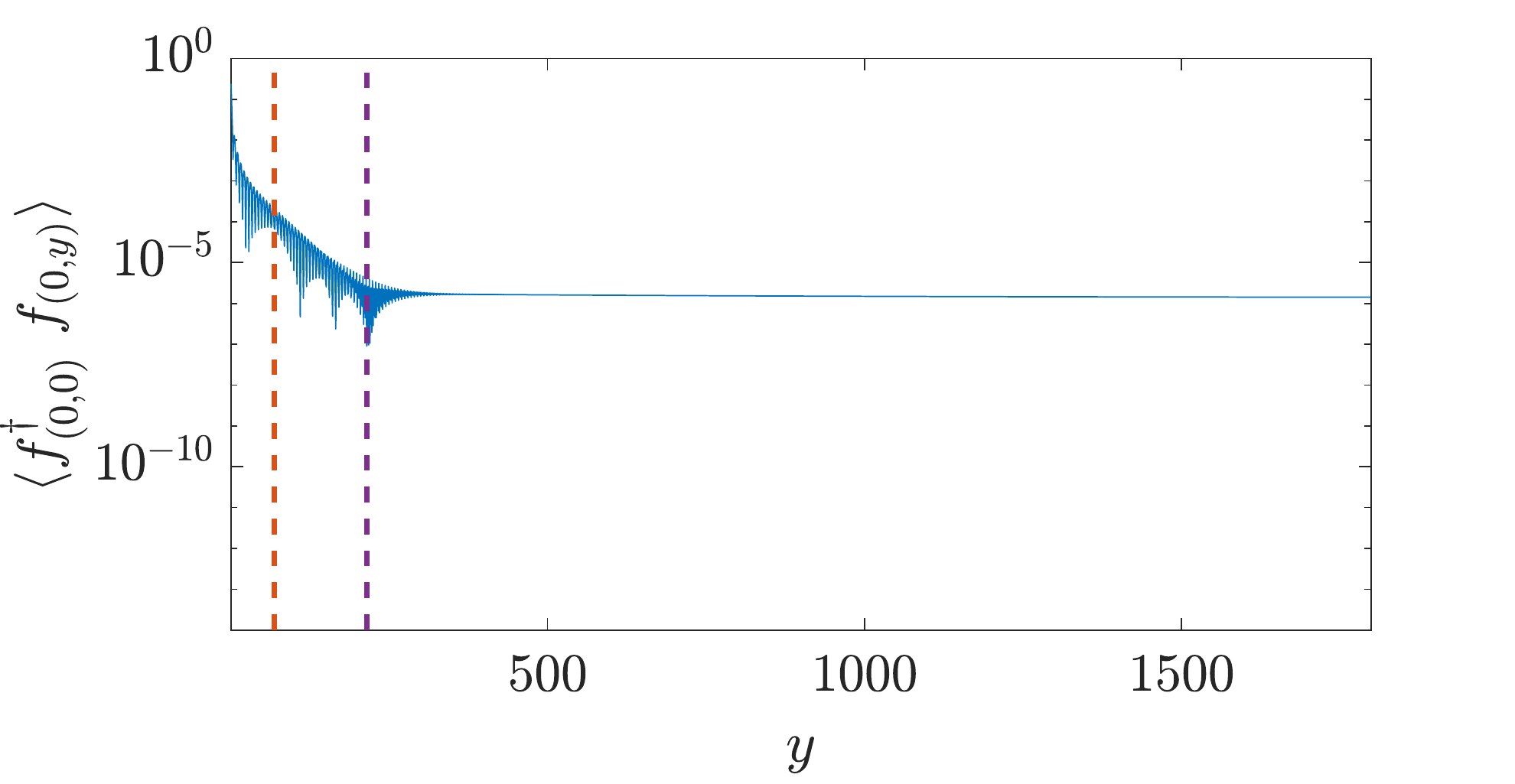}\\
		\raggedright
		(b) \hspace{8.2cm} (c)\\
		\centering
		$D=16$ \qquad $N_s = 500^2$ \hspace{5.5cm} $D=8$ \qquad $N_s = 500^2$ \qquad\\
		\includegraphics[scale=0.215]{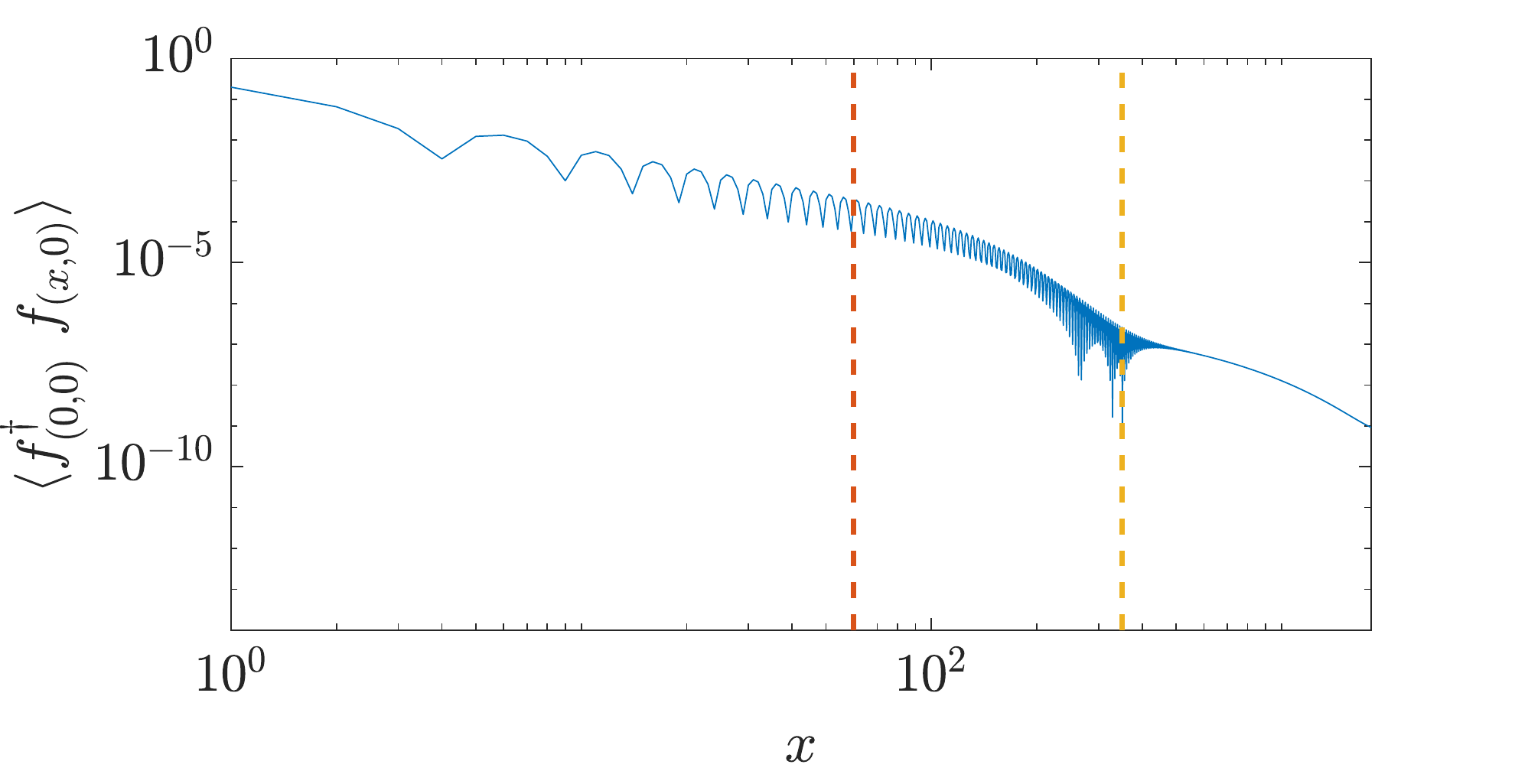} \includegraphics[scale=0.215]{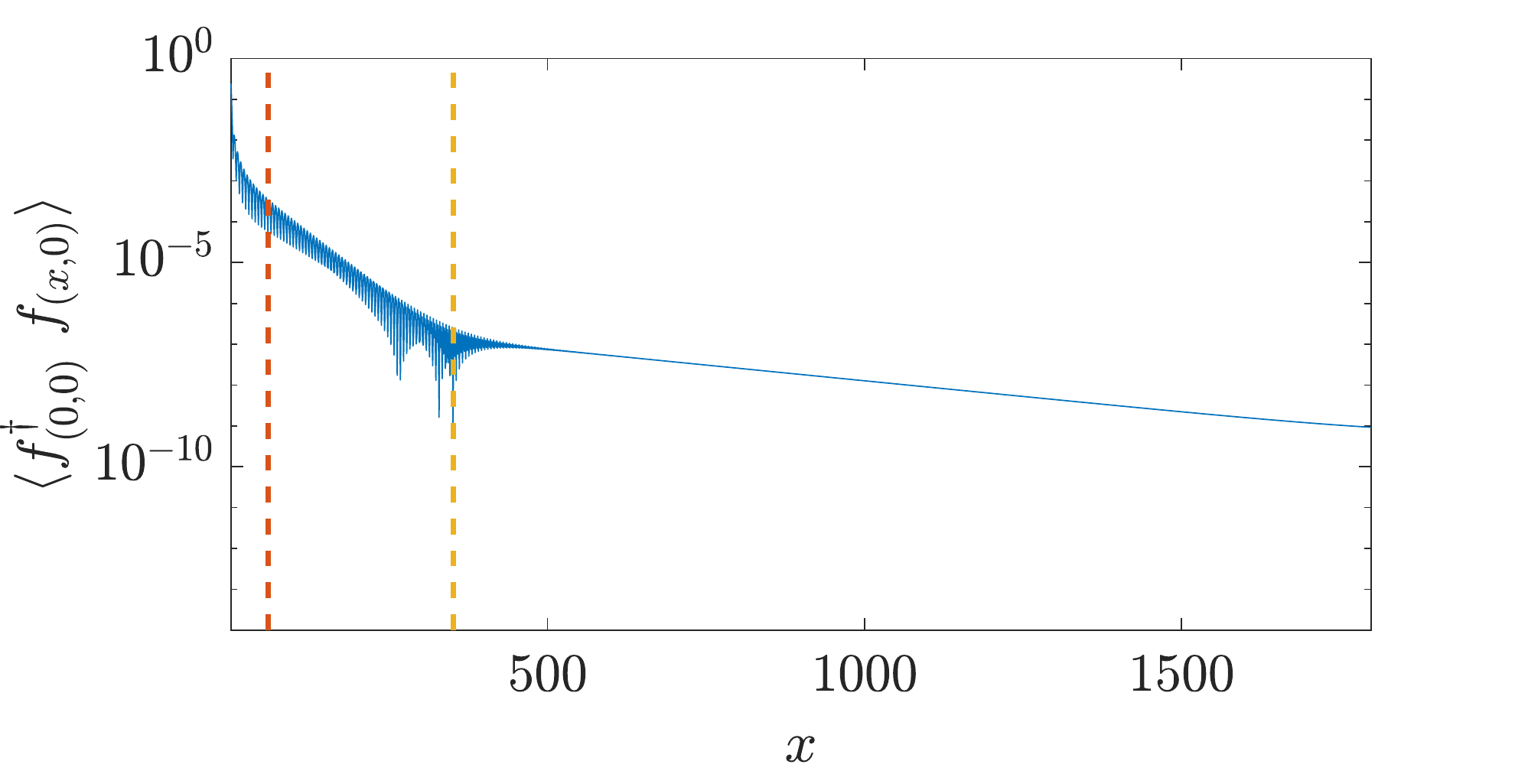} \includegraphics[scale=0.215]{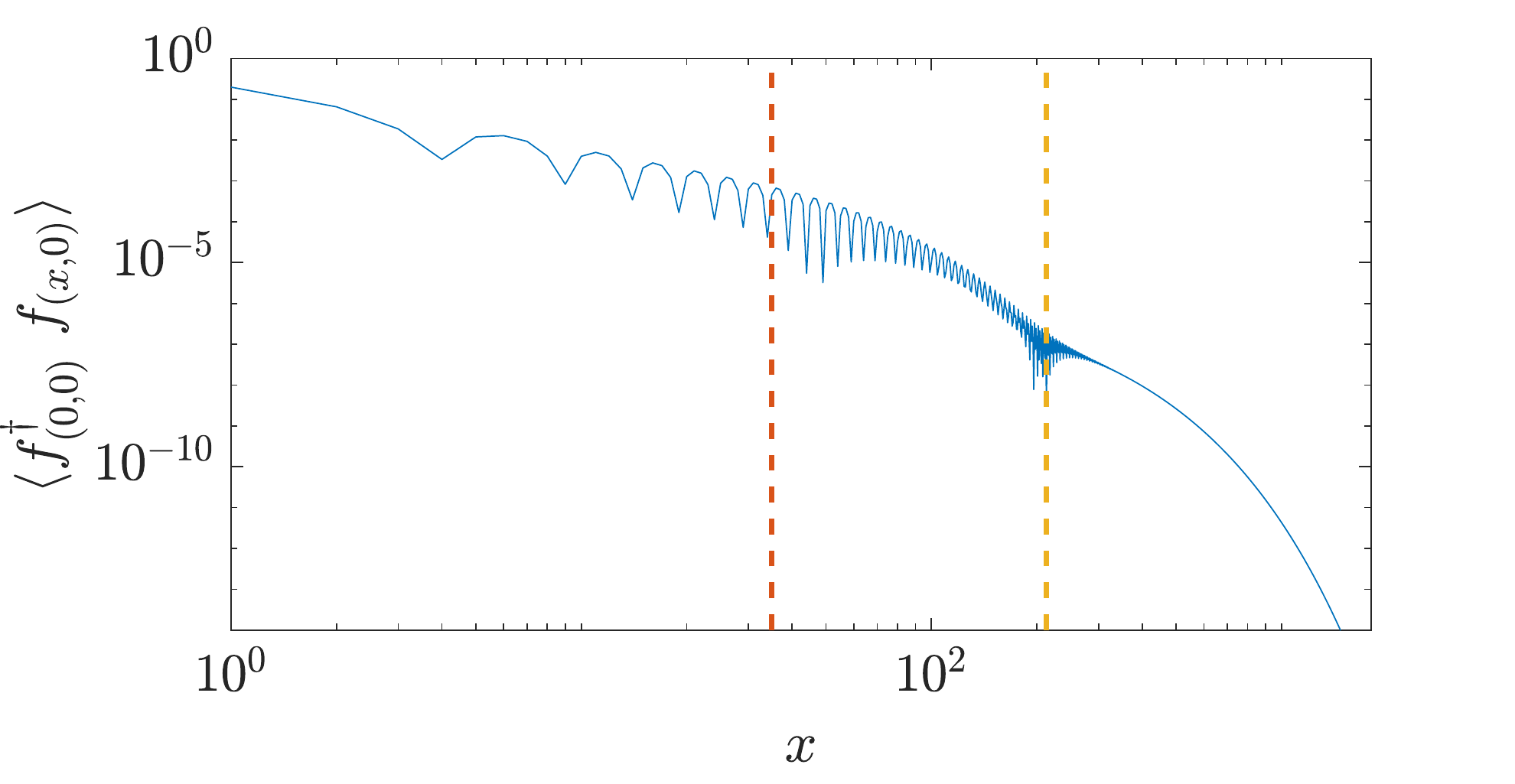} \includegraphics[scale=0.215]{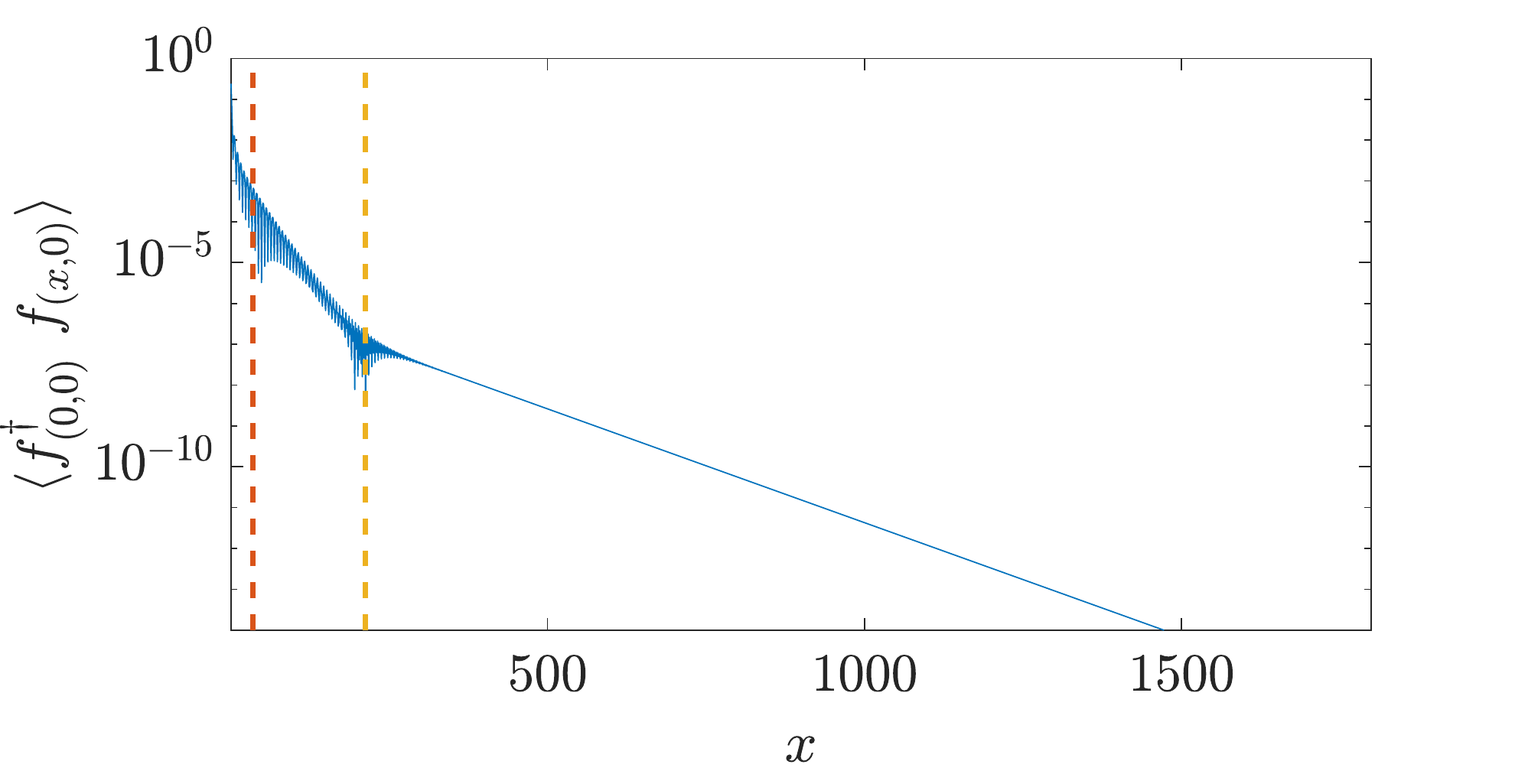}\\
		\includegraphics[scale=0.215]{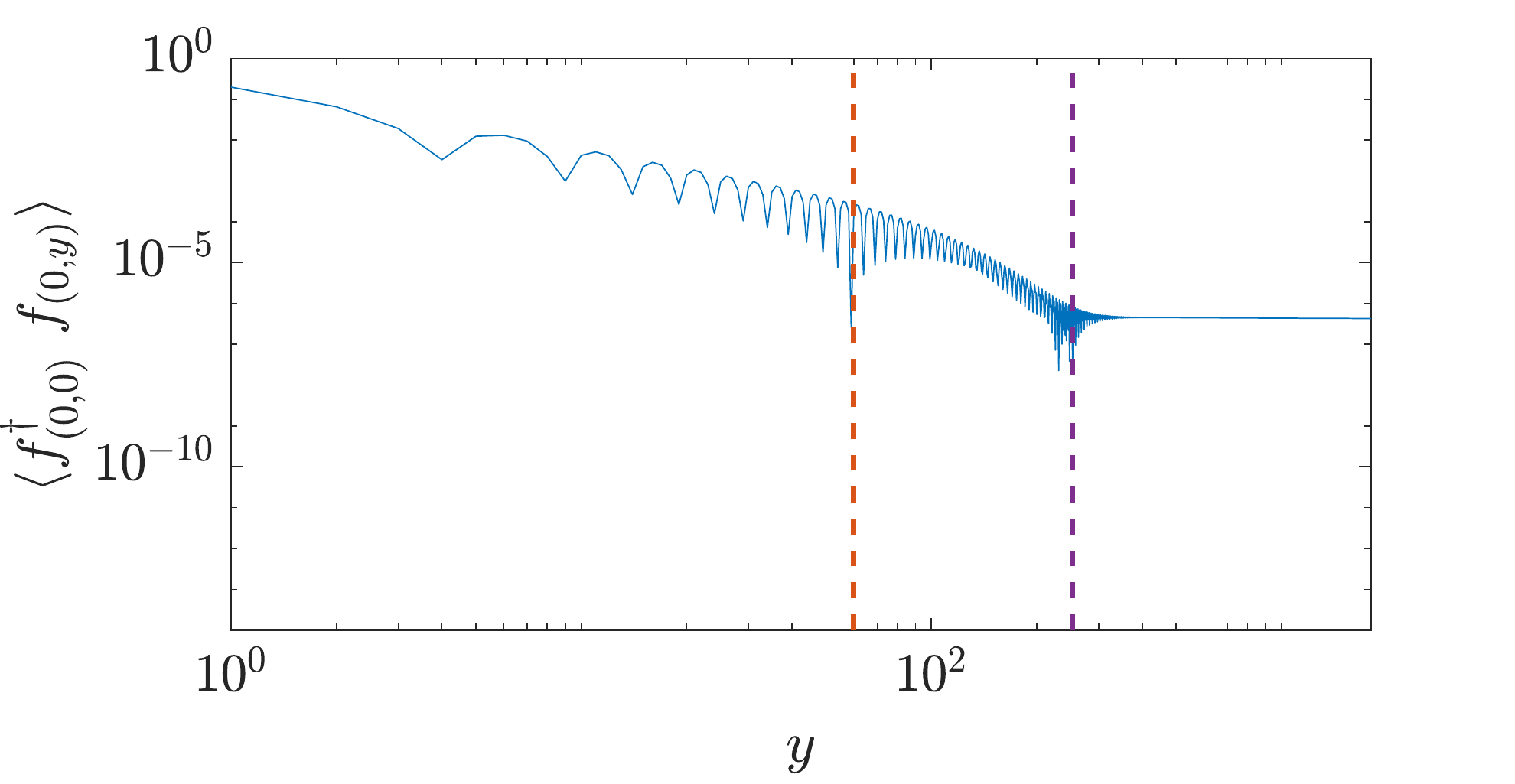} \includegraphics[scale=0.215]{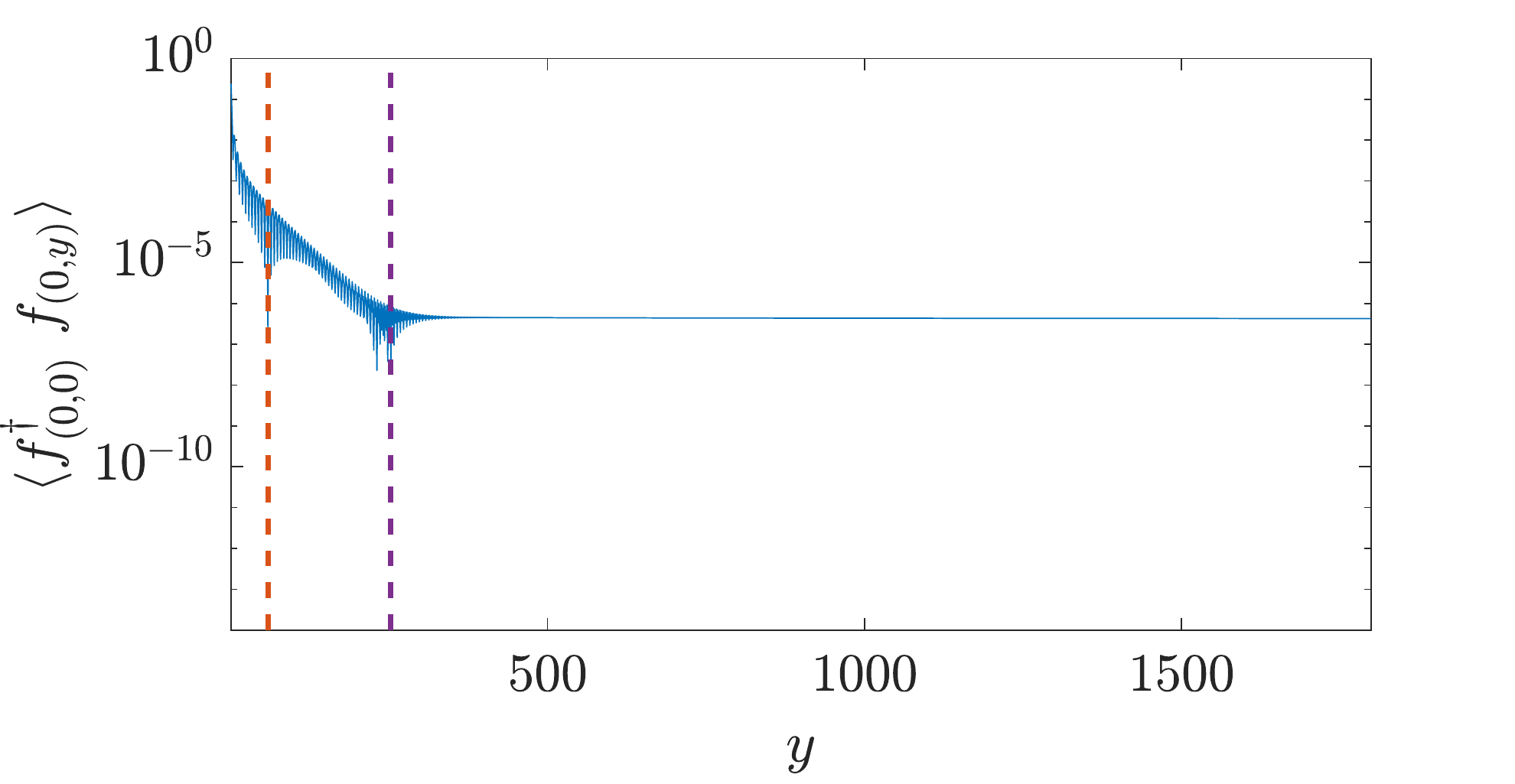} \includegraphics[scale=0.215]{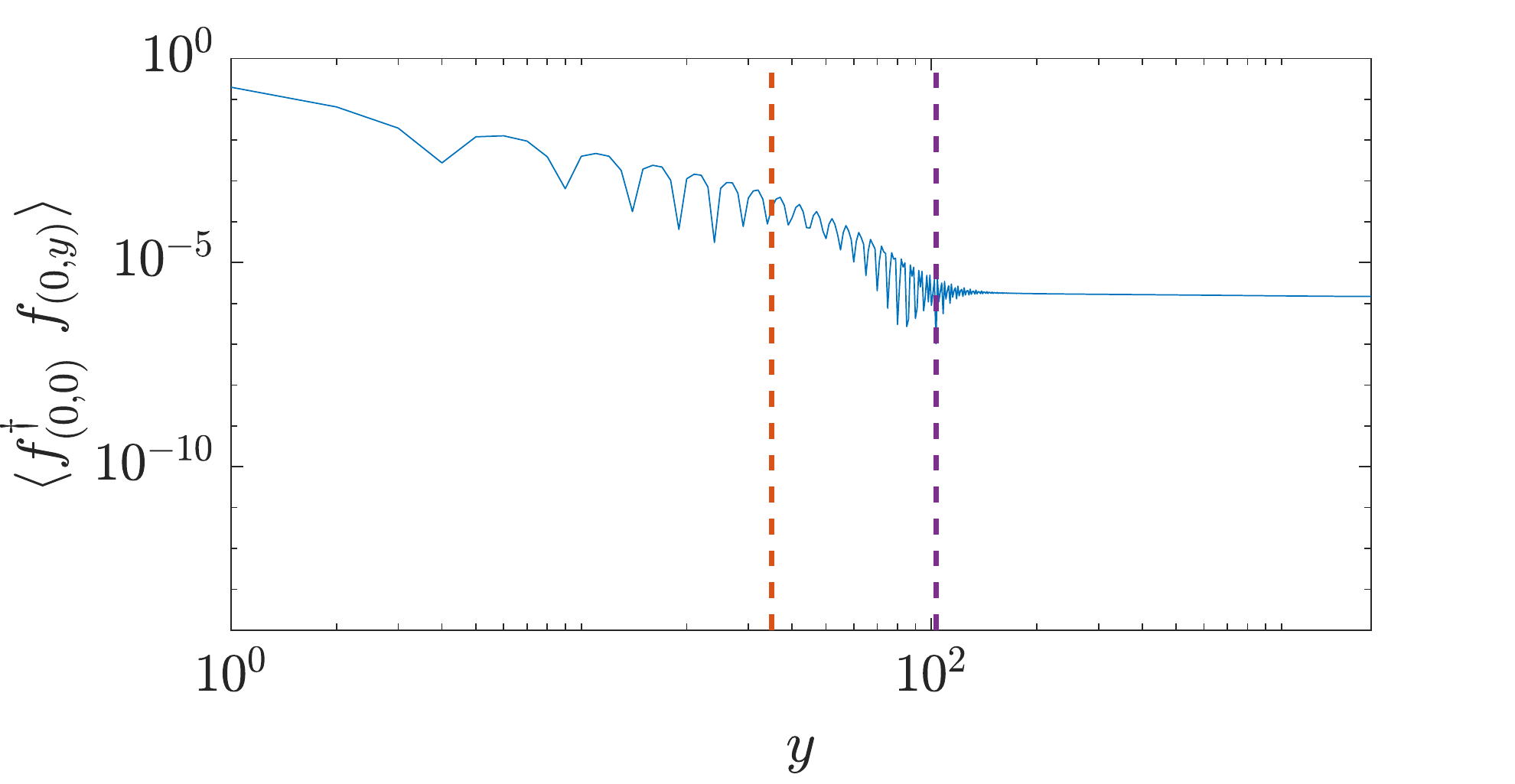} \includegraphics[scale=0.215]{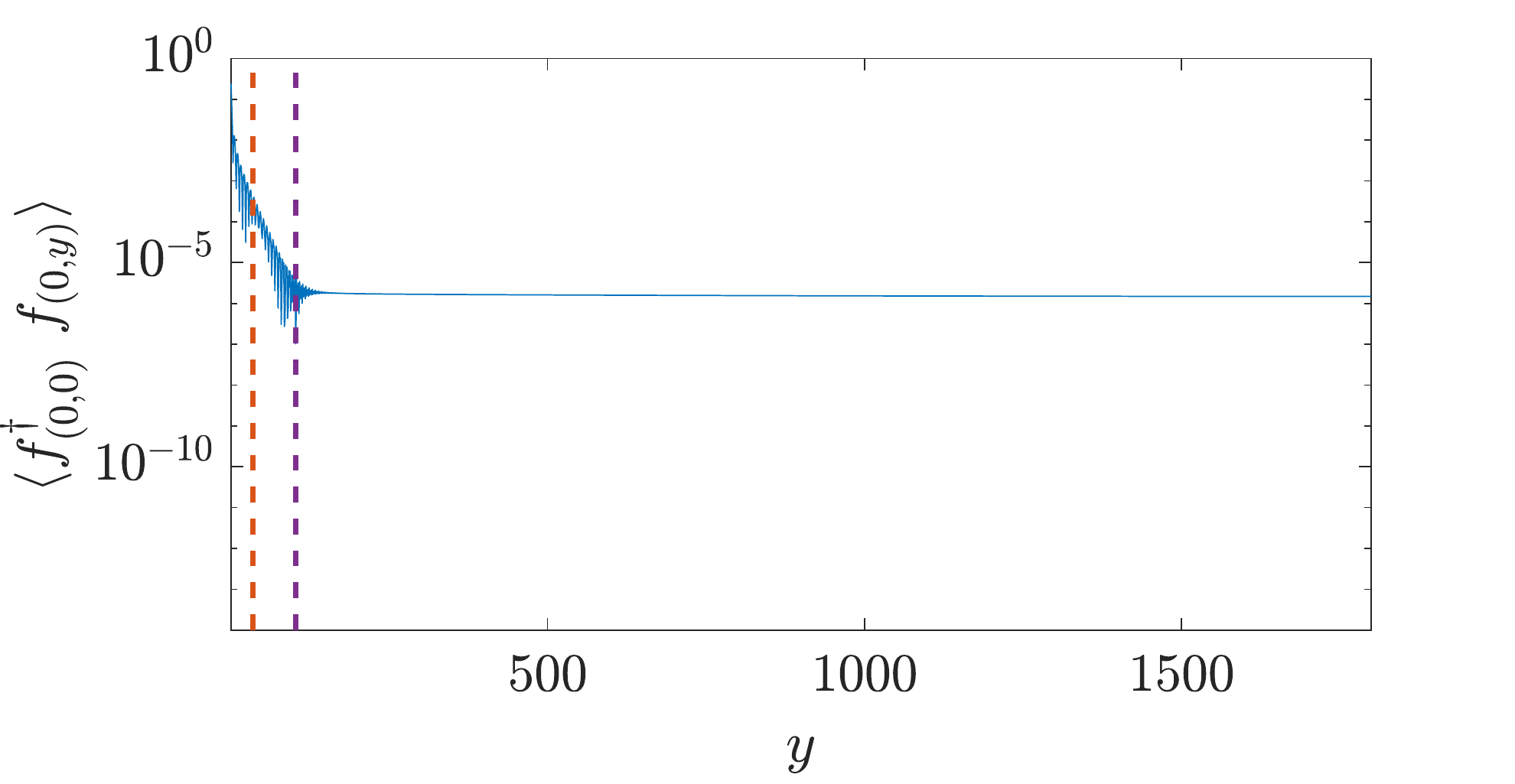}\\
		\caption{Real-space correlations $\langle f_\mathbf{0}^\dagger \, \, f_\mathbf{x} \rangle$ for GfTNS with the indicated bond dimension, $D$, optimised on a lattice with $N_s$ sites. Panel (a) displays the basic case with $D=16$ and $N_s=100^2$. In panel (b) the optimisation lattice is increased in size. Additionally, the bond dimension is decreased in panel (c). On each panel, the left (right) graphs have a log-log (lin-log) scale to discern between regions with algebraic and exponential decay of correlations. The bottom (top) graphs display the correlations in the $x$ ($y$) direction. The (envelope of the) exact power law (indicated with the dashed black line) is reproduced up to the bulk correlation length $\xi$ taken from Table~\ref{tab:corr_lengths} (region I). After that the decay becomes exponential with a fixed exponent  up to $\zeta$ (region II). Note that this length scale is different in the $x$ and $y$ direction, approaching each other when $N_s$ increases. For distances well above $\zeta$, a second exponential region sets in (region III) where the decay becomes flatter when $N_s$ increases. A higher bond dimension shifts $\xi$ and $\zeta$ to higher values and flattens out the tails as well.}
		\label{fig:correlations}
	\end{figure}
	 
	As our optimised GfTNS are not exactly degenerate, we do not expect power-law tails in real space correlations as in \cite{Wahl2013}. However, three regimes can still be discerned (see Fig.~\ref{fig:correlations}). For the smallest distances, the algebraic decay of correlations of the exact metallic ground state is reproduced up to the bulk correlation length $\xi$, corresponding to the superconducting gap near the Fermi surface. Fig.~\ref{fig:correlations} confirms that this bond dimension induced length scale is the same as in the EE scaling law and thus the most relevant for this work. After $\xi$, the state realises an exponential decay of correlations, characteristic to TNS with a finite bond dimension and similar to the situation for MPS approximations of one-dimensional critical states. In these two regimes (Friedel) oscillations are present (consistent with wavelength $\frac{2 \pi}{k_F}$). Indeed, the correlations are dominated by what happens near the Fermi surface (i.e.~at the Fermi momentum $k_F$). Finally, a third regime takes over at a very large length scale, $\zeta$. The decay in this regime is exponential as well but displays no oscillations. We understand this regime as originating from what happens near the zone center (hence the absence of oscillations). Indeed, the near-degenerate/chiral, sharply peaked behavior at $\k=0$ requires long-range tails in the real-space correlations. This is reminiscent of the long-range tails observed in \cite{hasik2022} for generic PEPS approximations of chiral states. We also obtained some qualitative insight in how these tails behave as function of the bond dimension $D$ and $N_s$, the system size used for the optimisation. For increasing $N_s$, GfTNS become more singular with a sharper peak near the zone center. As a result, the exponent in the long-range tails decreases, yielding a more flat profile in the third region. This effect can also be observed when examining the correlations in the $y$-direction. Indeed, the underoccupation region is already more squeezed in this direction for the smallest $N_s$, resulting in even more flat profiles setting in at smaller $\zeta$. Increasing $N_s$, it thus seems that the real-space correlations in both directions become more alike with an almost flat long-range tail setting in at $\zeta$ values that grow closer in both directions. The bond dimension plays a more involved role. As we know from the EE scaling, increasing $D$ increases $\xi$. This is reflected in a lower $\xi$ in Fig.~\ref{fig:correlations}(c) compared to (a) due to the smaller $D$. Furthermore, the smaller number of variational degrees of freedom, results in a smoother peak at $\k=0$ when compared to panel (b). As a result, correlations in the third region decay faster and the $\zeta$ values in both directions differ more. However, they are both smaller than the $\zeta$ values observed for $D=16$. Therefore, we expect that the transition between the second and third region also shifts to higher distances when the bond dimension is increased. We reserve a more rigorous study of the real-space correlations for a follow-up work.\\

	To summarize, optimised GfTNS are not strictly chiral as they are not degenerate and therefore do not have power-law tails in their real-space correlations. However, increasing the system size used during optimisation, they approach degeneracy/chirality increasingly well so that a minor change in the variational parameters can convert the states from non-chiral to chiral. Moreover, the long-range tails have no influence whatsoever on expectation values of local operators and the optimised states thus behave as their fully chiral counterparts for all practical purposes on finite lattices. That is why we classify the optimised GfTNS in the main text as approximate $p_x+ip_y$-superconductors. Alternatively, one could impose degeneracy (and thus chirality) as an extra constraint on the parametric manifold. However, this constraint is more involved  (Appendix A) to combine with the utilized optimisation schemes. Therefore, this manifestly chiral construction was not pursued here.

\section{Appendix H -- Connection between GfTNS and general fTNS defined via super vector spaces}
	
	The most commonly used TNS represent spin wave functions, and hence are bosonic, i.e.~the constituent tensors are simply arrays of complex numbers. To every index of the bosonic tensors we can associate a vector space with a particular choice of basis, and the arrays of complex numbers represent the components of the tensors in these basis.
	
	Fermionic TNS are defined as a natural extension of conventional bosonic TNS. In particular, to every index of a fermionic tensor one associates a \emph{super} vector space $V$ \cite{Bultinck2016}. A super vector space is a $\mathbbm{Z}_2$ graded vector space, which means that it comes with an operator $Z$ which squares to the identity (the fermion parity operator), and partly induces a canonical choice of basis such that basis vectors are eigenstates of $Z$. States which are eigenstates of $Z$ are called homogeneous states, and the parity of a homogeneous state $|i\rangle$ is denoted as	
	\begin{equation}
		Z|i\rangle = (-1)^{|i|}|i\rangle\,,\; |i|\in\{0,1\}\, .
	\end{equation}
	The subspace of $V$ spanned by the vectors which are even under $Z$ is denoted as $V^0$, and is called the even subspace. The odd subspace is denoted as $V^1$. The properties of a super vector space $V$ naturally carry over to its dual space $V^*$, which is also graded. The natural action of dual vectors on vectors gives rise to the evaluation map $\mathcal{C}$, and we can choose a canonical dual basis $\langle i |$ such that
	\begin{equation}\label{C}
		\mathcal{C}:V^*\otimes V\rightarrow \mathbbm{C}: \langle i|\otimes |j\rangle \rightarrow \delta_{ij} \, .
	\end{equation}
    It then also follows that $\langle i | Z = (-1)^{|i|} \langle i |$ so that $|i\rangle$ and $\langle i |$ have the same parity. The $\mathbbm{Z}_2$ grading induced by $Z$ becomes important when vectors are `reordered'. More precisely, when working with tensor products of super vector spaces, one always uses the following canonical isomorphism,	
	\begin{equation}\label{F}
		\mathcal{F}:V\otimes W\rightarrow W\otimes V: |i\rangle \otimes |j\rangle \rightarrow (1)^{|i||j|} |j\rangle \otimes |i\rangle\,,
	\end{equation}
	which encodes the fermionic anticommutation relations. The same reordering rule is used when one or two of the vectors involved are dual vectors. \emph{Tensor contraction} is then defined as the following sequence of steps: (1) take the tensor product of the tensors to be contracted, (2) use $\mathcal{F}$ to bring the vectors and dual vectors corresponding to the legs which are to be contracted next to each other, and (3) use the evaluation $\mathcal{C}$ as defined in Eq.~\eqref{C} to contract the legs. This procedure is unambiguous up to an innocuous overall minus sign as long as the tensors respect the superselection rule which comes with super vector spaces: all tensors need to have a well-defined fermion parity. For more details we refer to Ref. \cite{Bultinck2016}.
	
	The GfTNS introduced in the main text are a special case of the general fTNS defined in terms of super vector spaces. To unveil the connection, we have to translate the Gaussian formalism based on Grassmann numbers to the language of super vector spaces. This translation is based on an isomorphism between polynomials of $M$ Grassmann numbers and a super vector space of dimension $2^M$. To illustrate how this isomorphism works, consider the case of a single Grassmann number $\theta$. To every monomial we associate a basis state of the super vector space as follows,	
	\begin{equation}
		\theta^n \cong |n\rangle\, ,\quad n\in \{0,1\} \, .
	\end{equation}
	The dual space is isomorphic to polynomials of another Grassmann number $\bar{\theta}$,	
	\begin{equation}
		\bar{\theta}^n \cong \langle n|\,,\quad n\in\{0,1\} \, .
	\end{equation}
	The evaluation map is then given by the following Berezin integral, 	
	\begin{equation}
		\mathcal{C}: \langle n|\otimes |m\rangle \cong \bar{\theta}^n\theta^m\rightarrow \int \mathrm{d}\theta \int\mathrm{d}\bar{\theta} \,e^{\bar{\theta}\theta} \bar{\theta}^n\theta^m = \langle n|m\rangle = \delta_{nm} \, .
	\end{equation}
	This explains the presence of the factors $e^{\bar{\theta}_\x\theta_{\x+\e_{x/y}}}$ in the definition of the GfTNS in Eq.~\eqref{contraction}: they ensure that the Berezin integral implements the tensor contraction according to the conventional evaluation map of the super vector spaces associated with the legs of the fermionic tensors. The canonical isomorphism $\mathcal{F}$ defined in Eq.~\eqref{F} is implemented automatically via the anticommutation relations of Grassmann numbers. The mapping of monomials of Grassmann numbers to basis states of a super vector space generalizes straightforwardly to the case with more than one Grassmann number. 
 
	\section{Appendix I -- Connection between the Gu-Verstraete-Wen and Kraus-Schuch formalisms for GfTNS}
	
	In this work, two different Gaussian \textit{Ans\"atze} were applied to approximate critical fermion states with a Fermi surface. Here, we establish a connection between both in the 1D setting. To do so, it is important to reflect on the conceptual differences between a GVW state $\ket{\psi}$ (Eq.\,\eqref{contraction} in the main text) and the Kraus-Schuch \textit{Ansatz} $\rho_\text{out} = \mathcal{E}\left(\rho_\text{in}\right)$ (as defined in Appendix C). An immediate observation is that the latter allows for mixed states (when $X X^T < \mathbbm{1}$) while the former is a pure state by construction. As a result, we only intend to establish a connection in the pure case, i.e.~when $\rho_\text{out}=\ket{\psi_\text{out}} \bra{\psi_\text{out}}$. Another direct observation is that the GVW \textit{Ansatz} always contains a vacuum contribution in its local tensors due to the exponential in $\hat{T}_\mathbf{x}$ (Eq.\,\eqref{Tx} in the main text). As a result, $\ket{\psi}$ will also contain the vacuum, precluding a $\ket{\psi_\text{out}}\perp \ket{0}$ (e.g.~an odd-parity $\ket{\psi_\text{out}}$). This can be solved by the addition of virtual Grassmann numbers to the exponential in the local tensor or to the overall contraction (e.g.~as discussed in Appendix A). Here, we will only focus on cases where $\ket{\psi_\text{out}} \not\perp \ket{0}$ and is thus an even state. A final, more high-level difference between both formalisms lies in their general architecture. While the Kraus-Schuch formalism starts from an input state to which a local Gaussian projection is applied to realise the output state, the GVW \textit{Ansatz} starts from local entities and contracts these in a conventional way with the Berezin integral. Schematically, they can be contrasted as
	\begin{equation*}
		\ket{\psi_\text{out}} = \diagram{appendix}{1} \quad,\quad \text{respectively} \quad ,\quad \ket{\psi} = \diagram{appendix}{2}
	\end{equation*}
	where $A_\mathbf{x} \cong \hat{T}_\mathbf{x}$ in the sense of Appendix G. In these diagrams, the outward (inward) arrows differentiate between (co)vectors. Furthermore, we assumed purity of $\rho_\text{out}$ so that $T_\mathcal{E}$ is the single Kraus operator of the local channel $\mathcal{E}^\text{loc}$. To establish a connection, we rephrase the Kraus-Schuch formalism as in Ref.~\cite{Schuch2019}, where one starts from a local Gaussian state, $\rho^\text{loc}$, defined on both the virtual and physical level. Due to Gaussianity, this state can be characterized by a correlation matrix $\gamma^{ij} = \frac{i}{2} \text{tr} \left(\rho^\text{loc} \left[c^i,c^j\right]\right)$ with $\{c^i\}$ the local Majorana operators. Discerning between physical (placed first) and virtual (placed last) Majorana operators, we rewrite this real antisymmetric matrix as $\gamma=\begin{pmatrix}
		A & B\\
		C & D
	\end{pmatrix}$ with $\gamma\gamma^T = \mathbbm{1}$ (again assuming purity). Next, we make $\rho_\omega$, a simple Gaussian state on two fermions with correlation matrix $\gamma_\omega=\begin{pmatrix}
		0 & -\sigma_x\\
		\sigma_x & 0
	\end{pmatrix}$. This state is pure and $\rho_\omega=\ket{\omega}\bra{\omega}$ with $\ket{\omega} = \frac{1}{\sqrt{2}}\left(\ket{0}\ket{0}+\ket{1}\ket{1}\right)$. $N_s$ copies of $\rho^\text{loc}$ placed on a line can be described in Fourier space by $G_\rho = \bigoplus_\mathbf{k} \gamma$. Identifying the spaces of $\ket{\omega}$ with two neighbouring virtual spaces in $(\rho^\text{loc})^{\otimes N_s}$ and doing this for each neighbouring pair, $\rho_\omega^{\otimes (\frac{\chi}{2} N_s)}$ is described by $G_\omega = \bigoplus_k \begin{pmatrix}
		0 & - e^{-ik} \sigma_x\\
		e^{ik} \sigma_x& 0
	\end{pmatrix}^{\oplus \frac{\chi}{2}}$.
	One can then construct the partial trace over the virtual spaces of the composition of these two density operators, $\text{tr}_\text{virt}\left((\rho^\text{loc})^{\otimes N_s} {\rho_\omega}^{\otimes(\frac{\chi}{2} N_s)}\right)=|\braket{\omega^{\otimes(\frac{\chi}{2} N_s)}|(\psi^\text{loc})^{\otimes N_s}}_\text{virt}|^2$, yielding
	\begin{equation*}
		\diagram{appendix}{3} \quad ,
	\end{equation*}
	where the purity of $\rho^\text{loc}=\ket{\psi^\text{loc}}\bra{\psi^\text{loc}}$ was used to only consider the ket layer. Essentially, we use $\bra{\omega}$ to glue the local tensors together like the Berezin integral. Expressing $(\rho^\text{loc})^{\otimes N}$ and $\rho_\omega^{\otimes(\frac{\chi}{2} N)}$ in Grassmann numbers as in \cite{Bravyi2005}, this partial trace can be rewritten as a Gaussian integral, yielding
	\begin{equation}
		G(\mathbf{k}) = A - B(D+{G_\omega}(\mathbf{k}))^{-1}{C}
	\end{equation}
	for the resulting Gaussian state. Both formulations of the the Kraus-Schuch \textit{Ansatz} thus result in a very similar Schur complement formula and can be converted into each other by requiring ${G_\omega}(\mathbf{k})=-{G_\text{in}}(\mathbf{k})$. Some minor remarks are in order. While we discussed this reinterpretation in 1D, it can be readily extended to a general spatial dimension $d$. An odd number of virtual Majorana pairs per bond (i.e.~an odd $\chi_i$) constitutes an important exception. In this case, one cannot simply define $\ket{\psi^\text{loc}}$ with full ($D=2$) virtual orbitals and combine these with a conventional $\bra{\omega}$ as this would immediately lead to a doubling of the entangled Majorana pairs. 
	
	We conclude that a direct conversion of the Kraus-Schuch into the GVW \textit{Ansatz} is only possible for pure states, non-orthogonal to the vacuum and for bond dimensions that are powers of 2. Under these conditions, we can reinterpret the Kraus-Schuch \textit{Ansatz} and equate
	\begin{equation*}
		\diagram{appendix}{4} = \diagram{appendix}{5} \qquad \text{or equivalently} \qquad \diagram{appendix}{6} = \diagram{appendix}{7} = \frac{1}{Z} \exp\left(\frac{1}{2} (f^\dagger)^T A f^\dagger \right) \ket{0}
	\end{equation*}
	where $Z$ is a normalization constant and where $f^\dagger$ collects the creators (e.g.~$f^\dagger = \left(f^{\dagger^T}_p \quad f^{\dagger^T}_l \quad f^{\dagger^T}_r\right)^T$ in case we order the operators as in Eq.\,\eqref{Tx}, i.e.~physical, left,right). Relating both GfTNS constructions thus boils down to relating the pure (all ket) $\ket{\psi^\text{loc}}$ with correlation matrix $\gamma$ to the exponential creator parametrized by $A$ which is exactly the same $A$ as in Eq.\,\eqref{Tx} when the operator order is chosen accordingly. This final step can be taken relatively easily as states of the latter kind have been studied extensively \cite{Perelomov1986}, where for the hopping and pairing expectation values it is derived that
	\begin{equation}
		n_{ij} = \braket{f^\dagger_i f_j} = \left(A(1+A^\dagger A)^{-1}A^\dagger\right)_{ji} \qquad \qquad x_{ij} = \braket{f_i f_j} = -\left(A(1+A^\dagger A)^{-1}\right)_{ij}
	\end{equation}
	while $Z= \det \left(1+A A^\dagger\right)^{-\frac{1}{4}}$. Note that $n^T x = x n$ and $\left(n^T-\frac{1}{2}\right)^2-x\bar{x} = \frac{1}{4}$ while $n^T = -x A^\dagger$. The inverse is also true. When $n$ and $x$ satisfy these equations, the matrix $A$ can be obtained from the latter such that $n$ and $x$ can be expressed as a function of this $A$ or, equivalently, such that there exists a GVW parametrization for these states. It is straightforward to check that the former two equations for $n$ and $x$ just express purity while the latter can always be solved with the (pseudo)inverse of $x$ as long as the Gaussian state described by $n$ and $x$ is not orthogonal to the vacuum. Note that relating the states in this way, the main computational effort is the solution of the linear system. Hence, there is no exponential scaling in $\chi_i$, which would happen if the tensor entries would have been evaluated as in Appendix G.\\
	
	We conclude this Appendix by summarizing the recipe to convert a pure Kraus-Schuch state, non-othogonal to vacuum and with bond dimension $D=2^M$, to a GVW GfTNS (or vice-versa) and apply it to an example.
	\begin{itemize}
		\item $\rho_\text{out}$ is built from  $\mathcal{E}$ and $\rho_\text{in}$, respectively characterized by $X^\text{loc}$ and $G_\text{in}(\mathbf{k})$. Reinterpret this state as a contraction of local tensors, i.e.~in terms of a local $\gamma$ and $\ket{\omega}$.
		\item Extract from $\gamma$ the hopping and pairing expectation values $n$ and $x$.
		\item Solve $n^T = -x A^\dagger$ to find $A$ and hence $\hat{T}_\mathbf{x}$ from Eq.\,\eqref{Tx}.
	\end{itemize}
	
	As an example, consider a $D=2$ GfMPS, $\rho_\text{out}$, in the Kraus-Schuch formalism with
	\begin{equation}
		X^\text{loc} = \left(\begin{array}{cc|cccc}
			0 & 0 & -1 & 0 & 0 & 0\\
			0 & 0 & 0 & 1 & 0 & 0\\
			\hline
			1 & 0 & 0 & 0 & 0 & 0\\
			0 & -1 & 0 & 0 & 0 & 0\\
			0 & 0 & 0 & 0 & 0 & 1\\
			0 & 0 & 0 & 0 & -1 & 0\\
		\end{array}\right) = \left(\begin{array}{c|c}
			A^\text{loc} & B^\text{loc} \\
			\hline
			-{B^\text{loc}}^T & D^\text{loc}
		\end{array}\right)
	\end{equation}
	where we order the operators as physical, left, right and with $G_\text{in}(k)$ as in Eq.\,\eqref{Gin} where $\chi = 2$. Note that the state is pure as $X^\text{loc}{X^\text{loc}}^\dagger = \mathbbm{1}$. Its output correlation matrix is given by the Schur complement formula,
	\begin{equation}
		G_\text{out}(k)=A^\text{loc}+B^\text{loc}\left(D^\text{loc}-G_\text{in}(k)\right)^{-1}{B^\text{loc}}^T = \begin{pmatrix}
			0 & e^{ik}\\
			-e^{-ik} & 0
		\end{pmatrix} \, .
	\end{equation}
	Using that $G_\text{in}(k) = -P^T G_{\omega}(k)P$ with
	\begin{equation}
		G_{\omega}(k) = \begin{pmatrix}
			0 & - e^{-ik} \sigma_x\\
			e^{ik} \sigma_x& 0
		\end{pmatrix} \qquad \text{and} \qquad P = \begin{pmatrix}
			0 & 1 & 0 & 0\\
			0 & 0 & 0 & 1\\
			0 & 0 & -1 &0\\
			-1& 0 & 0 & 0	
		\end{pmatrix}
	\end{equation}
	we can rewrite this formula as
	\begin{equation}
		G_\text{out}(k)=A^\text{loc}+B^\text{loc} P^T \left(P D^\text{loc} P^T+G_{\omega}(k)\right)^{-1}P {B^\text{loc}}^T
	\end{equation}
	and thus reinterpret the \textit{Ansatz} as a contraction of local Gaussian tensors with correlation matrix
	\begin{equation}
		\gamma = \begin{pmatrix}
			A^\text{loc} & B^\text{loc} P^T\\
			P {B^\text{loc}}^T & P D^\text{loc} P^T
		\end{pmatrix} 
	\end{equation}
	by means of the conventional $\ket{\omega}$. From this $\gamma$ we obtain
	\begin{equation}
		n = \frac{1}{2} \begin{pmatrix}
			1 & \frac{1}{2} & -\frac{1}{2}\\
			\frac{1}{2} & 1 & \frac{1}{2}\\
			-\frac{1}{2} & \frac{1}{2} & 1
		\end{pmatrix} \qquad \text{and} \qquad x = \frac{1}{4} \begin{pmatrix}
			0 & -1 & -1\\
			1 & 0 & -1\\
			1 & 1 & 0
		\end{pmatrix}
	\end{equation}
	allowing to solve for the GVW matrix $A$ from $n^T = -x A^\dagger$. While $x$ is not invertible, application of its pseudoinverse yields $A = -4x$, solving the system and signaling that the original state was non-orthogonal to the vacuum. Note that this parameter matrix is the same as in Eq.\,\eqref{degenexp} so that for the GVW state we immediately obtain,
	\begin{equation}
		\begin{split}
			n(k) &= \frac{|g_k|^2}{1+|g_k|^2} = \frac{\sin^2 k}{2(1-\cos k )}\\
			x(k) &= \frac{g_k}{1+|g_k|^2} = -\frac{i}{2} \sin k		
		\end{split} \qquad \text{since} \qquad g_k = -i \frac{\sin k}{1- \cos k} = -i \, \text{cotan}\left(\frac{k}{2}\right)\, .
	\end{equation}
	Compare these to their Kraus-Schuch analogues,
	\begin{equation}
		\begin{split}
			n(k) &= \frac{1}{2} + \frac{i}{4} W^T G^T_\text{out}(k) \overline{W} = \frac{1}{2}(1+ \cos k)\\
			x(k) &= -\frac{i}{4} W^\dagger G_\text{out}(k) \overline{W} =-\frac{i}{2} \sin k		
		\end{split}\, .
	\end{equation}
	Indeed, both Gaussian \textit{Ans\"atze} yield exactly the same correlation matrices and are hence equivalent.
			
\end{document}